\documentclass[a4paper,11pt]{article}

\usepackage{jheppub} 
\usepackage{tikz}
	
\usepackage{amsmath}
\usepackage{amssymb}
\usepackage{braket}
\usepackage{comment}
\usepackage[many]{tcolorbox}
\usepackage{enumerate}
\usepackage{enumitem}
\usepackage{hyperref}
\usepackage[normalem]{ulem}

\newcommand{\llangle}{\langle\!\langle}
\newcommand{\rrangle}{\rangle\!\rangle}

\newtcolorbox{empheqboxed}{colback=gray!30, 
 colframe=white,
 width=\textwidth,
 sharpish corners,
 top=-2mm, 
 bottom=0mm
}

\newcommand{\TT}{T\overline{T}}

\newcommand*{\Scale}[2][4]{\scalebox{#1}{\ensuremath{#2}}}%

\title{$\boldsymbol{T\overline{T}}$ correlators from tensionless strings}

\abstract{Motivated by earlier approaches, we develop a worldsheet framework for computing correlation functions in the single trace $\TT$-deformed tensionless AdS$_3$/CFT$_2$ duality. By describing the deformed bulk theory as a Berkovits-Vafa $\mathcal N=4$ topological string, we obtain a consistent definition of physical states and correlation functions, yielding a tractable setup for testing aspects of holography beyond AdS/CFT. We construct deformed physical vertex operators and compute their tree-level two-point functions exactly. We discuss the relation of our results to previous proposals for $\TT$-deformed two-point functions obtained from alternative worldsheet approaches, JT gravity, and perturbative field theory computations.}

\author[a]{Andrea Dei,}
\author[b]{Kiarash Naderi}
\affiliation[a]{\vskip 0.01cm
Leinweber Institute for Theoretical Physics \& Enrico Fermi Institute \& Kadanoff Center for Theoretical Physics, University of Chicago, 5640 S. Ellis Ave., Chicago, IL 60637, USA}
\affiliation[b]{Abdus Salam Centre for Theoretical Physics, The Blackett Laboratory,
Imperial College London, Prince Consort Road, London, SW7 2AZ, UK}
\emailAdd{adei@uchicago.edu}
\emailAdd{k.naderi@ic.ac.uk}

\begin{document}

\maketitle

\section{Introduction and summary of results}

\noindent \underline{\it Motivation}

\smallskip

Holography is the principle that a quantum theory of gravity in a higher-dimensional space-time can be recast as a lower-dimensional quantum field theory where dynamical gravity is absent. In its original formulation by ’t Hooft and Susskind \cite{tHooft:1993dmi, Susskind:1994vu}, the holographic principle is quite general, and one may expect it to hold in any quantum theory of gravity containing black holes. In practice, much of the literature over the last three decades has focused on the specific incarnation where the bulk geometry is Anti-de Sitter and the boundary field theory is conformal \cite{Maldacena:1997re}. This focus reflects the fact that the AdS/CFT correspondence provides an especially well-defined and controllable framework, in which many computations can be carried out explicitly and the duality can be tested in detail. 

Two of the main hurdles one may encounter when exploring holography beyond the AdS/CFT paradigm are the reduced amount of symmetry and the lack of locality of the boundary theory.
String theory is typically under good theoretical control for a limited class of highly symmetric backgrounds or deformations thereof. Moreover, conformal symmetry, often accompanied by supersymmetry, frequently facilitates the computation of boundary observables.

A local $d$-dimensional QFT has, at high temperature, entropy scaling with volume $V$ and temperature $T$ as $S \sim V T^{d-1}$, and hence a microcanonical growth at high energies $E$ of the form $S \sim E^{\frac{d-1}{d}}$. This behavior is compatible with the scaling of large black holes in asymptotically AdS backgrounds. By contrast, in backgrounds with a faster growth of the density of states, such as asymptotically flat or asymptotically linear dilaton geometries, the possibility of an ordinary local QFT dual is ruled out by this mismatch in the density of states.

\bigskip

In recent years, the rapid development of two areas of the high energy literature has provided a new handle on the study of the holographic principle beyond the AdS/CFT paradigm. The AdS$_3$/CFT$_2$ correspondence (see \cite{Kovensky:2026usc} for a recent review) has now reached a level of understanding at which non-protected observables can be computed independently in the bulk and on the boundary and shown to agree. The matching is particularly impressive in the tensionless $k=1$ regime, relating \cite{Eberhardt:2018ouy,Eberhardt:2019ywk}
\vspace{8pt}
\begin{equation}
\begin{tikzpicture}[baseline = -2ex]
\node[inner sep=0pt] at (1.5,0)
{Pure NS-NS strings on}; 
   \node[inner sep=0pt] at (1.5,-0.5)
{$\text{AdS}_3 \times \text{S}^3 \times \mathbb T^4 $ with $k=1$}; 
   
\node[inner sep=0pt] at (5.7,-0.3)
   {$\Scale[2]{\iff} $};    
 
\node[inner sep=0pt] at (8.6,-0.2)
   {{$\text{Sym}^N (\mathbb T^4)$ }};  

\end{tikzpicture}
\label{ads3/cft2} \ .
\end{equation}
\vskip 0.1in
\noindent In fact, the large number of precise holographic matches indicates that in this regime the AdS$_3$/CFT$_2$ correspondence is solved, at least at tree level in the string coupling.\footnote{Although the localization of tensionless higher-genus correlators has been argued for in \cite{Eberhardt:2020akk,Knighton:2020kuh}, an exact holographic match of their full functional dependence has not yet been achieved.} This degree of control suggests that deformations of the tensionless duality that break conformal symmetry on the boundary may now be within reach as never before. 

The other rapid recent development is the so-called $\TT$ deformation of two-dimensional quantum field theory~\cite{Zamolodchikov:2004ce, Smirnov:2016lqw, Cavaglia:2016oda}, see \cite{Jiang:2019epa, He:2025ppz, Guica:2025jkq} for reviews. This is an irrelevant deformation for which the deformed spectrum can be written in closed form, not only for small values of the deformation parameter, but exactly as a function of it. One may then wonder whether these two developments can be combined, and whether a $\TT$ deformation of the CFT$_2$ dual to tensionless AdS$_3$ strings might lead to an exact realization of holography beyond AdS/CFT, one that can be solved in full and that may shed light on the inner workings of holography when the boundary theory is neither conformal nor local. 

\bigskip 

Holographic applications of $\TT$ deformations of two-dimensional CFTs can, at a rough level, be divided into two classes. One is the \emph{double-trace} deformation, which perturbs the full boundary orbifold and, at leading order in the deformation parameter, is of the form 
\begin{equation}
    \sum_{i = 1}^N \sum_{j=1}^N T_i \overline{T}_j \ , 
\end{equation}
where $T_i$ and $\overline{T}_i$ are the holomorphic and anti-holomorphic components of the stress-tensor of $\mathbb T^4$, respectively. See e.g.~\cite{McGough:2016lol,Kraus:2018xrn,Cottrell:2018skz,Guica:2019nzm,Hirano:2020nwq,Kawamoto:2023wzj,Apolo:2023vnm,Blacker:2024rje} for interesting recent developments. The other is the \emph{single-trace} deformation, which acts instead on the orbifold \emph{seed} theory $\mathbb{T}^4$ and, at first order, takes the form
\begin{equation}
    \sum_{i=1}^N T_i \overline{T}_i \ . 
\end{equation}
In recent years, various pieces of evidence have suggested a holographic relation between $\TT$ and linear dilaton backgrounds \cite{Giveon:2017nie,Giveon:2017myj,Chang:2023kkq,Dei:2025ryd}, and between the single-trace deformation of the symmetric orbifold of $\mathbb T^4$ and current-current or TsT deformations of AdS$_3$ strings at generic tension $k>1$ \cite{Giveon:2017nie, Giveon:2017myj, Asrat:2017tzd, Araujo:2018rho, Chakraborty:2019mdf, Hashimoto:2019wct, Hashimoto:2019hqo, Apolo:2019zai, Chakraborty:2020yka, Apolo:2021wcn, Demise:2021cfx, Georgescu:2022iyx}. At the same time, it has by now become clear that the CFT$_2$ dual to AdS$_3$ strings at $k>1$ is \emph{not} a symmetric orbifold \cite{Balthazar:2021xeh,Eberhardt:2021vsx}, since, for example, three-point function selection rules do not match \cite{Dei:2021xgh}. As a result, the very definition of single-trace deformation loses meaning away from the tensionless $k=1$ regime, where the boundary theory is indeed the symmetric orbifold of $\mathbb T^4$ and therefore admits a consistent definition of the single-trace $\TT$ deformation.\footnote{See \cite{Dei:2024sct} for a more extensive discussion of this point.} Nevertheless, some of the internal structure of $k>1$ AdS$_3$ strings is shared by the tensionless $k=1$ regime. This leads us to believe that several results originally derived for bosonic $k>1$ strings, see in particular \cite{Cui:2023jrb}, although not precisely defined away from the symmetric orbifold point, may still capture the correct physics and admit a precise formulation at $k=1$, where the duality becomes exact. 

As already mentioned, $\TT$-deforming the tensionless duality may provide an exactly solvable incarnation of holography beyond AdS/CFT, despite the lack of conformal symmetry and locality in the boundary theory. This expectation was the main driving force behind \cite{Dei:2024sct}: together with our co-authors and building on previous $k>1$ literature \cite{Giveon:2017nie,Giveon:2017myj,Asrat:2017tzd}, we proposed the holographic correspondence
\vspace{8pt}
\begin{equation}
\begin{tikzpicture}[baseline = -0.6ex]
\node[inner sep=0pt] at (1.5,0.5)
   {\large{$J^+\bar{J}^+$ deformation of}};
\node[inner sep=0pt] at (1.5,0)
{pure NS-NS strings on}; 
   \node[inner sep=0pt] at (1.5,-0.5)
{$\text{AdS}_3 \times \text{S}^3 \times \mathbb T^4 $ with $k=1$}; 
   
\node[inner sep=0pt] at (5.2,-0.1)
   {$\Scale[2]{\iff} $};    

\node[inner sep=0pt] at (8.6,+0.3)
   {{Single-trace $\TT$-deformed}};  
\node[inner sep=0pt] at (8.6,-0.3)
   {{$\text{Sym}^N (\mathbb T^4)$ }};  

\end{tikzpicture}    \quad \ ,  
\label{ads3/cft2-deformed}
\end{equation}
\vskip 0.1in
\noindent where $J^+$ denotes the raising operator of the $\mathfrak{sl}(2,\mathbb R)_1$ affine algebra on the worldsheet and $\bar J^+$ its anti-holomorphic analogue. 

In \cite{Dei:2024sct}, the duality \eqref{ads3/cft2-deformed} was tested at the level of the spectrum. Using the holographic dictionary of the tensionless string, it was shown that the single-trace $\TT$ operator is indeed dual to the $J^+ \bar J^+$ deformation on the worldsheet, and that the bulk string theory partition function reproduces the torus partition function of the $\TT$-deformed boundary field theory. 

In the present paper, we take a further step toward a complete solution of the bulk string theory and we investigate the duality \eqref{ads3/cft2-deformed} at the level of correlation functions.

 \bigskip

\noindent \underline{\it $\TT$ correlation functions and holography}

\smallskip

The UV interpretation of $\TT$-deformed theories has been approached from several complementary viewpoints. One class of approaches relates the deformation to two-dimensional gravity \cite{Dubovsky:2017cnj, Cardy:2018sdv, Dubovsky:2018bmo, Tolley:2019nmm}, while in holographic single-trace settings the deformation is closely tied to string theory in asymptotically linear dilaton backgrounds and to Little String Theory \cite{Aharony:1998ub,Giveon:2017nie,Giveon:2017myj}. A holographic interpretation of the deformation may help sharpen this picture, and could provide a first-principles definition through the holographic dictionary and the stringent consistency conditions of string theory.

Independently of how one views the high-energy completion of $\TT$-deformed theories, a viable bulk dual should satisfy non-trivial consistency requirements. In particular, any bulk string theory dual to the single-trace $\TT$-deformed symmetric orbifold should reproduce several characteristic features of the deformation in order to qualify as such and to furnish a plausible holographic definition of $\TT$.

First, in the untwisted sector the bulk string theory should reproduce the square-root formula for the deformed energy levels characteristic of $\TT$ deformed CFTs \cite{Zamolodchikov:2004ce, Smirnov:2016lqw, Cavaglia:2016oda}. This has been achieved in the $J^+ \bar J^+$ deformation of $k>1$ AdS$_3$ \cite{Giveon:2017myj}, in the TsT-deformed $k>1$ AdS$_3$ approach of \cite{Apolo:2019zai}, and in the single brane string theory of \cite{Dei:2025ilx}. A more stringent requirement, which in particular implies the previous one, is that a putative string dual to the single-trace deformation of the symmetric orbifold reproduce the $\TT$-deformed one-loop partition function \cite{Cardy:2018sdv, Dubovsky:2018bmo, Hashimoto:2019wct, Hashimoto:2019hqo, Apolo:2023aho} and capture, directly from the worldsheet, the modular properties characteristic of $\TT$-deformed CFTs \cite{Datta:2018thy, Aharony:2018ics}. This was indeed achieved in \cite{Dei:2024sct}, thereby providing strong support for the duality \eqref{ads3/cft2-deformed}.

Second, one expects the bulk string theory on the left-hand-side of \eqref{ads3/cft2-deformed}, or more generally any candidate string dual to the single-trace deformation, to reproduce boundary $\TT$-deformed correlation functions. A consistency check is whether these deformed correlators are compatible with the $\TT$ flow. In particular, by resumming leading logarithmic divergences, Cardy derived in \cite{Cardy:2019qao} a partial differential equation, valid to all orders in perturbation theory, which we refer to as Cardy's Callan-Symanzik equation, that $\TT$-deformed correlators in momentum space are expected to satisfy.\footnote{Given the non-local nature of the $\TT$ deformation, it is often convenient to study correlation functions in momentum space.} A concrete criterion for testing a putative string theory dual to the single-trace $\TT$-deformed symmetric orbifold is therefore to verify whether its correlation functions satisfy Cardy's differential equation.

Cardy's Callan-Symanzik equation, however, does not determine correlation functions uniquely, not even in the case of two-point functions. In fact, several incompatible proposals have appeared in the literature for the $\TT$-deformed two-point function of two-dimensional CFTs \cite{Giribet:2017imm, Asrat:2017tzd, Cardy:2019qao, Callebaut:2019omt, Hirano:2020nwq, Cui:2023jrb, Aharony:2023dod, Giveon:2023gzh}. Abstracting from the details of the particular operators studied in each proposal, the resulting momentum space two-point functions share a common structure that may be written schematically as\footnote{For all these proposals, once Fourier transformed, \eqref{generic-two-pointf} reduces to the corresponding two-point function in the undeformed CFT in the limit $\mu\to 0$.}
\begin{equation}  \label{generic-two-pointf}
    \left\langle \mathcal O_{h^0}^\mu(p, \bar p) \mathcal O_{h^0}^\mu(-p, -\bar p) \right\rangle_\mu
    =
    U(h^0,\mu^2 p \bar p)
    (p \bar p)^{2H(h^0, \, \mu^2 p \bar p) -1} \ ,
\end{equation}
where $(h^0,h^0)$ are the undeformed conformal weights of the quasi-primary field $\mathcal{O}_{h^0}^0$.
Here $U$ and $H$ are functions of two variables, $p$ and $\bar p$ are the holomorphic and anti-holomorphic components of the two-dimensional momentum and $\mu^2$ is the dimensionless $\TT$ deformation parameter. It is straightforward to check that, in order for eq.~\eqref{generic-two-pointf} to solve Cardy's differential equation, the function $H$ must be linear in both arguments and given by
\begin{equation} \label{H-is-linear}
    H(x,y) = x + y \ . 
\end{equation}
This observation appears in tension with the proposals of \cite{Giribet:2017imm,Asrat:2017tzd}, see \cite{Giveon:2023gzh, Chakraborty:2025wvr} for related discussions.

Cardy's Callan-Symanzik equation does not, however, fix the function $U$ entering eq.~\eqref{generic-two-pointf}. Indeed, in \cite{Cui:2023jrb} it was shown that the ansatz \eqref{generic-two-pointf} for a fixed value of $h^0$, with $H$ chosen as in \eqref{H-is-linear}, solves Cardy's differential equation for any smooth function $U$. The solution originally proposed by Cardy in \cite{Cardy:2019qao} corresponds to a particular choice within this family, namely $U(x,y) =1$. By contrast, the TsT-deformed AdS$_3$ analysis of \cite{Cui:2023jrb} for $k>1$ selects a different solution in the same class, namely
\begin{equation} \label{two-point-Cui}
    U(x,y) = \frac{\pi \, 2^{1-4x-4y} \, \Gamma(1-2x-2y)}{\Gamma(2x+2y)} \ , 
\end{equation}
leading to a distinct proposal for the $\TT$-deformed two-point function. 

In \cite{Aharony:2023dod}, Aharony and Barel emphasized the different large momentum behavior of the two-point function proposals of \cite{Cardy:2019qao} and \cite{Cui:2023jrb}: in the large momentum limit, the former diverges, whereas the latter generically tends to zero. They also showed that the large momentum behavior of $\TT$-deformed two-point functions can be derived from the relation between the $\TT$ deformation and JT gravity \cite{Dubovsky:2017cnj, Dubovsky:2018bmo}. Their result is compatible with the two-point function of \cite{Cui:2023jrb}, but not with the large  momentum behavior of the proposal of \cite{Cardy:2019qao}. 

Finally, a series of perturbative computations \cite{Cardy:2019qao,He:2019vzf,He:2019ahx,He:2020qcs,He:2020udl,Hirano:2020ppu,He:2023kgq,Hirano:2024eab,Ebert:2020tuy,Rosenhaus:2019utc,Menskoy:2024vqv} culminated in the derivation of the position space $\TT$-deformed genus-zero two-point function to all orders in the deformation parameter $\mu^2$ \cite{Hirano:2025tkq,Li:2026ecl}. This result is reported in eq.~\eqref{eq:perturbative-formula-ttbar}.\footnote{More accurately, eq.~\eqref{eq:perturbative-formula-ttbar} only captures the leading logarithmic contributions, see Section~\ref{sec:TT-correlator-pos-and-mom} for a discussion on this point.}

\bigskip

Given the existence of mutually incompatible proposals in the literature for $\TT$-deformed two-point functions, together with the scarcity of formulae for $\TT$-deformed higher-point functions that are exact in $\mu^2$,\footnote{See, however, \cite{He:2019ahx,He:2020qcs,Hirano:2020ppu,He:2023kgq,Hirano:2025tkq} for perturbative computations of three-point functions, and \cite{Callebaut:2019omt} for a proposal for an all order three-point function.} we believe that constructing an exactly solvable string theoretic framework may be particularly illuminating. Starting from the exact duality \eqref{ads3/cft2}, and extending the deformed correspondence \eqref{ads3/cft2-deformed} beyond the level of the spectrum, may help clarify aspects of the $\TT$ deformation, identify which of the existing two-point function proposals is compatible with the holographic interpretation of the single-trace $\TT$ deformation, and yield closed-form formulae for higher-point functions. Such a framework would also furnish an exactly solvable incarnation of holography beyond AdS/CFT, in which the non-locality of the boundary theory can be quantified directly through correlation functions.

\bigskip 

\noindent \underline{Towards the worldsheet dual of single-trace $\TT$}

\smallskip

The definition of the string dual to the single-trace deformation in eq.~\eqref{ads3/cft2-deformed}, given in terms of the exactly marginal worldsheet deformation that perturbs the tensionless worldsheet action, does not by itself provide a direct prescription for a computation of correlation functions that is exact in $\mu^2$. A natural route to such a prescription would be to construct a deformed string theory with the following ingredients:
\begin{itemize}
    \item An explicit condition that defines the set of physical states in terms of the fundamental worldsheet fields,
    \item A consistent definition of correlation functions that is compatible with the physical state conditions.
\end{itemize}    
These ingredients were not needed for the computation and holographic matching of the partition function in \cite{Dei:2024sct}, but they seem necessary before a worldsheet prediction for $\TT$-deformed correlators can be formulated. This is the problem addressed in the present work: we formulate the physical state conditions and correlation function prescriptions directly at the tensionless point, where the boundary theory is a symmetric orbifold CFT and the single-trace $\TT$ deformation has a direct definition.

Given that the proposed dual is formulated as a deformation of the tensionless string, which is most naturally defined in the hybrid formalism \cite{Berkovits:1999im,Eberhardt:2018ouy}, we adopt a topological string description of the string dual of the single-trace deformation \cite{Witten:1988xj,Berkovits:1993xq,Berkovits:1994vy,Ooguri:1995cp}. More specifically, the structures listed above are built out of a worldsheet topologically twisted $\mathcal N=4$ algebra with central charge $\mathtt c = 6$ \cite{Berkovits:1993xq,Berkovits:1994vy}. 

\bigskip 

\noindent \underline{Summary of results}

\smallskip

Let us briefly summarize our findings and discuss how the paper is organized. We begin in Section~\ref{sec:tensionless} with a brief review of several ingredients entering the construction of the tensionless string, which are used throughout the paper. Readers familiar with this literature may skip this section.

In Section~\ref{sec:auxiliary-duality}, we introduce an auxiliary construction that will serve as a convenient worldsheet framework for formulating the deformed theory and for defining its correlation functions. More precisely, we construct an `auxiliary' string theory obtained by enlarging the field content of the tensionless string worldsheet CFT by an additional zero central charge sector. This additional zero central charge sector is naturally motivated by the idea of supplementing the worldsheet fields with two free bosons, a feature common to many previous worldsheet approaches to $\TT$, together with the requirement of preserving worldsheet supersymmetry. We propose that this auxiliary string theory is holographically dual to the symmetric orbifold of $\mathbb T^4 \times \mathcal A_0$, where $\mathcal A_0$ denotes a free field conformal field theory with vanishing central charge.  While the auxiliary duality proposed here has a different purpose from the earlier worldsheet constructions of \cite{Giveon:2017myj, Martinec:2017ztd, Gaberdiel:2022als, Dei:2025ilx, Eberhardt:2025sbi}, some of its ingredients have close counterparts in those works.
In particular, the extra zero central charge sector on the worldsheet coincides with the one introduced in \cite{Gaberdiel:2022als} and \cite{Dei:2025ilx}. Moreover, through a suitable bosonization map, it can be related to the zero central charge sector built from four free symplectic bosons and fermions discussed in Section~4 of \cite{Eberhardt:2025sbi}. In \cite{Giveon:2017myj}, two auxiliary bosons are also introduced; however, their $k>1$ construction is not supersymmetric and no additional auxiliary fields are included. Finally, in \cite{Martinec:2017ztd}, auxiliary bosons and auxiliary fermions are introduced in the context of gauged Wess-Zumino-Witten models at $k>1$.

To support the proposed auxiliary duality, we match the bulk and boundary partition functions and construct, from the worldsheet, the DDF operators associated with each of the fields defining the symmetric orbifold of $\mathbb T^4 \times \mathcal A_0$. We conclude this section by explaining how the tensionless duality embeds into this auxiliary holographic duality: every physical operator in the tensionless string can be mapped unambiguously to a physical operator of the auxiliary string theory, and similarly every vertex operator of the symmetric orbifold of $\mathbb T^4$ can be associated to a vertex operator of the auxiliary boundary CFT. 

The utility of the auxiliary duality becomes apparent in Section~\ref{sec:deforming}, where we deform the bulk theory by the worldsheet exactly marginal deformation $J^+ \bar J^+$ and propose a string dual to the single-trace deformation of the symmetric orbifold of $\mathbb T^4 \times \mathcal A_0$. Since the tensionless duality is naturally embedded in the undeformed auxiliary duality of Section~\ref{sec:auxiliary-duality}, we expect that this construction provides a framework to compute from the worldsheet observables in the single-trace $\TT$ deformation of the symmetric orbifold of $\mathbb T^4$, at least at tree level in the string coupling. In particular, we construct an $\mathcal N=4$ algebra on the worldsheet, give an explicit construction for the vertex operators dual to the $\TT$-deformed boundary ground state and space-time $\mathcal{R}$-symmetry generators, and investigate the space-time global symmetries from the worldsheet. As mentioned, some elements of our construction are reminiscent of earlier $k>1$ constructions, in particular those of \cite{Giveon:2017myj, Asrat:2017tzd, Apolo:2019zai, Cui:2023jrb}.  

In Section~\ref{sec:TTbar-correlators}, we investigate $\TT$-deformed correlation functions from the boundary perspective. This section presents results that may be relevant to the broader study of $\TT$ deformations, independently of holographic applications and without relying on string-theoretic inputs. In Section~\ref{sec:TT-correlator-pos-and-mom}, we review the position space two-point function derived in \cite{Cardy:2019qao,Hirano:2025tkq,Li:2026ecl} to all orders in perturbation theory in the deformation parameter; see eq.~\eqref{eq:perturbative-formula-ttbar}. We then show that the Fourier transform of the momentum space two-point function proposed in \cite{Cui:2023jrb} agrees with eq.~\eqref{eq:perturbative-formula-ttbar}. Together with its consistency with the analysis of \cite{Aharony:2023dod}, this further strengthens the proposal of \cite{Cui:2023jrb} and shows that the position space perturbative solution of \cite{Hirano:2025tkq} satisfies Cardy's Callan-Symanzik equation. In Section~\ref{sec:KZ-solution} we speculate on the form of genus-zero $\TT$-deformed higher-point correlators. We argue that a particularly simple solution to Cardy's differential equation can be constructed from any momentum space genus-zero correlator of quasi-primary fields in a two-dimensional CFT by shifting the conformal weights of the operators as $h^0 \mapsto h^0 + \mu^2 p \bar p$.

Although the two-point function of \cite{Cui:2023jrb} passes various perturbative checks, and is compatible with the analysis of \cite{Aharony:2023dod}, the string theory derivation there was performed away from the tensionless \(k=1\) point. Since the CFT$_2$ dual to AdS$_3$ strings at \(k>1\) is known not to be a symmetric orbifold CFT, that computation does not by itself constitute a derivation of the correlator in the single-trace \(T\overline T\) deformation of a symmetric orbifold. It is therefore important to derive the same observable directly at \(k=1\), where the undeformed boundary theory is the symmetric orbifold and the single-trace deformation has a direct definition.

In Section~\ref{sec:corr}, we define correlation functions, explain how exact states decouple and how they become independent of picture assignment. A key difference with the approach of \cite{Cui:2023jrb} is that our framework first identifies the relevant external operators as physical superstring states in the proposed BRST cohomology, and then computes their correlators using the correlation function prescription defined in Section~\ref{sec:corr}. 

After some warm-up examples, we compute from the worldsheet the tree-level two-point function of $\TT$-deformed states identified in Section~\ref{sec:vertex}. We find that the exact $k=1$ result precisely reproduces eqs.~\eqref{generic-two-pointf}--\eqref{two-point-Cui}. Thus the \(k=1\) computation selects the same two-point function as \cite{Cui:2023jrb}; it also has the large momentum behavior discussed in \cite{Aharony:2023dod} and is compatible with the analysis of \cite{Chen:2025jzb}.

Several appendices spell out our conventions, review aspects of the tensionless duality, and explain some of the more technical ingredients of our construction.

\bigskip

\noindent \underline{\it Future directions}

\smallskip 

Let us briefly highlight a few future directions opened by our work, as well as some aspects that require further investigation.

\paragraph{Asymptotic symmetries.} It was argued in \cite{Georgescu:2022iyx,Chakraborty:2023mzc} that asymptotically linear dilaton backgrounds exhibit an asymptotic symmetry algebra given by a non-linear deformation of Virasoro. It would be interesting to understand whether this symmetry is present at the level of the string theory we discuss in Section~\ref{sec:deforming}. 
Useful insights in this direction may come from the worldsheet analysis of asymptotic symmetries developed in \cite{Du:2024bqk,Du:2024tlu}.

\paragraph{Higher-point functions.} The definitions we laid down in Section~\ref{sec:corr} provide all the necessary ingredients to compute higher-point correlation functions. In the two-point function computed in Section~\ref{sec:two-pt-deformed}, the background charge constraints imply that only the first terms in \(\Psi^L\) and $\Psi^R$ contribute. As a result, the final answer agrees with what one would obtain by simply shifting the spacetime weight in the undeformed two-point function. We suspect that this simplification is special to the two-point computation: for higher-point functions, background charge conservation does not in general decouple the additional  contributions we identify in \eqref{eq:psil-psir}. Can one then deduce from the worldsheet computation a closed-form expression for $\TT$-deformed correlators? Can such a result be tested against perturbative $\TT$ predictions \cite{He:2019ahx,He:2020qcs,Hirano:2020ppu,He:2023kgq,Hirano:2025tkq,Callebaut:2019omt}? We plan to return to this question in future work \cite{higher-spectrum}.

\paragraph{Twisted sectors.} In Section~\ref{sec:deforming} we construct a few worldsheet vertex operators, all of which lie in the untwisted $w=1$ sector. It would be interesting to extend this construction to twisted sectors and to understand whether a $w$-twisted analogue of Cardy's Callan-Symanzik equation exists.

\paragraph{Spectrum and non-perturbative states.} Beyond the examples constructed in Section~\ref{sec:vertex}, it would be interesting to understand how a generic state of the undeformed tensionless string deforms when $\mu \neq 0$. Moreover, it was argued in \cite{Dei:2024sct} that the string theory on the left-hand side of \eqref{ads3/cft2-deformed} contains states whose energy diverges as $\mu \to 0$, and therefore decouple in this limit. As a result, these states are absent in the tensionless theory at $\mu=0$. One may aim to construct worldsheet vertex operators for such states and to compute their correlation functions, which should in turn shed light on non-perturbative aspects of the $\TT$ deformation. We plan to return to these questions in future work \cite{higher-spectrum}.

\paragraph{Null gauging.} A number of asymptotically linear dilaton backgrounds have been studied in string theory using gauged Wess-Zumino-Witten models \cite{Witten:1991yr, Dijkgraaf:1991ba,  Forste:1994wp, Giveon:1999px, Giveon:1999tq, Hanany:2002ev, Quella:2002fk, Israel:2003ry, Giveon:2017myj,  Martinec:2017ztd, Martinec:2018nco, Martinec:2019wzw, Martinec:2020gkv, Bufalini:2021ndn, Bufalini:2022wyp, Bufalini:2022wzu, Martinec:2022okx, Martinec:2025xoy}, and a framework for null gauging of the tensionless string has recently been proposed in \cite{Dei:2025ilx}. While our construction shares structural features with this literature, the string theory developed in Section~\ref{sec:deforming} is not formulated in terms of (null) gauging. It would be interesting to understand whether such a formulation is possible in our case, and more generally which space-time irrelevant deformations admit a description in terms of gauged worldsheet CFTs, and which instead fall outside this framework.

\paragraph{Higher genus correlators.} In Appendix~\ref{app:corr}, we explain that our definition of single-trace $\TT$-deformed tree-level correlators can be extended to include higher-genus contributions on the worldsheet. Similarly, it would be interesting to understand whether one can reproduce space-time higher genus correlation functions discussed in \cite{He:2022jyt,He:2024xbi}.

\paragraph{Integrability and S-matrix.} It would be interesting to connect our results with the integrability literature. Two-dimensional CFTs are closely related to integrable systems in several ways. Examples include the quantum KdV charges of \cite{Bazhanov:1994ft, Bazhanov:1996dr, Bazhanov:1998dq, Bazhanov:1996aq, Negro:2016yuu} and the magnon S-matrix description recently discussed for symmetric orbifold CFTs in \cite{Gaberdiel:2023lco}. If the $T\overline T$-deformed four-point function can be extracted from our formalism, one could ask whether an LSZ-type prescription identifies an associated two-body S-matrix in the relevant sector. Since $\TT$ deformations of integrable two-dimensional QFTs preserve factorized scattering and act by a CDD dressing of the S-matrix \cite{Smirnov:2016lqw}, one may expect the resulting S-matrix to differ from the undeformed one by the universal $\TT$ CDD factor. Extending this analysis to higher-point functions would furthermore provide a direct test of factorization and hence of the underlying integrable structure.

\section{The tensionless string toolkit} 
\label{sec:tensionless}

In this section we briefly summarize some of the key ingredients entering the construction of the $k=1$ $\text{AdS}_3 \times \text{S}^3 \times \mathbb T^4$ string, dual to the symmetric orbifold of $\mathbb T^4$ \cite{Gaberdiel:2018rqv, Eberhardt:2018ouy, Eberhardt:2019ywk}. We do not attempt a thorough review and we restrict attention to the elements that will be used in this manuscript.

\subsection[{The worldsheet CFT}]{The worldsheet CFT}

The left-moving sector of the worldsheet CFT$_2$ of the tensionless $\text{AdS}_3 \times \text{S}^3 \times \mathbb T^4$ string is generated by a collection of free fields, factorizing into different sectors \cite{Berkovits:1999im, Eberhardt:2018ouy, Dei:2020zui, Dei:2023ivl},
\begin{equation}
    \mathcal F_\text{T} = \mathcal F_{\beta \gamma} \otimes \mathcal F_{p\theta} \otimes \mathcal F_{\mathbb T^4} \otimes \mathcal F_{\rho \sigma} \ ,  
    \label{F}
\end{equation}
which we discuss in turn. An analogous discussion applies to the right-moving sector, whose fields are denoted by a bar over the corresponding left-moving fields. We therefore restrict our attention to the left-moving sector here.

The $\beta \gamma$ sector $\mathcal F_{\beta \gamma}$ consists of a $\beta \gamma$ system $(\beta, \gamma)$ of conformal dimensions $(1,0)$, with the action  
\begin{equation}
    S_{\beta \gamma} = \frac{1}{2 \pi} \int \, \text d^2 z \, (\beta \bar \partial \gamma + \bar \beta \partial \bar \gamma) \ , 
    \label{S0}
\end{equation}
and obeying the OPE
\begin{equation}
    \beta(z) \gamma(w) \sim - \frac{1}{z-w} \,,
    \label{betagamma}
\end{equation}
where $\sim$ denotes equality up to regular terms. Two $bc$ systems $(p_a, \theta^a)$ of dimensions $(1,0)$, obeying the OPE 
\begin{equation}
 p_a(z) \theta^b(w) \sim \frac{\delta_a^b}{z-w} \ , \qquad a, b \in \{1,2\} \ ,  
  \label{ptheta}
\end{equation}
define the $\mathcal F_{p\theta}$ sector and together with the fields \eqref{betagamma} generate the affine superalgebra $\mathfrak{psu}(1,1|2)_1$ \cite{Beem:2023dub,Dei:2023ivl}, see eqs.~\eqref{psu-comm-rel} and \eqref{psu-free-field}. For later use, we will bosonize $(p_a,\theta^a)$ as follows
\begin{equation} \label{eq:pa-thetaa-bosonization}
    p_1=e^{-if_1} \ , \qquad \theta^1 = e^{if_1} \ , \qquad p_2 = e^{if_2} \ , \qquad \theta^2 = e^{-if_2} \ .
\end{equation}
The $\mathcal F_{\mathbb T^4}$ sector is composed of four free compact bosons and four topologically twisted fermions. Together, they form a topologically twisted $\mathcal N=4$ algebra with central charge $\mathtt c=6$, which we describe in detail in Appendix~\ref{app:details-tensionless}. The topologically twisted fermions are bosonized in terms of two chiral bosons, 
\begin{equation} \label{Hij-OPE}
    H^i(z) H^j(w) \sim - \delta^{ij} \ln(z-w) \ , \qquad i,j \in \{1, 2 \} \ , 
\end{equation}
see eq.~\eqref{T4-fermions}, so that their sum obeys\footnote{We will frequently make use of the operators $e^{+iH}$ and $e^{-iH}$. Their OPE depends on the choice of cocycles, which we fix by writing them as $e^{iH^1}e^{iH^2}$ and $e^{-iH^1}e^{-iH^2}$ respectively.}
\begin{equation} \label{H-def-and-OPE}
    H = H^1 + H^2 \ , \qquad H(z) H(w) \sim - 2\ln (z-w) \ .  
\end{equation}
Finally, the ghost sector  $\mathcal F_{\rho \sigma}$ consists of the free chiral bosons
\begin{equation}
    \rho(z) \rho(w) \sim - \ln (z-w) \ , \qquad \sigma(z) \sigma(w) \sim - \ln (z-w) \ , 
    \label{rho-sigma}
\end{equation}
with central charge $\mathtt c =28$ and $\mathtt c =-26$ respectively. Considering all the contributions, the central charge of the worldsheet CFT$_2$ \eqref{F} vanishes, as required to define a consistent worldsheet string theory. 

\subsection[\texorpdfstring{The $\mathcal{N}=2$ and $\mathcal{N}=4$ algebras on the worldsheet}{The N=2 and N=4 algebras on the worldsheet}]
{\texorpdfstring{The $\boldsymbol{\mathcal{N}=2}$ and $\boldsymbol{\mathcal{N}=4}$ algebras on the worldsheet}{The N=2 and N=4 algebras on the worldsheet}}

As we review in the following sections, the tensionless string is defined as a Berkovits-Vafa topological string theory where  physical observables are defined in terms of a small $\mathcal N=4$ topologically twisted superconformal algebra on the worldsheet \cite{Berkovits:1994vy,Berkovits:1999im,Dei:2020zui}.

\paragraph{The $\boldsymbol{\mathcal N=2}$ algebra and its similarity transformation.} The fields discussed above form a topologically twisted $\mathcal N=4$ algebra with $\mathtt c=6$, which in turn is constructed starting from the $\mathcal N=2$ algebra\footnote{In eq.~\eqref{eq:n=2-free-before} fields are normal ordered according to the so-called `conformal normal ordering' prescription, see \cite{Polchinski:1998rq,Gaberdiel:2022als}. Equivalently, if adopting the usual `radial normal ordering' prescription of \cite{DiFrancesco:1997nk},  products of fields should be normal ordered from right to left. For example, we write $\partial(\rho+i\sigma) \partial(\rho+i\sigma)e^{i \sigma}$ to denote ${(\partial(\rho+i\sigma) (\partial(\rho+i\sigma)e^{i \sigma}))}$.\label{footnote-normal}}
\begin{subequations} \label{eq:n=2-free-before}
	\begin{align}
        T &= T_{\mathfrak{psu}} + T_\rho + T_\sigma + T_{\mathbb T^4}\ , \\
		G^{+} &=e^{-\rho} Q + T_{\mathfrak{psu}} e^{i \sigma} -\tfrac{1}{2} \partial(\rho+i\sigma) \partial(\rho+i\sigma)e^{i \sigma} +\tfrac{1}{2} \partial^2(\rho+i\sigma)e^{i \sigma}   + G^{+}_{\mathbb T^4} \ , \\
		G^{-} &=e^{-i \sigma}+G^{-}_{\mathbb T^4} \ , \\ 
		J &= \tfrac{1}{2}\partial(\rho+i\sigma)+J_{\mathbb T^4} \ ,         
	\end{align}
\end{subequations}
where 
\begin{align} \label{eq:def-stress-tensors}
    T_{\mathfrak{psu}} &= - \beta \partial \gamma  - p_a \partial \theta^a \ , &
    T_\rho &= -\tfrac{1}{2} (\partial \rho)^2 + \tfrac{3}{2} \partial^2 \rho \ , &
    T_\sigma &= -\tfrac{1}{2} (\partial \sigma)^2 + \tfrac{3}{2} i \partial^2 \sigma \ , 
\end{align}
and 
\begin{equation} \label{eq:q-free}
    Q = p_1 p_2 \partial \gamma \ .  
\end{equation}
The generators $T_{\mathbb T^4}$, $G^{+}_{\mathbb T^4}$, $G^{-}_{\mathbb T^4}$ and $J_{\mathbb T^4}$ form a topologically twisted $\mathcal{N}=2$ algebra with central charge $\mathtt c =6$, see Appendix~\ref{app:details-tensionless}.  
The generators in eqs.~\eqref{eq:n=2-free-before} close into a topologically twisted $\mathcal N=2$ algebra, with conventions given in eq.~\eqref{N=2-conventions}. This was shown in \cite{Berkovits:1999im,Gaberdiel:2022als} and verified here both by
hand and using the \texttt{Mathematica} package \cite{Thielemans:1991uw}.

\smallskip

It is often useful to work in terms of an equivalent (twisted) $\mathcal N=2$ algebra, which is obtained by a similarity transformation, where each generator $F \in \{T, G^+, G^-, J \}$ is mapped to 
\begin{equation}
\label{eq:similarity-tensionless}
    F \mapsto e^{R}Fe^{-R} \ , 
\end{equation}
with
\begin{equation} \label{R}
    R = \oint \text dz \, e^{i\sigma} G^-_{\mathbb{T}^4}  \ . 
\end{equation}
Since this operation preserves the (anti-)commutation relations of the algebra, the two constructions are equivalent. As we review in detail in Appendix~\ref{app:similarity-transformation}, the similarity transformation \eqref{eq:similarity-tensionless} maps the generators \eqref{eq:n=2-free-before} to\footnote{In eq.~\eqref{eq:n=2-free-after}, conformal normal ordering is assumed. Equivalently, the same expression can be translated in terms of radial normal ordering by replacing $Te^{i\sigma} $ with $(T_{\mathfrak{psu}}+ T_{\rho} + T_{\mathbb T^4})e^{i\sigma} + \tfrac{1}{2}(j_\sigma (j_\sigma c))+\frac{3}{2} (\partial j_\sigma c) $ where $j_\sigma = (cb)=i\partial \sigma$, see eq.~\eqref{eq:def-stress-tensors}.}
\begin{subequations} \label{eq:n=2-free-after}
	\begin{align}
        T &= T_{\mathfrak{psu}} + T_{\rho} + T_{\sigma} + T_{\mathbb T^4} \ , \\
		G^+ &= e^{-\rho} Q + T e^{i\sigma} - \partial[e^{i\sigma} (2J-i \partial\sigma)] +G^+_{\mathbb T^4} \label{G+-tensionless-after}\ , \\
		G^- &= e^{-i\sigma} \ , \\ 
		J &= \tfrac{1}{2}\partial(\rho+i\sigma)+J_{\mathbb T^4} \ .     \label{eq:n=2-free-after-J}     
	\end{align}
\end{subequations}
We will refer to the algebras \eqref{eq:n=2-free-before} and \eqref{eq:n=2-free-after} respectively as the $\mathcal N=2$ algebra \emph{before} and \emph{after} the similarity transformation.\footnote{Readers familiar with the hybrid and tensionless string literature will notice that we swapped the roles of `before' and `after' the similarity transformation relatively to the way they were historically described in \cite{Berkovits:1999im}. We hope that, for the purpose of this work, this makes it easier to follow our derivations.} While a given argument may be easier in one formulation or the other, any derivation carried out in either of the two realizations can be easily translated into the other.

\paragraph{The $\boldsymbol{\mathcal N=4}$ algebra.}

The topologically twisted $\mathcal N=2$ algebra \eqref{eq:n=2-free-after} can be extended to a topologically twisted $\mathcal N=4$ algebra as described in \cite{Berkovits:1994vy}. The $\mathfrak{u}(1)$ generated by $J$ in \eqref{eq:n=2-free-after-J} is promoted to an $\mathfrak{su}(2)_1$ by defining 
\begin{equation} \label{Jppmm-tensionless}
    J^{++} = e^{\rho + i \sigma + i H} \ , \qquad J^{--} = e^{-(\rho + i \sigma + i H)} \ , 
\end{equation} 
and the additional generators $\widetilde G^\pm$ are obtained by acting with $J^{\pm\pm}_0$ on $G^{\mp}$, see eq.~\eqref{OPE-J-G-generating}. One obtains\footnote{In eqs.~\eqref{eq:n=4-free} conformal normal ordering is assumed. Equivalently, in terms of radial normal ordering the term $Te^{-\rho-iH}$ should be understood as $(T_{\mathfrak{psu}}+T_\sigma+ \partial X^j \partial \bar X^j)e^{-\rho-iH}-\frac12 (\partial \rho (\partial \rho \, e^{-\rho-iH})) - \frac12 (\partial H^j (\partial H^j e^{-\rho-iH})) + \frac{3}{2} \partial^2 \rho \,  e^{-\rho-iH}  + \frac{i}{2} \partial^2 H e^{-\rho-iH}$.}
\begin{subequations} \label{eq:n=4-free}
	\begin{align}
        \widetilde G^+ &= e^{\rho + i H}\ , \label{eq:free-gtildep}\\
        \widetilde G^- & = e^{-2 \rho - i \sigma - iH}Q- Te^{- \rho - i H} -  \widetilde G^-_{\mathbb T^4} e^{- \rho - i \sigma}\nonumber  \\
        & \quad \ \ + [\partial(i \sigma) \partial(\rho + i H) + \partial^2(\rho + i H)]e^{- \rho - i H} \ .  
	\end{align}
\end{subequations}
In a similar way, one can extend the $\mathcal N=2$ algebra \eqref{eq:n=2-free-before} to an $\mathcal N=4$ algebra, see eq.~\eqref{eq:before-additioan-n=4}.

\subsection{Physical state conditions and correlation functions} 
\label{sec:tensionless-physical}

The $\mathcal N=4$ algebra reviewed in the previous section provides the structure necessary to define a topological string theory according to the prescription of Berkovits and Vafa \cite{Berkovits:1994vy,Berkovits:1999im,Dei:2020zui}. Let us review how the physical state conditions and correlation functions are defined. 

\paragraph{The physical state conditions.} A physical vertex operator $\phi$ is defined in terms of the $\mathcal N=4$ generators by the conditions
\begin{equation} \label{eq:physical-state-conditions}
    G^+_0 \phi = \widetilde G^+_0 \phi = (J_0-\tfrac{1}{2})\phi = T_0 \phi = 0 \ , \qquad \phi \sim \phi + G^+_0 \widetilde G^+_0 \psi \ , 
\end{equation}
where $\sim$ denotes equivalence up to exact states. Although an explicit solution of this (double) cohomology is technically challenging, the holographic match of the partition function \cite{Eberhardt:2018ouy,Eberhardt:2020bgq} and the explicit construction of DDF operators for the free fields of the boundary CFT \cite{Eberhardt:2019qcl, Naderi:2022bus} provide strong evidence that the tensionless string is dual to the symmetric orbifold of~$\mathbb{T}^4$.

\paragraph{The $\boldsymbol x$-basis.} To encode the dependence of correlation functions on the space-time insertion points, it is convenient to introduce worldsheet vertex operators in the so-called `$x$-basis', obtained by conjugating the worldsheet vertex operators with the space-time translation generator $\mathcal L_{-1} = J^+_0$. We define $x$-basis vertex operators as 
\begin{equation} \label{eq:x-basis-definition}
    V(x,z) = e^{x J^+_0} V(0,z) e^{-xJ^+_0} \ , \qquad V(0,z) \equiv V(z) \ .
\end{equation}

\paragraph{Twisted sector ground states.} Let us review the worldsheet duals of some boundary CFT vertex operators. We begin with the operators corresponding to the single-cycle $w$-twisted ground states of the space-time theory. Their explicit form depends on the parity of the spectral flow parameter $w$. For $w$ odd they read \cite{Dei:2020zui,Dei:2023ivl}
\begin{equation} \label{eq:w-odd-tensionless}
    \Phi_w(x,z) = -\exp{\left[\tfrac{w+1}{2}(if_1-if_2)\right]} \left(\frac{\partial^w \gamma}{w!}\right)^{-m_w} \delta_w(\gamma(z)-x) \, e^{2\rho+i\sigma+iH} \ ,
\end{equation}
where
\begin{equation} \label{eq:mw}
    m_w = -\frac{(w-1)^2}{4w} \ .
\end{equation}
For $w$ even, the $w$-twisted ground states form an $\mathfrak{su}(2)$ doublet,\footnote{There are two additional ground states as well, see e.g.~\cite{Gaberdiel:2015uca}. Here we are focusing on those ground states that are compactification independent, see \cite{Berkovits:1999im,Gerigk:2012lqa,Gaberdiel:2021njm}.}
\begin{subequations}  \label{eq:w-even-tensionless}
\begin{align}
    \Phi^{+}_w(x,z) &= -\exp{\left[ \frac{w+2}{2}if_1- \frac{w}{2}if_2\right]} \left(\frac{\partial^w \gamma}{w!}\right)^{-m^{+}_w} \delta_w(\gamma(z)-x) \, e^{2\rho+i\sigma+iH} \ , \\
    \Phi^{-}_w(x,z) &= -\exp{\left[\frac{w}{2}if_1-\frac{w+2}{2}if_2\right]} \left(\frac{\partial^w \gamma}{w!}\right)^{-m^{-}_w} \delta_w(\gamma(z)-x) \, e^{2\rho+i\sigma+iH} \ ,
\end{align}
\end{subequations}
where
\begin{equation} \label{eq:mw-pm}
    m^{\pm}_w = -\frac{w-2}{4} \ .
\end{equation}
Here $\delta_w(\gamma(z)-x)$ is defined as \cite{Verlinde:1987sd,Witten:2012bh,Dei:2023ivl}\footnote{Notice that these objects can also be defined by bosonization, see \cite{Verlinde:1987sd}. However, for the purpose of this paper, we will use the form \eqref{eq:delta-def}.}
\begin{equation} \label{eq:delta-def}
    \delta_w(\gamma(z)-x) = \delta(\gamma(z)-x) \prod_{j=1}^{w-1} \delta(\partial^j \gamma(z)) \ ,
\end{equation}
where $\delta(x)$ denotes the Dirac delta function. As discussed in detail in \cite{Dei:2023ivl}, the states in eq.~\eqref{eq:delta-def} are in the $w$-spectrally flowed sector of the $\beta\gamma$ and $(p_a,\theta^a)$ free fields. We will denote a generic state in the $w$-spectrally flowed sector as $[\ket{m}]^{\sigma^w}$ where $m$ labels representations of $\beta\gamma$: as shown in Appendix~A of \cite{Dei:2023ivl}, the value $m$ is directly related to the exponent of $\left(\frac{\partial^w \gamma}{w!}\right)$. In particular, the values in \eqref{eq:mw} and \eqref{eq:mw-pm} are set by the mass-shell condition $T_0=0$ on the worldsheet and guarantee that the fields \eqref{eq:w-odd-tensionless} and \eqref{eq:w-even-tensionless} have the correct space-time conformal dimensions and $\mathcal R$-charge to be identified with $w$-twisted symmetric orbifold ground states.

\paragraph{DDF operators.} DDF (Del Giudice, Di Vecchia, Fubini) operators  \cite{DelGiudice:1971yjh} are operators that (anti-)commute with the $\mathcal N=4$ generators that define the physical state conditions. More specifically, $\mathcal{D}$ is a DDF operator if it satisfies\footnote{Here the brackets should be understood as graded
commutators, with the appropriate sign determined by the statistics of the operators.}
\begin{equation} \label{eq:DDF-physical-state-conditions}
    [G^+_0, \mathcal{D}] = [\widetilde G^+_0,\mathcal{D}] = [J_0,\mathcal{D}] = [T_0,\mathcal{D}] = 0 \ , \qquad \mathcal{D} \sim \mathcal{D} + [G^+_0,[\widetilde G^+_0,\mathcal{D}^{\prime}]] \ ,
\end{equation}
where by $\sim$ we mean equivalence up to exact terms and in the following we will make use of the $\mathcal N=4$ generators \eqref{eq:n=2-free-after} and \eqref{eq:n=4-free} to define DDF operators. The DDF operators of the symmetric orbifold free bosons and fermions read \cite{Naderi:2022bus,Dei:2023ivl}
\begin{subequations} \label{eq:t4-ddf-tensionless}
    \begin{align}
        \partial \bar{\mathcal{X}}^j_n &= \oint \text dz \, \partial \bar X^j \gamma^{n} \ , \quad & \Lambda^{-,j}_r &= -\oint \text dz \, \theta^2 \gamma^{r-\tfrac{1}{2}} e^{\rho+iH^j} \ , \\
        \partial \mathcal X^j_n & = \oint \text  dz \, \left[ \partial X^j \gamma^n+n \gamma^{n-1} e^{\rho+iH^j} \theta^2 \theta^1\right] \ , \quad & \Lambda^{+,j}_r &= \oint \text dz \, \theta^1 \gamma^{r-\frac{1}{2}} e^{\rho+iH^j}   \ ,
    \end{align}
\end{subequations}
for $j\in\{1,2\}$. The DDF operators \eqref{eq:t4-ddf-tensionless} obey 
\begin{align}
    [ \partial \mathcal X^i_m, \partial \bar{\mathcal{X}}^j_n ] & = m \, \delta^{ij} \, \delta_{m+n,0} \, \mathcal I \ , \\
    \{ \Lambda^{\alpha, j} \Lambda^{\beta, \ell} \} &= \epsilon^{\alpha \beta} \epsilon^{\ell j} \delta_{r+s,0} \, \mathcal I \ , 
\end{align}
where $\alpha, \beta \in \{ +,-\}$ and 
\begin{equation}
    \mathcal I = \oint \text dz \, \gamma^{-1} \partial \gamma 
\end{equation}
denotes the space-time identity. The spectrum is then generated by acting with the fractional modes of these fields on the $w$-twisted ground states. In particular, the ground state of the space-time theory arises in the $w=1$ sector and is explicitly given by $\Omega^0(x,\bar x,z,\bar z)$ where for later convenience we introduce the notation 
\begin{equation} \label{eq:omega-n-def}
    \Omega^m(x,\bar x,z,\bar z) = \Omega^{m}_L(z) \Omega^m_R(\bar z) \delta^{(2)}(\gamma(z)-x) \ , \qquad m\geq 0 \ , 
\end{equation}
together with
\begin{equation} \label{eq:omega-n-L-R}
    \Omega^m_L(z) = e^{if_1-if_2} (\partial \gamma)^{-m} e^{2\rho+i\sigma+iH} \ , \qquad  \Omega^m_R(z) = e^{i\bar f_1-i\bar f_2} (\bar \partial \bar \gamma)^{-m} e^{2\bar \rho+i\bar \sigma+i\bar H}  \ .
\end{equation}

\paragraph{Three- and higher-point functions.} Given $n$ physical fields $V_j$, $j=1, \dots, n$, their tree-level correlation function is defined as \cite{Berkovits:1994vy,Berkovits:1999im,Dei:2020zui,Gaberdiel:2021njm,Dei:2023ivl}\footnote{More generally, the prescription involves a sum over all values of $M$ weighted by a corresponding chemical potential. In the present setup, however, the only non-vanishing contribution comes
from the value of $M$ fixed by eq.~\eqref{eq:N-covering-cond}. See \cite{Dei:2023ivl} for further details.}
\begin{equation} \label{eq:corr}
    \int \text d^{2M} \zeta_\ell \, \text{d}^2 z_4 \cdots \text d^2 z_n \, \big<\prod_{\ell=1}^{M} D\overline D(\zeta_\ell) \widetilde{V}_1(z_1,\bar z_1) \, V_2(z_2,\bar z_2) \, V_3(z_3,\bar z_3) \prod_{j=4}^n G^-_{-1} \bar{G}^-_{-1} V_j(z_j,\bar z_j) \big> \ ,
\end{equation}
where the sum of the pictures of all the operators is assumed to be
\begin{equation} \label{picture-sum}
    \sum_{j=1}^n P_j = \sum_{j=1}^n \overline P_j = -4 \,.
\end{equation}
For each $V_j$ the associated picture $P_j$ is effectively defined as the eigenvalue under $(\partial \rho)_0$, see the discussion around eq.~\eqref{eq:picture-accurate} for a more accurate definition. For each $V_j$ the associated field $\widetilde V_j$ entering \eqref{eq:corr} is defined by solving
\begin{equation} \label{eq:vtilde}
    V_j = \widetilde{G}^+_0 \overline{\widetilde{G}^+_0} \widetilde{V}_j \ .  
\end{equation}
It is often useful to introduce the field $\xi$, defined by 
\begin{equation} \label{eq:xi-def-relation}
    \{\widetilde G^+_n,\xi_m\}=\delta_{n+m,0} \ ,  
\end{equation}
so that eq.~\eqref{eq:vtilde} is solved explicitly as 
\begin{equation} \label{eq:picture-explicit}
    \widetilde{V}_j = \xi_0 \bar\xi_0 V_j \ .
\end{equation}
For the tensionless string, $\xi$ reads
\begin{equation} \label{eq:xi-def}
    \xi = -e^{-\rho-iH} \ ,
\end{equation}
and similarly for the right-moving field $\bar \xi$. The notation we have adopted is not accidental, since the field $\xi$ plays the same role of the dimension zero field appearing in the bosonization of the superdiffeomorphism ghosts of the RNS formalism, see e.g.~\cite{Blumenhagen:2013fgp}. Indeed, the field $\xi$ enters the definition of the picture raising operator $P_+$, as 
\begin{equation} \label{eq:picture raising-def}
    P_+ V_j = G^+_0 \xi_0 V_j \ .
\end{equation}
In Appendix~\ref{app:corr}, following \cite{Berkovits:1999im}, we review the argument showing that the correlation function
\eqref{eq:corr} is independent of the picture assignment of each individual vertex operator, provided that the total picture satisfies
eq.~\eqref{picture-sum}. Equivalently, the correlator is invariant under permutations of the picture raising operators among the various vertex operators $V_j$. We also show that exact states decouple.

The operators $D$ and $\overline{D}$ in \eqref{eq:corr} are given by \cite{Dei:2023ivl}
\begin{equation} \label{eq:def-d}
    D = p_1 p_2 \delta(\beta) \ , \qquad \overline{D} = \bar p_1 \bar p_2 \delta(\bar \beta) \ .
\end{equation}
They are screening operators for the $\mathfrak{psu}(1,1|2)_1$ free fields and their insertion is necessary in order to obtain non-vanishing correlation functions \cite{Dei:2023ivl}. Moreover, for fields $V_j$ in the $w_j$-spectrally flowed sectors, it can be shown that $M$ is fixed to be
\begin{equation} \label{eq:N-covering-cond}
    M=1+\sum_{j=1}^{n}\frac{w_j-1}{2} \ ,
\end{equation}
as the degree of the corresponding (connected) covering map.

\paragraph{Two-point functions.} The two-point function cannot be defined by setting $n=2$ in eq.~\eqref{eq:corr}, as this would yield a vanishing result \cite{Berkovits:1994vy}. In fact, the background charge of the diffeomorphism $bc$ ghosts would not be saturated. The correct definition is given by \cite{Berkovits:1994vy, Maldacena:2001km, Erbin:2019uiz, Gaberdiel:2021njm}
\begin{equation} \label{eq:two-point-free}
    \int \text d^{2M}\zeta_\ell \, \big< \prod_{\ell=1}^{M} D\overline D(\zeta_\ell) [(e^{i\sigma+i\bar \sigma})_0 \widetilde{V}_1](z_1,\bar z_1) \, V_2(z_2,\bar z_2) \big> \ .
\end{equation}
As it will be useful later on, we will calculate the two-point function of \eqref{eq:omega-n-def}. More specifically, we will compute
\begin{equation}
    \left< \Omega^m \Omega^m \right> = \int \text d^{2}\zeta \, \big< D\overline D(\zeta) [(e^{i\sigma+i\bar \sigma})_0 \widetilde{\Omega}^m](x_1,\bar x_1,z_1,\bar z_1) \, \Omega^m(x_2,\bar x_2,z_2,\bar z_2) \big> \ ,
\end{equation}
where here and in the following we often omit the explicit dependence on the worldsheet and space-time insertion points. This two-point function factorizes into the correlators of the different sectors given in \eqref{F}. Schematically, 
\begin{equation} \label{eq:tensionless-vacuumtwo-point}
    \left< \Omega^m \Omega^m \right> = \left< \Omega^m \Omega^m \right>_{\mathfrak{psu}} \left< \Omega^m \Omega^m \right>_{\text{rest}} \ ,
\end{equation}
where the subscripts indicate the corresponding contributions. With a little effort, one can see that $\left< \Omega^m \Omega^m \right>_{\text{rest}}$ is independent of the space-time coordinates $x_j$, see eqs.~\eqref{eq:omega-n-def} and \eqref{eq:omega-n-L-R}, and its contribution can be absorbed into the normalization of vertex operators. For this reason, we will focus on the $\beta\gamma$ and $(p_a,\theta^a)$ contributions. Their contribution can e.g.\ be computed by making use of worldsheet global Ward identities, see \cite{Eberhardt:2019ywk,Bertle:2020sgd,Gaberdiel:2021njm}. One obtains
\begin{equation} \label{eq:free-two-point-xbasis}
    \left< \Omega^m \Omega^m \right> = \frac{C}{|x_1-x_2|^{4m}} \ ,
\end{equation}
where $C$ is an $m$-independent normalization constant.

\section{The auxiliary holographic duality}
\label{sec:auxiliary-duality}

Inspired by the $k>1$ constructions of \cite{Giveon:2017nie, Giveon:2017myj, Apolo:2019zai, Apolo:2021wcn, Cui:2023jrb}, we explain in Section~\ref{sec:deforming} how the $J^+\bar J^+$ deformation in eq.~\eqref{ads3/cft2} can be naturally implemented by augmenting the worldsheet theory with two free bosons $X^\pm$. We develop this `auxiliary' string theory in detail in Section~\ref{sec:auxiliary-string}, where we also clarify the role of the additional free fields. The purpose of this enlargement is not to modify the physical content of the undeformed tensionless duality, but to obtain a framework in which the $J^+ \bar{J}^+$ deformation can be treated using the same Berkovits-Vafa topological string formalism.
As reviewed in Section~\ref{sec:tensionless}, the tensionless string is described by a Berkovits-Vafa topologically twisted $\mathcal N=4$ algebra with $\mathtt c=6$ on the worldsheet. In order to preserve this algebraic structure, and in particular maintain worldsheet $\mathcal N=4$ supersymmetry, we are led to introduce additional fields together with $X^\pm$. We collectively refer to these as the \emph{auxiliary fields}.
We then study the string theory obtained by combining the tensionless sector with these auxiliary fields, which we refer to as the \emph{auxiliary string theory}. In Section~\ref{sec:auxiliary-space-time-CFT}, we propose a holographic dual description in terms of a symmetric orbifold of $\mathbb T^4$ tensored with an \emph{auxiliary space-time seed theory} $\mathcal A_0$ of central charge $\mathtt c=0$. As evidence, in Sections~\ref{sec:partition-functions-match} and \ref{sec:physical-states} we match torus partition functions following \cite{Eberhardt:2018ouy} and explicitly construct the relevant DDF operators along the lines of \cite{Naderi:2022bus}. Finally, in Section~\ref{sec:embedding} we explain how the tensionless duality naturally embeds in the auxiliary duality we described in the previous pages. 

We note that related ideas connecting tensionless strings to symmetric orbifolds with a vanishing central charge sector have appeared previously in \cite{Eberhardt:2025sbi,Dei:2025ilx}.

\subsection{The auxiliary string theory}
\label{sec:auxiliary-string}

In this section, we construct an auxiliary string theory by supplementing the field content of the tensionless string with a collection of free fields, which  will prove useful in Sections~\ref{sec:deforming} for the construction of the deformed $\mathcal N=4$ algebra on the worldsheet.

\paragraph{Field content and $\boldsymbol{\mathcal N=2}$ algebra.} The field content of the auxiliary string theory can be summarized as
\begin{equation} \label{F-aux}
    \mathcal F = \mathcal F_\text{T} \otimes \mathcal F_{X} \otimes \mathcal F_{\psi \pi} \otimes  \mathcal F_{\text g} \ , 
\end{equation}
where $\mathcal F_\text{T}$ denotes the field content of the tensionless string, see eq.~\eqref{F}, while the remaining factors on the right-hand-side denote the auxiliary fields that we introduce below.

Inspired by the constructions of \cite{Giveon:2017nie, Giveon:2017myj, Cui:2023jrb}, we find it useful to introduce auxiliary free bosons $X^+$ and $X^-$ obeying the OPEs
\begin{equation}
\label{Xpm-trivial}
    X^+(z,\bar z) X^-(w,\bar w) \sim -\ln|z-w|^2 \ , \qquad X^{\pm}(z,\bar z) X^{\pm}(w,\bar w) \sim 0 \ ,   
\end{equation}
which constitute the $\mathcal F_X$ sector of the worldsheet CFT. Let us now briefly explain the motivation for introducing the remaining auxiliary fields in eq.~\eqref{F-aux}, beyond the free bosons $X^\pm$. Recall that we would like to formulate the auxiliary string theory as an $\mathcal N=4$ topological string, which requires the corresponding worldsheet structure with central charge $\mathtt c=6$. In particular, the self-OPE of the twisted stress tensor contains no fourth-order pole; see eq.~\eqref{n=2-TT-OPE}. In order to preserve this structure while obtaining a space-time supersymmetric theory, we introduce a pair of topologically twisted fermions. These form a $bc$ system $(\psi,\pi)$ of conformal dimensions $(1,0)$, obeying
\begin{equation}
    \psi(z) \pi(w) \sim \frac{1}{z-w} \ ,  
    \label{psi-pi}
\end{equation}
and defining the $\mathcal F_{\psi \pi}$ sector. The resulting stress-tensor has no forth order pole in its self-OPE and can be straightforwardly extended to an $\mathcal N=2$ topologically twisted algebra by defining the generators
\begin{subequations} \label{auxiliary-n=2-b}
\begin{align}
    T_{\text{aux}|b} &= -\partial X^+ \partial X^- - \psi \partial \pi \ , \\
    G^+_{\text{aux}|b} &= \partial X^+ \pi \ , \label{eq:added-g+-b}\\
    G^-_{\text{aux}|b} &= -\partial X^- \psi \ , \\
    J_{\text{aux}|b} &= -\tfrac12 \psi \pi \ .     
\end{align}
\end{subequations}
The central charge of the $\mathcal N=2$ algebra can be read off from the self-OPE of the $\text{U}(1)$ Cartan current, see eq.~\eqref{eq:n=2-jj-ope}, and one finds $\mathtt c =3$. One may then proceed by combining these generators with the $\mathcal{N}=2$ generators of tensionless string, thereby obtaining a topologically twisted $\mathcal{N}=2$ algebra with central charge $\mathtt c = 9$. This, however, does not coincide with the critical value $\mathtt c = 6$ required in the Berkovits-Vafa construction, necessary to admit an extension to a small $\mathcal N=4$ topological string.\footnote{Nevertheless, it is interesting to note that the critical central charge of $\mathcal{N}=2$ topological string theory is indeed $\mathtt c=9$, and one may therefore wonder whether the auxiliary string theory could equivalently be constructed as an $\mathcal N=2$ topological string theory by adding only the auxiliary fields $X^\pm$ and $(\psi,\pi)$. Here, we prefer to mirror as close as possible the structure of the tensionless string and therefore follow a different route and formulate the auxiliary string theory as an $\mathcal{N}=4$ topological string theory.} We are therefore led to introduce additional fields. More specifically, we introduce a $\beta \gamma$ system $(\beta^{(\text g)}, \gamma^{(\text g)})$ of dimensions $(1,0)$ and a further $bc$ system $(b^{\text{(g)}},c^{\text{(g)}})$ of dimensions $(1,0)$, obeying
\begin{align}
    \beta^{(\text g)}(z) \gamma^{(\text g)}(w) &\sim - \frac{1}{z-w} \ , \label{betan-gamman} \\
    b^{\text{(g)}}(z) c^{\text{(g)}}(w)  &\sim \frac{1}{z-w} \ , \label{bg-cg-ope}
\end{align}
which constitute the ghost sector $\mathcal F_{\text g}$ entering eq.~\eqref{F-aux}. 
In the following, we will refer to the fields introduced in eqs.~\eqref{Xpm-trivial}, \eqref{psi-pi}, \eqref{betan-gamman} and \eqref{bg-cg-ope} collectively as \textit{auxiliary} fields. They form a topologically twisted $\mathcal{N}=2$ with $\mathtt{c}=0$ given by
\begin{subequations} \label{auxiliary-n=2}
\begin{align}
    T_{\text{aux}} &=   - b^{(\text{g})} \partial c^{(\text{g})} -  \beta^{(\text{g})} \partial \gamma^{(\text{g})} -\partial X^+ \partial X^- -\psi \partial \pi \ , \\
    G^+_{\text{aux}} &= \gamma^{(\text{g})} b^{\text{(g)}} + \partial X^+ \pi\ , \label{eq:added-g+}\\
    G^-_{\text{aux}} &= -\beta^{(\text{g})} \partial c^{\text{(g)}} - \partial X^- \psi \ , \label{eq:g--aux-left}\\
    J_{\text{aux}} &= - \tfrac12 \beta^{(\text{g})} \gamma^{(\text{g})} -\tfrac12 \psi \pi \ .     
\end{align}
\end{subequations}
We should mention that this structure is not new: related ingredients have appeared previously in \cite{Figueroa-OFarrill:1990eum,Figueroa-OFarrill:1993whc,
Figueroa-OFarrill:1995qkv}. More recently, the algebra \eqref{auxiliary-n=2} was studied in \cite{Dei:2025ilx} and used to construct the worldsheet string theory of a single decoupled NS5-brane.\footnote{Our notation here is related to theirs as $(X^+, X^-) \mapsto (X_0, X^*_0)$, $(\psi, \pi) \mapsto (\Psi_0^*, \Psi_0)$, $(\beta^{(\text g)}, \gamma^{(\text g)}) \mapsto (\hat{\boldsymbol{\beta}}_n, \hat{\boldsymbol{\gamma}}_n) $ and $(b^{\text{(g)}}, c^{\text{(g)}}) \mapsto (\boldsymbol{b}_n, \boldsymbol{c}_n)$.} 

Combining the tensionless string generators \eqref{eq:n=2-free-before} with the ones defined in eqs.~\eqref{auxiliary-n=2} we obtain a topologically twisted $\mathcal{N}=2$ algebra with central charge $\mathtt c=6$:
\begin{subequations} \label{eq:n=2-auxiliary-before}
\begin{align}
T &=T_{\mathfrak{psu}}+T_{\rho}+T_{\sigma}+ T_{\mathbb T^4} + T_{\text{aux}} \ , \\
G^+ &= e^{-\rho} Q + e^{i \sigma} \left[ T_{\mathfrak{psu}}-\frac{1}{2} [\partial(\rho+i\sigma)]^2 +\frac{1}{2} \partial^2(\rho+i\sigma) \right] + G^+_{\mathbb T^4} + G^+_{\text{aux}} \ ,\\
G^- &= e^{-i \sigma} + G^-_{\mathbb T^4} + G^-_{\text{aux}} \ , \\ 
J &= \tfrac{1}{2}\partial(\rho+i\sigma) + J_{\mathbb T^4} + J_{\text{aux}} \ . 
\end{align}
\end{subequations}
The right-moving $\mathcal{N}=2$ algebra is defined similarly, combining the anti-holomoprhic $\mathcal{N}=2$ tensionless string generators with the auxiliary $\mathcal N=2$ algebra 
\begin{subequations} \label{auxiliary-n=2-right}
\begin{align}
    \overline T_{\text{aux}} &=   - \bar b^{(\text{g})} \bar \partial \bar c^{(\text{g})} -  \bar \beta^{(\text{g})} \bar \partial \bar \gamma^{(\text{g})} -\bar \partial X^+ \bar \partial X^- - \bar \psi \bar \partial \bar \pi \ , \\
    \bar G^+_{\text{aux}} &= \bar \gamma^{(\text{g})} \bar b^{\text{(g)}} + \bar \partial X^- \bar \pi\ , \label{eq:added-g+-right}\\
    \bar G^-_{\text{aux}} &= -\bar \beta^{(\text{g})} \bar \partial \bar c^{\text{(g)}} - \bar \partial X^+ \bar \psi \ , \label{eq:g--aux-right} \\
    \bar J_{\text{aux}} &= - \tfrac12 \bar \beta^{(\text{g})} \bar \gamma^{(\text{g})} -\tfrac12 \bar \psi \bar \pi \ .     
\end{align}
\end{subequations}
Notice that, for later convenience, the roles of $\bar \partial X^+$ and $\bar \partial X^-$ in the right-moving sector are interchanged relatively to those of $\partial X^+$ and $\partial X^-$ in the left-moving sector. 

In the following, it will prove useful to bosonize the fields \eqref{psi-pi} and \eqref{betan-gamman} as 
\begin{equation}  \label{eq:bosonizations-added-ghosts}
\begin{aligned}
    \psi &= e^{-i\chi} \ , & \qquad \quad  \beta^{(\text g)} &= e^{u+iv} \partial(iv) \ ,  \\
    \pi &= e^{i\chi} \ ,  &  \qquad \quad \gamma^{(\text g)} &= e^{-u-iv} \,,
\end{aligned}
\end{equation}
where
\begin{equation}
    u(z) u(w) \sim -\ln(z-w) \ , \quad v(z) v(w) \sim -\ln(z-w) \ , \quad \chi(z) \chi(w)\sim -\ln(z-w) \ .
\end{equation}
We note that the bosons $i\chi$ and $u$ have background charges equal to $1$, see eq.~\eqref{eq:background-charge-definition} for our conventions. The right-moving fields are bosonized similarly.

\paragraph{Similarity transformation and $\boldsymbol{\mathcal N=4}$ algebra.} As discussed in Section~\ref{sec:tensionless}, it will be useful to also construct the $\mathcal N=2$ generators \textit{after} the similarity transformation \eqref{eq:similarity-tensionless}, where the similarity operator $R$ is now given by 
\begin{equation}
    R = \oint \text dz \, e^{i\sigma} [G^-_{\mathbb{T}^4}+G^-_{\text{aux}}] \ . 
\end{equation}
Under this similarity transformation, the algebra \eqref{eq:n=2-auxiliary-before} is mapped to
\begin{subequations} \label{eq:n=2-lambda0-after}
	\begin{align}
        T &= T_{\mathfrak{psu}} + T_{\rho} + T_{\sigma} + T_{\mathbb T^4} +T_{\text{aux}}\ , \\
		G^+ &= e^{-\rho} Q + e^{i\sigma} T - \partial[e^{i\sigma} (2J-i \partial\sigma)] +G^+_{\mathbb T^4}+G^+_{\text{aux}} \label{G+-after-lambda0}\ , \\
		G^- &= e^{-i\sigma} \ , \\ 
		J &= \tfrac{1}{2}\partial(\rho+i\sigma)+J_{\mathbb T^4}+J_{\text{aux}}=\tfrac{1}{2}\partial(\rho+i\sigma+iH+u+i\chi)\ ,     \label{eq:n=2-lambda0-after-J}     
	\end{align}
\end{subequations}
see Appendix~\ref{app:similarity-transformation} for a detailed derivation.

Since the central charge of the topologically twisted $\mathcal{N}=2$ algebra is $\mathtt c=6$, the Cartan current $J$ defined in \eqref{eq:n=2-lambda0-after-J} can be used to construct an $\mathfrak{su}(2)_1$ $\mathcal{R}$-symmetry by defining 
\begin{align} \label{eq:lambda0-su2}
    J^{++} &= e^{\rho+i\sigma+iH+u+i\chi} \ , &  J^{--} &= -e^{-(\rho+i\sigma+iH+u+i\chi)} \ .
\end{align}
Acting with the $\mathcal{R}$-symmetry generators on $G^\pm$ we obtain the following additional supercharges
\begin{subequations} \label{eq:n=4-lambda0}
	\begin{align}
        \widetilde G^+ &= e^{\rho + i H+u+i\chi}\ , \label{eq:lambda0-gtildep}\\
        \widetilde G^- & = -e^{-2 \rho - i \sigma - iH-u-i\chi}Q + T e^{- \rho - i H-u-i\chi} + \widetilde G^-_{\mathbb T^4} e^{- \rho - i \sigma-u-i\chi}  \nonumber \\ 
        & \quad \ \ + \partial X^+ e^{-\rho-i\sigma-iH-u} + b^{\text{(g)}} e^{-\rho-i\sigma-iH-2u-i\chi-iv}  \nonumber \\ 
        & \quad \ \ - [\partial(i \sigma) \partial(\rho + i H+u+i\chi) + \partial^2(\rho + i H+u+i\chi)]e^{- \rho - i H-u-i\chi} \ .
	\end{align}
\end{subequations}
Using these fields, we thus obtain a topologically twisted $\mathcal{N}=4$ algebra with $\mathtt c=6$. Following Berkovits and Vafa's prescription, the physical operators are defined by
\begin{equation} \label{eq:auxiliary-physical-state-conditions}
    G^+_0 \phi = \widetilde G^+_0 \phi = (J_0-\tfrac{1}{2})\phi = T_0 \phi = 0 \ , \qquad \phi \sim \phi + G^+_0 \widetilde G^+_0 \psi \ , 
\end{equation}
where the generators $G^+, \widetilde G^+, J$ and $T$ are now those in eqs.~\eqref{eq:n=2-lambda0-after} and \eqref{eq:n=4-lambda0}.
Together with the definition of correlation functions, to be discussed in Section~\ref{sec:corr}, this yields an $\mathcal N=4$ topological string theory that we call \textit{auxiliary} string theory.

\smallskip 

In the following section, we propose a boundary dual for the string theory constructed above.

\subsection{The auxiliary space-time CFT}
\label{sec:auxiliary-space-time-CFT}

There is by now a vast literature showing that the tensionless string is dual to the symmetric orbifold of $\mathbb T^4$. In this section, we propose a boundary dual for the auxiliary string theory defined in Section~\ref{sec:auxiliary-string}. We conjecture that it is again given by a symmetric orbifold, namely
\begin{equation} \label{eq:symm-dual-C0}
    \text{Sym}^N\left(\mathbb{T}^4 \times \mathcal A_0 \right) \ ,
\end{equation}
where $\mathcal A_0$ denotes a free field CFT with vanishing central charge, which we now describe.\footnote{Strictly speaking, as discussed in \cite{Kim:2015gak, Eberhardt:2020bgq, Aharony:2024fid}, the dual theory is the grand canonical ensemble of symmetric orbifolds.}

The seed theory $\mathcal A_0$ is a non-unitary CFT with vanishing total central charge. Its left-moving sector is the direct sum of two subsectors with central charges $\mathtt c=3$ and $\mathtt c=-3$.
The $\mathtt c=3$ subsector is generated by two free bosons,
\begin{subequations} \label{eq:t2-added-aux}
\begin{equation}
    [ \partial \mathcal X^+_n, \partial \mathcal X^-_m]
    = - n \, \delta_{n+m,0} \ ,
\end{equation}
and by two real free fermions of conformal dimensions
$(\tfrac12,\tfrac12)$,
\begin{equation}
    \{\Upsilon_r,\Pi_s\} = \delta_{r+s,0} \ .
\end{equation}
\end{subequations}
The $\mathtt c=-3$ subsector is generated by a $bc$ system of conformal
weights $(1,0)$,
\begin{subequations} \label{eq:ghost-added-aux}
\begin{equation}
    \{\mathcal B_r,\mathcal C_s\} = \delta_{r+s,0} \ ,
\end{equation}
and by a $\beta\gamma$ system of conformal weights
$(\tfrac12,\tfrac12)$,
\begin{equation}
    [\Sigma_n,\Gamma_m] = -\delta_{n+m,0} \ .
\end{equation}
\end{subequations}
In what follows, we will occasionally bosonize the fields $(\Sigma,\Gamma)$ and $(\Upsilon,\Pi)$ as
\begin{subequations} \label{bosonized-auxiliary-ghosts}
\begin{align}
    \Sigma &= e^{\mathcal Y} \partial \Xi \ ,  & \qquad \Gamma& = \mathcal N e^{-\mathcal Y}   \ , \\
    \Upsilon &= e^{-i \mathcal K} \ , & \qquad \Pi &= e^{i \mathcal K} \ , 
\end{align} 
\end{subequations}
where
\begin{equation}
    \mathcal{Y}(x_1)\mathcal{Y}(x_2)\sim -\ln(x_1-x_2) \ , \quad \mathcal{K}(x_1)\mathcal{K}(x_2)\sim -\ln(x_1-x_2) \ ,
\end{equation}
and
\begin{equation}
    \Xi(x_1)\,\mathcal{N}(x_2)\sim \frac{1}{(x_1-x_2)} \ .
\end{equation}
The right-moving sector of the $\mathcal A_0$ CFT is constructed analogously. These fields form an (untwisted) $\mathcal{N}=2$ superconformal algebra with central charge $\mathtt c=0$:
\begin{subequations} \label{auxiliary-n=2-space-time}
\begin{align}
    \mathcal T_{\text{aux}} &=   - \mathcal{B} \partial \mathcal{C} - \frac{1}{2} \Sigma \partial \Gamma + \frac{1}{2}\Gamma\partial \Sigma -\partial \mathcal{X}^+ \partial \mathcal{X}^- -\frac{1}{2}\Upsilon \partial \Pi -\frac{1}{2} \Pi \partial \Upsilon \ , \\
    \mathcal G^+_{\text{aux}} &= \Gamma \mathcal{B} + \partial \mathcal{X}^+ \Pi\ , \label{eq:added-g+-space-time}\\
    \mathcal{G}^-_{\text{aux}} &= -\Sigma \partial \mathcal{C} - \partial \mathcal{X}^- \Upsilon \ , \\
    \mathcal{J}_{\text{aux}} &= - \tfrac12 \Sigma\Gamma -\tfrac12 \Upsilon \Pi \ ,    \label{eq:aux-space-time-j3}
\end{align}
\end{subequations}
see eq.~\eqref{N=2-conventions-untwisted} for our conventions. We have a similar construction also in the right-moving sector. Adding the generators \eqref{auxiliary-n=2-space-time} to the $\mathcal N=2$ generators of $\mathbb T^4$, see eq.~\eqref{N=2-T4-untwisted}, we obtain an untwisted $\mathcal N=2$ superconformal algebra with $\mathtt c =6$, which can be extended to an untwisted $\mathcal N=4$ superconformal algebra. We spell out these generators in Appendix~\ref{app:dual}.

\subsection{Holographic match of partition functions}
\label{sec:partition-functions-match}

In this section, we show that the partition function of the auxiliary string theory described in Section~\ref{sec:auxiliary-string} reproduces the torus partition function of the space-time CFT \eqref{eq:symm-dual-C0} introduced in the previous section. The corresponding holographic match in the tensionless duality was derived in \cite{Eberhardt:2018ouy, Eberhardt:2020bgq}. Here we closely follow the analysis of \cite{Eberhardt:2018ouy}, which computes the
single-cycle contribution to the symmetric orbifold partition function, and apply it to the symmetric orbifold in eq.~\eqref{eq:symm-dual-C0}.

\subsubsection{The boundary torus partition function}

The symmetric orbifold partition function is calculated using Dijkgraaf-Moore-Verlinde-Verlinde (DMVV) formula \cite{Dijkgraaf:1996xw}. In particular, the single-cycle spectrum is discussed in detail in \cite{Gaberdiel:2018rqv}. Given the universality of the DMVV formula with respect to the seed theory, we will mostly omit the details related to the calculation of symmetric orbifold partition function and focus on the seed theory $\mathbb{T}^4 \times \mathcal{A}_0$. We write the space-time torus partition function as
\begin{equation}
    \mathcal{Z}_{\text{ST}}(t) = \text{Tr}_{\mathcal H_{\text{ST}}}\Bigl[ e^{2 \pi i t \, (\mathcal L_0-\frac{1}{4})} e^{-2 \pi i \bar t \, (\bar{\mathcal {L}}_0-\frac{1}{4})}\Bigr] = \mathcal{Z}_{\mathbb{T}^4}(t) \mathcal{Z}_{\text{aux}}(t) \ ,
\end{equation}
where $t$ is the torus modular parameter, and $\mathcal{Z}_{\text{aux}}(t)$ denotes the partition function of the auxiliary boundary CFT discussed in Section~\ref{sec:auxiliary-space-time-CFT}. The $\mathbb{T}^4$ theory has $4$ free bosons and $4$ free fermions, so we get
\begin{equation} \label{ZT6}
    \mathcal{Z}_{\mathbb{T}^4}(t) = \Theta(t) \frac{|\vartheta_3(0;t)|^4}{|\eta(t)|^{12}} \ ,
\end{equation}
where $\vartheta_i$ for $i=1, \dots, 4$ denote Jacobi theta functions. In eq.~\eqref{ZT6}, $\Theta(t)$ denotes the Narain theta function describing the winding and momenta along $\mathbb{T}^4$,
\begin{equation}
\label{Theta-t}
    \Theta(t) = \sum_{(p,\bar p) \in \Gamma_{4,4}} e^{i \pi t p^2} e^{-i \pi \bar t \bar p^2} \ ,
\end{equation}
where $\Gamma_{4,4}$ is the Narain lattice. Recall that the auxiliary space-time theory consists of a $\mathtt c=3$ and a $\mathtt c=-3$ sector. Having this, we write
\begin{equation}
    \mathcal{Z}_{\text{aux}}(t) = \mathcal{Z}_{\mathtt c=3}(t) \mathcal{Z}_{\mathtt c=-3}(t) \ ,
\end{equation}
where $\mathcal{Z}_{\mathtt c=\pm 3}(t)$ denotes the corresponding partition function. The $\mathtt c=3$ sector consists of $2$ non-compact bosons $\mathcal{X}^\pm$ and a pair of free fermions $(\Upsilon,\Pi)$, so we have
\begin{equation}
    \mathcal{Z}_{\mathtt c=3}(t) = \frac{1}{2 t_2} \frac{|\vartheta_3(0;t)|^2}{|\eta(t)|^{6}} \ .
\end{equation}
The $\mathtt c=-3$ sector consists of $(\mathcal{B},\mathcal{C})$ and $(\Sigma,\Gamma)$, which cancel out $2$ bosonic and $2$ fermionic degrees of freedom, respectively. The associated partition function reads\footnote{The $\mathtt c=-3$ sector, both on the worldsheet and in space-time, have infinitely many ground states given by $e^{m\mathcal{Y}}$ with $m \in \mathbb{Z}$, see eq.~\eqref{bosonized-auxiliary-ghosts}. In \eqref{eq:ghost-pt-space-time} we focus on $m=0$. One can refine the partition function by introducing an additional chemical potential to count the contribution of the other ground states both in the bulk and in space-time. Moreover, $2t_2$ in \eqref{eq:ghost-pt-space-time} appears as an integration over the zero modes of $(\mathcal{B},\mathcal{C})$, with the convention that the measure is $\text{d}z \wedge \text{d}\bar{z}$.\label{footnote:ghost-ground-states}}
\begin{equation} \label{eq:ghost-pt-space-time}
    \mathcal{Z}_{\mathtt c=-3}(t) = 2 t_2 \frac{|\eta(t)|^6}{|\vartheta_3(0;t)|^2} \ .
\end{equation}
As a result, we get
\begin{equation}
    \mathcal{Z}_{\text{ST}}(t) = \Theta(t) \frac{|\vartheta_3(0;t)|^4}{|\eta(t)|^{12}} \ .
\end{equation}
We note that this is exactly the same partition function as the $\mathbb{T}^4$ theory. We now reproduce the partition function of the symmetric orbifold \eqref{eq:symm-dual-C0} from the worldsheet.

\subsubsection{The string partition function}

The worldsheet partition function
\begin{equation}
    Z_{\text{string}}(\tau,t) = \text{Tr}_{\mathcal H}\Bigl[e^{2 \pi i \tau T_0} e^{-2 \pi i \bar \tau \bar T_0} e^{2 \pi i t J^3_0} e^{-2 \pi i \bar t \bar J^3_0}\Bigr]
\end{equation}
of the auxiliary string theory factorizes into different sectors 
\begin{equation} \label{Z-string-factorized}
    Z_{\text{string}}(\tau,t) = Z_{\mathfrak{psu}}(\tau,t) Z_{\rho\sigma}(\tau) Z_{\mathbb{T}^4}(\tau) Z_{\text{aux}}(\tau) \ ,
\end{equation}
which we now describe in turn. Here $\tau$ is the worldsheet torus modular parameter, while $t$ is the chemical potential for $J^3_0$.

\paragraph{The $\boldsymbol{\mathfrak{psu}(1,1|2)}_{\boldsymbol{1}}$ spectrum.} The contribution of the sectors $\mathcal F_{\beta \gamma}$ and $\mathcal F_{p \theta}$, which generate the affine algebra $\mathfrak{psu}(1,1|2)_1$ has been computed in \cite{Eberhardt:2018ouy} and reads
\begin{equation} \label{eq:psu-ptfunc}
    Z_{\mathfrak{psu}}(\tau,t) = \sum_{w,m \in \mathbb Z} |q|^{w^2} \left| \frac{\vartheta_2(\frac{t}{2};\tau)}{\eta(\tau)^2} \right|^4 \delta^{(2)}(t-w \tau +m) \ , 
\end{equation}
where $w$ labels the sum over different spectral flow sectors.

\paragraph{The $\boldsymbol \rho$ and $\boldsymbol \sigma$ spectrum.} The free bosons $\rho$ and $\sigma$ cancel the contribution of two pairs of topologically twisted fermions and of two free bosons. The associated partition function reads \cite{Eberhardt:2018ouy}
\begin{equation}
    Z_{\rho\sigma}(\tau) = \frac{|\eta(\tau)|^8}{|\vartheta_2(0,\tau)|^4} \ ,
\end{equation}

\paragraph{The $\mathbb{T}^4$ spectrum.} The partition function $Z_{\mathbb{T}^4}$ in eq.~\eqref{Z-string-factorized} captures the contribution of the four free bosons and two pairs of topologically twisted fermions associated with the tensionless string. It is then given by 
\begin{equation}
    Z_{\mathbb{T}^4}(\tau) = \Theta(\tau) \frac{|\vartheta_2(0;\tau)|^4}{|\eta(\tau)|^{12}} \ , 
\end{equation}
where 
\begin{equation}
    \Theta(\tau) = \sum_{(p,\bar p) \in \Gamma_{4,4}} q^{\frac{p^2}{2}} \bar q^{\frac{\bar p^2}{2}} 
\end{equation}
denotes the Narain theta function and $\Gamma_{4,4}$ is the Narain lattice of momenta and winding numbers and $q=e^{2\pi i \tau}$, $\bar q=e^{-2\pi i \bar \tau}$.   

\paragraph{The auxiliary string spectrum.} As before, the auxiliary string theory consist of a $\mathtt c=3$ and a $\mathtt c=-3$ sector and we write\footnote{As mentioned in Section~\ref{sec:auxiliary-string}, the central charge here refers to the central term in the Cartan of $\text{U}(1)$.}
\begin{equation}
    Z_{\text{aux}}(\tau) = Z_{\mathtt c=3}(\tau) Z_{\mathtt c=-3}(\tau) \ .
\end{equation}
The $\mathtt c=3$ sector consists of two non-compact bosons $X^\pm$ and a pair of topologically twisted fermions $(\psi,\pi)$ with weights $(1,0)$. Therefore, the partition function reads
\begin{equation}
    Z_{\mathtt c=3}(\tau) = \frac{1}{2\tau_2} \frac{|\vartheta_2(0;\tau)|^2}{|\eta(\tau)|^{6}} \ .
\end{equation}
We finally come to the contribution of the $\mathtt c=-3$ sector consisting of $(b^{\text{(g)}},c^{\text{(g)}})$ and $(\beta^{\text{(g)}},\gamma^{\text{(g)}})$. As will become more clear in the next section, the latter cancels out the contributions of a pair of topologically twisted fermions, while the former cancels out two free bosons, so we have\footnote{The degeneracy that is discussed in footnote~\ref{footnote:ghost-ground-states} also exists on the worldsheet, as $(\beta^{\text{(g)}},\gamma^{\text{(g)}})$ is a $\beta\gamma$-system.}
\begin{equation}
    Z_{\mathtt c=-3}(\tau) = 2\tau_2 \, \frac{|\eta(\tau)|^6}{|\vartheta_2(0,\tau)|^2} \ . 
\end{equation}

\paragraph{Assembling all the contributions.} Putting all contributions together, we obtain
\begin{equation}  \label{Z-string-final}
    Z_{\text{string}}(\tau,t) = Z_{\mathfrak{psu}}(\tau,t) \frac{\Theta(\tau)}{|\eta(\tau)|^4} \ .
\end{equation}
As in the boundary theory analysis of the previous section, eq.~\eqref{Z-string-final} coincides exactly with the partition function of the tensionless string. It then follows from the analysis of \cite{Eberhardt:2018ouy} that the string partition function agrees with the single-cycle spectrum of the symmetric orbifold in eq.~\eqref{eq:symm-dual-C0}.\footnote{We note that eq.~(5.5) of \cite{Eberhardt:2018ouy} remains unchanged in our setting and follows from the mass-shell condition $T_0=0$.}

Because the auxiliary sector has vanishing total central charge, the partition function match is not fully constraining: the auxiliary fields cancel in the final expression for the torus partition function. The DDF operator construction in the next subsection provides a more detailed test of the proposed auxiliary duality, since it identifies the individual free-field excitations of the auxiliary boundary theory directly on the worldsheet.

\subsection{Physical states}
\label{sec:physical-states}

In this section, we present several examples of physical states and explain how the spectrum of the auxiliary boundary CFT \eqref{eq:symm-dual-C0} can be expressed in terms of the worldsheet fields~\eqref{F-aux}.

\smallskip The spectrum of the symmetric orbifold of $\mathbb T^4$ is generated by repeated applications of the fractionally moded oscillators of four free bosons and fermions on twisted sector ground states. In Section~\ref{sec:tensionless} we reviewed how the latter are realized on the worldsheet, see eqs.~\eqref{eq:w-odd-tensionless} and \eqref{eq:w-even-tensionless}. The action of the free bosons and fermions of the boundary CFT is encoded in the bulk by the DDF operators \eqref{eq:t4-ddf-tensionless}, which satisfy eq.~\eqref{eq:DDF-physical-state-conditions} and map physical states to physical states. 

Similarly, as further evidence of the auxiliary duality described in Sections~\ref{sec:auxiliary-string} and \ref{sec:auxiliary-space-time-CFT}, we provide below worldsheet expressions for the twisted sector ground states of the symmetric orbifold \eqref{eq:symm-dual-C0} and construct DDF operators for all free fields that generate the auxiliary boundary CFT spectrum.

\paragraph{Twisted sector ground states.} We begin by considering
\begin{equation} \label{eq:ground-ansatz-auxiliary}
    \Phi_w(x,z) \, e^{(q+1)(u+i\chi)} \ ,
\end{equation}
where, with a little abuse of notation, $\Phi_w$ is given in eqs.~\eqref{eq:w-odd-tensionless} and \eqref{eq:w-even-tensionless}, depending on the parity of $w$. For any $q \in \mathbb Z_{\geq -1}$, the fields \eqref{eq:ground-ansatz-auxiliary} satisfy eqs.~\eqref{eq:auxiliary-physical-state-conditions}, with the $\mathcal N=4$ generators given by eqs.~\eqref{eq:n=2-lambda0-after} and \eqref{eq:n=4-lambda0}, and are therefore physical in the auxiliary string theory. Note that the fields \eqref{eq:ground-ansatz-auxiliary} have space-time conformal dimension 
\begin{equation} \label{eq:weights}
\begin{aligned}
&\frac{w^2-1}{4w}  \qquad \qquad & \text{for } & w \text{ odd} \ , \\
& \frac{w}{4} \qquad \qquad & \text{for } & w \text{ even} \ , 
\end{aligned}
\end{equation}
for all $q \in \mathbb Z_{\geq -1}$, which is the space-time conformal dimension expected for $w$-twisted symmetric orbifold ground states. One can check that \eqref{eq:ground-ansatz-auxiliary} also have, for any $q \in\mathbb Z_{\geq -1}$, the correct space-time $\mathcal{R}$-charge under $K^3_0$ to be identified with symmetric orbifold ground states. For each $w \in \mathbb Z_{> 0}$, it is then natural to ask which value of $q$ corresponds to the $w$-twisted sector ground state, and how the remaining values should be interpreted. To answer these questions, we notice that picture raising the vertex operator~\eqref{eq:ground-ansatz-auxiliary} with $q=0$ and $w=1$ yields the space-time identity in picture $P=0$, 
\begin{equation}
    \mathcal{I} = \oint \text dz \, \gamma^{-1}\partial \gamma \ .
\end{equation}
Here we have used the procedure discussed in \cite{Kutasov:1999xu,Naderi:2024wqx} to relate the space-time modes of a physical vertex operator to DDF operators, see in particular Section~4.3 in \cite{Naderi:2024wqx}. It is therefore natural to identify \eqref{eq:ground-ansatz-auxiliary} with $q=0$ and $w=1$ with the space-time CFT vacuum, and more generally to identify
\begin{equation} \label{twisted-ground-states-aux}
    \ket{0}_w \equiv \Phi_w(x,z) e^{(u+i\chi)}
\end{equation}
as the ground states of the $w$-twisted sectors. As a further check, in Section~\ref{sec:corr} we will show  that, due to background charge conservation, only for $q=0$ do the fields in \eqref{eq:ground-ansatz-auxiliary} have a constant non-vanishing two-point function, as expected for twisted sector ground states of the boundary CFT. 

It remains to determine the interpretation of the states \eqref{eq:ground-ansatz-auxiliary} for the remaining values of $q$. Owing to the presence of the auxiliary free fields \eqref{eq:ghost-added-aux}, the seed CFT $\mathbb T^4 \times \mathcal A_0$ possesses infinitely many ground states. A subset of these can be represented in terms of the bosonized fields \eqref{bosonized-auxiliary-ghosts} as
\begin{equation}
    e^{f(\mathcal Y + i \mathcal K)} \ , 
\end{equation}
where $f \in \mathbb Z$. The special feature of these fields is that they have vanishing
conformal weights for any $f\in\mathbb Z$. As a result, the corresponding symmetric product orbifold contains infinitely many states with the same conformal dimension and $\mathcal R$-charges as the $w$-twisted ground states, see Appendix~\ref{app:dual}. It is therefore natural to identify
\eqref{eq:ground-ansatz-auxiliary} with
\begin{equation} \label{eq:degeneracy-twisted-ground-states}
    \left( e^{f(q)(\mathcal Y + i \mathcal K)} \right)_0 \ket{0}_w \equiv \Phi_w(x,z) e^{(q+1)(u+i\chi)} \ , 
\end{equation}
where $f=f(q)$ should be thought as a function of $q$ to be determined, and where $e^{f(q)(\mathcal Y + i \mathcal K)}$ should be understood as the associated DDF operator. From the analysis of $n$-point functions that we will present in Section~\ref{sec:corr}, one finds 
\begin{equation} \label{eq:q-n-conditions}
    \sum_{j=1}^n f(q_j) = 0 \ , \qquad \text{for } \sum_{j=1}^n q_j = 0 \ , 
\end{equation}
which implies that $f(q)$ is a $\mathbb{Z}$--linear and odd function. This leads to the natural conclusion that $f(q) = q$, and hence to the identification
\begin{equation} \label{DDF-twisted-vacuum-id}
    \left( e^{q(\mathcal Y + i \mathcal K)} \right)_0 \ket{0}_w = \Phi_w(x,z) e^{(q+1)(u+i\chi)} \,.
\end{equation}
Finally, in Section~\ref{sec:corr} we provide a further check of the identification \eqref{DDF-twisted-vacuum-id} by computing the correlation functions of the vertex operators appearing on the two sides of eq.~\eqref{DDF-twisted-vacuum-id} and showing exact agreement.  

\paragraph{DDF operators.} As anticipated, we now construct DDF operators for all the free fields of the auxiliary CFT. As in the tensionless string case, DDF operators satisfy 
\begin{equation} \label{eq:DDF-physical-state-conditions-auxiliary}
    [G^+_0,\mathcal{D}] = [\widetilde G^+_0,\mathcal{D}] = [J_0,\mathcal{D}] = [T_0,\mathcal{D}] = 0 \ , 
\end{equation}
where the $\mathcal N=4$ generators are now given by eqs.~\eqref{eq:n=2-lambda0-after} and \eqref{eq:n=4-lambda0}. 

Let us begin with the DDF operators associated with the free bosons and fermions of $\mathbb{T}^4$, which are given by a slight generalization of eqs.~\eqref{eq:t4-ddf-tensionless},
\begin{subequations} \label{eq:t4-ddf}
    \begin{align}
        \partial \bar{\mathcal{X}}^j_n &= \oint \text dz \, \partial \bar X^j \gamma^{n} \ , & \qquad \partial \mathcal X^j_n &= \oint \text dz \, \left[ \partial X^j \gamma^n+n \gamma^{n-1} e^{\rho+iH^j} \theta^2 \theta^1\right] \ , \label{eq:t4-ddf-bosons} \\
        \Lambda^{+,j}_r &= \oint \text dz \, \theta^1 \gamma^{r-\frac{1}{2}} e^{\rho+iH^j} \ , & \qquad \Lambda^{-,j}_r &= -\oint \text dz \, \theta^2 \gamma^{r-\frac{1}{2}} e^{\rho+iH^j} e^{u+i\chi} \ ,  \label{eq:t4-ddf-fermions}
    \end{align}
\end{subequations}
with $j\in\{1, 2\}$. One can verify that the DDF operators given above satisfy eqs.~\eqref{eq:DDF-physical-state-conditions-auxiliary}, obey\footnote{In order to confirm \eqref{eq:fermions-ddf-tensionless-anti-commutator} explicitly, one may picture-raise one of the fermions.}
\begin{subequations}
\begin{align}
    [\partial \mathcal X^i_m, \partial  \bar{\mathcal{X}}^j_n] &= m \, \delta^{ij} \, \delta_{m+n,0} \, \mathcal{I} \ , \\
    \{\Lambda^{\alpha,j}_r,\Lambda^{\beta,l}_s\} & = \epsilon^{\alpha\beta} \epsilon^{lj} \delta_{r+s,0} \, \mathcal{I} \ , \label{eq:fermions-ddf-tensionless-anti-commutator}
\end{align}
\end{subequations}
and (anti-)commute among themselves. 

Let us now turn to the excitations of $\mathcal{A}_0$ in \eqref{eq:symm-dual-C0}. The DDF operators associated with the space-time fields $\mathcal X^{\pm}$ and $(\Upsilon,\Pi)$ can be constructed by exploiting the symmetry that maps the worldsheet $\mathbb T^4$ bosons to the auxiliary free bosons $\partial X^\pm$, and rotates the worldsheet $\mathbb T^4$ fermions into the auxiliary fields $(\psi,\pi)$. In this way, one finds 
\begin{subequations} \label{eq:ddf-aux1}
\begin{align}
        \partial \mathcal X^+_n &= \oint \text dz \, \partial X^+ \gamma^{n} \ , & \qquad \partial \mathcal X^-_n &= \oint \text dz \, \left[ \partial X^- \gamma^n-n \gamma^{n-1} e^{\rho+i\chi} \theta^2 \theta^1\right] \ . \label{eq:ddf-aux1-X}\\
        \Pi_r &= \oint \text dz \, \theta^1 \gamma^{r-\frac{1}{2}} e^{\rho+i \chi} \ , & \qquad \Upsilon_r &= -\oint \text dz \, \theta^2 \gamma^{r-\frac{1}{2}} e^{\rho+u+iH} \ .  \label{Upsilon-Pi-DDF}
\end{align}
\end{subequations}
Again, one can verify that the DDF operators \eqref{eq:ddf-aux1} satisfy eq.~\eqref{eq:DDF-physical-state-conditions-auxiliary}, (anti-)commute with all the previously defined DDF operators and close into the algebra
\begin{subequations}
\begin{align}
    [\partial \mathcal{X}^+_m,\partial \mathcal X^-_n] &= - m \, \delta_{n+m,0} \, \mathcal{I} \ , \\
    \{ \Upsilon_r, \Pi_s \} &= \delta_{r+s,0} \, \mathcal I \ . 
\end{align}
\end{subequations}
For the DDF operators associated with $(\mathcal{B},\mathcal{C})$, we find
\begin{subequations} \label{eq:ddf-aux2}
    \begin{equation}
        \mathcal{B}_n = \oint \text  dz \, b^{\text{(g)}} \, \gamma^{n} \ , \qquad \mathcal{C}_n = \oint \text dz \, \left[ c^{\text{(g)}} \gamma^{n-1} \partial \gamma - \gamma^{\text{(g)}} e^{\rho} \theta^2 \theta^1 \gamma^{n-1} \right] \ ,
    \end{equation}
\end{subequations}
which satisfy
\begin{subequations} \label{DDF-B-C}
    \begin{equation}
        \{ \mathcal{B}_m,\mathcal{C}_n \} = \delta_{m+n,0} \, \mathcal{I} \ .
    \end{equation}
\end{subequations}
To identify the DDF operators associated with the auxiliary ghosts $(\Sigma, \Gamma)$, it is convenient to use their bosonized expressions \eqref{bosonized-auxiliary-ghosts} and construct separetely the DDF operators corresponding to the fields $e^{-\mathcal Y}$, $e^{\mathcal Y}$, $\Xi$ and $\mathcal N$. The DDF operators associated with $\mathcal N$ and $\Xi$ are
\begin{equation}
 \mathcal{N}_n = \oint \text dz \, \gamma^n e^{-iv} \ , \qquad \Xi_n = \oint \text dz \, \left[ e^{iv} \gamma^{n-1} \partial \gamma -  e^{-u} b^{\text{(g)}} \gamma^{n-1} e^{\rho}\theta^2 \theta^1 \right] \ ,
\end{equation}
where we recall that $iv$ is the free boson entering the bosonization of $\gamma^{\text{(g)}}$, see eq.~\eqref{eq:bosonizations-added-ghosts}. We have checked that $\mathcal{N}_n$ and $\Xi_n$ satisfy eq.~\eqref{eq:DDF-physical-state-conditions-auxiliary}, (anti-)commute with all the DDF operators introduced so far and obey
\begin{equation}
    \{\mathcal{N}_m,\Xi_n\} = \delta_{m+n,0} \, \mathcal{I} \ .
\end{equation}
Finally, to construct the DDF operators associated with the fields $e^{\pm \mathcal Y}$, it is sufficient to act with the DDF operators \eqref{Upsilon-Pi-DDF} on the ground states \eqref{DDF-twisted-vacuum-id} with $q=\pm 1$. In this way, we find
\begin{subequations}
\begin{align}
        (e^{\mathcal Y})_r &= \oint \text{d}z \, \theta^1 \partial \theta^2 \theta^2 \gamma^{r-\frac{3}{2}} e^{3\rho+2iH+3u+2i\chi} \ , \label{eYDDF}\\
        (e^{-\mathcal Y})_r &= \oint \text{d}z \, \partial \theta^1 \theta^1 \theta^2 \gamma^{r-\frac{3}{2}} e^{3\rho+iH+i\chi} \ .
\end{align}
\end{subequations}
By construction, these are DDF operators as well; they (anti-)commute with all the other DDF operators and close into
\begin{equation}
    \{(e^{\mathcal Y})_r ,(e^{-\mathcal Y})_s \} = 0 \ ,
\end{equation}
as expected.

\subsection{Embedding of the tensionless duality}
\label{sec:embedding}

The tensionless duality reviewed in Section~\ref{sec:tensionless} can be embedded in the auxiliary duality described above. In fact, both the tensionless string theory and the symmetric orbifold of $\mathbb T^4$ can be regarded as subsectors of the auxiliary string theory and of the symmetric orbifold \eqref{eq:symm-dual-C0}, respectively. 
Indeed, since $\mathbb{T}^4$ and $\mathbb T^4 \times \mathcal A_0$ have the same central charge, every state of the symmetric orbifold $\text{Sym}^N(\mathbb{T}^4)$ can be mapped to a state of the symmetric orbifold $\text{Sym}^N(\mathbb T^4 \times \mathcal A_0)$ by: 1) writing it in terms of free fermionic and bosonic oscillators acting on a twisted sector ground state; 2) mapping each such oscillator and ground state to the corresponding free field and twisted sector ground state in $\text{Sym}^N(\mathbb T^4 \times \mathcal A_0)$.

A similar construction in principle can be mirrored out on the worldsheet: one first rewrites the physical states in terms of the DDF operators \eqref{eq:t4-ddf-tensionless} and twisted sector ground states \eqref{eq:w-odd-tensionless}, \eqref{eq:w-even-tensionless} and then maps them to the corresponding expressions~\eqref{eq:t4-ddf} and \eqref{twisted-ground-states-aux} in the auxiliary string theory.\footnote{Here we have implicitly assumed that any physical state of the tensionless string can be written in terms of the DDF operators \eqref{eq:t4-ddf-tensionless} acting on the twisted sector ground states \eqref{eq:w-odd-tensionless} and \eqref{eq:w-even-tensionless}. Although, strictly speaking, this has not been proven explicitly, the analysis of \cite{Naderi:2022bus}, which puts a lower bound on the string spectrum, together with the counting analysis of \cite{Eberhardt:2018ouy}, essentially shows that this is the case.}

For later purposes, it is useful to describe this embedding from a complementary viewpoint. Focusing on the untwisted sector, the states in $\text{Sym}^N(\mathbb T^4 \times \mathcal A_0)$ that lie in the image of $\text{Sym}^N(\mathbb T^4)$ under the map defined above can equivalently be characterized by the absence of auxiliary excitations, namely
\begin{subequations} \label{eq:null-gauging-ST}
\begin{equation}
    \partial \mathcal X^+_n \phi = 0 \ ,   \qquad \bar \partial \mathcal X^-_n \phi = 0 \ ,   \qquad \text{for } n \geq 0 \ , \label{mathcaldXp=0}
\end{equation}
and
\begin{align}
    \mathcal B_n \phi  &= 0 \ ,   \qquad \overline{\mathcal{B}}_n \phi  = 0 \ ,   \qquad \text{for } n \geq 1 \ , \\
    \Gamma_r \phi  &= 0 \ ,   \qquad \overline{\Gamma}_r \phi = 0 \ ,   \qquad \text{for } r \geq \frac{1}{2} \,,
    \\
    \Pi_r \phi  &= 0 \ ,   \qquad \overline{\Pi}_r \phi  = 0 \ ,   \qquad \text{for } r \geq \frac{1}{2} \,.
\end{align}
\end{subequations}
These conditions restrict us to excitations that involve only the $\mathbb{T}^4$, together with conjugate fields of auxiliary fields, e.g.\ the conjugate of $\mathcal{B}$ is $\mathcal{C}$. Since the conjugate fields have trivial OPEs between each other, the gauging imposed by the conditions in \eqref{eq:null-gauging-ST} is in fact a null gauging and effectively removes all the $\mathcal{A}_0$ excitations.\footnote{We do not impose the zero-mode conditions $\mathcal{B}_0 \phi=\overline{\mathcal{B}}_0 \phi=0$. Indeed, in the
space-time theory, the $(\mathcal{B},\mathcal{C})$ system carries a non-zero background charge, so both the vacuum and $\mathcal{C}$ insertions are needed for correlation functions to be non-vanishing.} Therefore, imposing eqs.~\eqref{eq:null-gauging-ST} leaves us with the $\mathbb{T}^4$ excitations.

By rewriting the null gauging conditions \eqref{eq:null-gauging-ST} in terms of the corresponding DDF operators discussed in Section~\ref{sec:physical-states}, one expects this null gauging to have a worldsheet realization that removes the bulk auxiliary fields. A detailed derivation, however, appears challenging, since it would require a picture independent formulation of the gauging in terms of the DDF operators entering eqs.~\eqref{eq:null-gauging-ST}.

\section{Deforming the duality} 
\label{sec:deforming}

The goal of this section is to identify the conditions under which a worldsheet vertex operator is physical in the deformed string theory of eq.~\eqref{ads3/cft2-deformed}. As reviewed in Section \ref{sec:tensionless} for the tensionless string, within the hybrid formalism of Berkovits, Vafa and Witten, physical states are defined by the conditions \eqref{eq:physical-state-conditions}, and $G^+, \widetilde G^+, T$ and $J$ generate a topologically twisted algebra. In the previous section we explained how the tensionless duality can be embedded into the auxiliary holographic duality. This suggests that the deformed string theory \eqref{ads3/cft2-deformed} should likewise admit a description in terms of a suitable deformation of the auxiliary string theory. In the present section, we construct such a deformation of the auxiliary string $\mathcal N=4$ superconformal algebra on the worldsheet and propose it as the framework for defining physical vertex operators in the deformed string theory~\eqref{ads3/cft2-deformed}. 

The construction proceeds in the following steps. First, in Section \ref{sec:path-integral-N=2-N=4} we discuss the path integral formulation of the worldsheet CFT. Next, in Section~\ref{N=2-and-N=4-deformed} we deform the $\mathcal N=2$ algebra \eqref{eq:n=2-lambda0-after}, which is then extended to an $\mathcal N=4$ algebra, providing the algebraic structure needed to define the physical vertex operators and correlation functions. We then test the resulting proposal from two complementary viewpoints. In Section~\ref{sec:vertex}, we construct physical vertex operators of the deformed string theory. This sets the stage for the computation of correlation functions in Section~\ref{sec:corr}, which provides a further test of our proposal. In Section~\ref{sec:global}, we analyze the global symmetries of the resulting space-time theory and explain how the deformation breaks boundary conformal symmetry.

\subsection{Path integral} 
\label{sec:path-integral-N=2-N=4}

Before turning into the technical details, we begin by developing the relevant intuition through a reformulation of the path integral of the deformed worldsheet CFT. This path-integral representation was employed in \cite{Dei:2024sct} to demonstrate the agreement between the bulk and boundary partition functions, and we rewrite it here in a different but equivalent form. Our discussion here is similar in some respects to the analysis of \cite{Apolo:2019zai,Cui:2023jrb}.

\smallskip

In \cite{Dei:2024sct} the deformation dual to the single-trace $\TT$-deformation of the symmetric orbifold of $\mathbb T^4$ was shown to take the form   
\begin{equation} \label{beta-betabar}
    \int \text d^2 z \, \beta \bar \beta \ , 
\end{equation}
where the integration is performed over the string worldsheet and $\bar \beta$ denotes the anti-holomorphic analogue of $\beta$, see eq.~\eqref{F}. More specifically, in the deformed worldsheet theory, the free-field action \eqref{S0} is promoted to
\begin{equation} \label{eq:deformed-action-section4}
    S_{\beta \gamma}^\mu = S_{\beta \gamma} - \frac{\mu^2}{2\pi} \int \text d^2 z \, \beta \bar \beta \ , 
\end{equation}
where $\mu^2$ denotes the worldsheet deformation parameter.\footnote{More specifically, the parameter $\mu^2$ is related to the deformation parameter $\lambda$ used in \cite{Dei:2024sct} via $\lambda=\mu^2$. Also, $\mu^2=2\tilde{\lambda}$ is the relation to the deformation parameter used in \cite{Cui:2023jrb}.} The associated path integral\footnote{The worldsheet path integral also depends on the additional free fields appearing in the tensionless string construction; see eq.~\eqref{F}. Since these fields are unaffected by the worldsheet deformation \eqref{beta-betabar}, we suppress them in the following in order to streamline the notation.}
\begin{equation} \label{Bob-Sav-path-integral}
    \int \mathcal D \gamma \, \mathcal D \beta \, \exp \Bigl[- S_{\beta \gamma} + \frac{\mu^2}{2\pi} \int \text d^2 z \, \beta \bar \beta \Bigr]
\end{equation}
was used in \cite{Dei:2024sct} to compute the string partition function on the massless BTZ geometry, also known as the `cusp AdS$_3$ geometry', and to reproduce the expected boundary torus partition function. In the tensionless duality \eqref{ads3/cft2}, the worldsheet field $\gamma$ is identified with the space-time position coordinate $x$, and the twisted boundary conditions appropriate to the massless BTZ quotient are implemented as
\begin{equation} \label{BTZ-quotient}
    \gamma \sim \gamma +1 \sim \gamma + t \ , 
\end{equation}
so that this identification on the worldsheet captures the quotient of the space-time plane to a torus of modular parameter $t$.

\smallskip

Taking into account the auxiliary fields \eqref{Xpm-trivial}, we promote the path integral \eqref{Bob-Sav-path-integral} to\footnote{Again, since they are not affected by the deformation, we suppress other auxiliary fields.}   
\begin{equation} \label{path-integral-with-X}
    \int \mathcal D \gamma \, \mathcal D \beta \, \mathcal D X^+ \, \mathcal D X^- \, \exp \Bigl[- S_{\beta \gamma}-S_X + \frac{\mu^2}{2\pi} \int \text d^2 z \, \beta \bar \beta \Bigr]  \ , 
\end{equation}
where the action
\begin{equation}
    S_X = \frac{1}{2\pi} \int \text d^2z \, \partial X^+ \bar \partial X^-
\end{equation}
is the standard free boson action.

Guided by \cite{Alday:2005ww,Apolo:2019zai,Cui:2023jrb}, we now notice a simple but important point: after a change of variables, the action \eqref{path-integral-with-X} takes a free-field form. Define\footnote{$\int \beta$ and $\int \bar \beta$ are defined as solutions to $\partial \mathcal{G}(z,\bar z)=\beta(z,\bar z)$ and $\bar \partial \bar{\mathcal{G}}(z,\bar z)=\bar\beta(z,\bar z)$, respectively. On a worldsheet with a non-trivial topology, \eqref{eq:sbeta-free} holds modulo zero modes of $\beta$ and $\bar \beta$.}
\begin{subequations} \label{path-integral-change-variables}
\begin{align}
    \gamma_\mu & = \gamma - \mu X^- \ , & \qquad \bar \gamma_\mu & = \bar \gamma + \mu X^+ \ ,  \\
    X^+_\mu &= X^+ +\mu \int \beta \ , & \qquad X^-_\mu & = X^- -\mu \int \bar \beta \ . 
\end{align}
\end{subequations}
Then the action entering the path integral \eqref{path-integral-with-X} can be rewritten as\begin{equation} \label{eq:sbeta-free}
    S_{\beta \gamma} + S_{X} -\frac{\mu^2}{2\pi} \int \text{d}^2 z \, \beta \bar \beta =\frac{1}{2\pi} \int \text{d}^2 z \, \big[ \beta \bar \partial \gamma_\mu + \bar \beta \partial \bar \gamma_\mu + \partial X^+_\mu \bar \partial X^-_\mu\big] \ .
\end{equation}
Thus, at least at the level of the local worldsheet action, the variables $(\gamma_\mu,\bar\gamma_\mu,X^+_\mu,X^-_\mu)$ reveal a free-field structure: the deformed theory takes the form of a free $\beta\gamma$ system coupled to two free bosons. Moreover, one can check that the transformation \eqref{path-integral-change-variables} has unit Jacobian, so that the path integral measure is unchanged. Notice that the $\beta \bar \beta$ term entering the exponential in eq.~\eqref{path-integral-with-X} has been reabsorbed in the change of variables and is no longer present in the RHS of eq.~\eqref{eq:sbeta-free}.
This suggests that the deformation \eqref{eq:deformed-action-section4} of \cite{Dei:2024sct} can be understood as implementing the field redefinition \eqref{path-integral-change-variables}, while keeping the boundary condition \eqref{BTZ-quotient} fixed. In this sense, adding the auxiliary bosons \(X^\pm\) is what makes the deformation locally tractable: it allows the \(\mu^2\beta\bar\beta\) interaction to be rewritten in terms of free fields, while the non-trivial effect of the deformation remains encoded in the global data, in particular in the boundary conditions. The equality \eqref{eq:sbeta-free} might seem to imply that the $\mu^2\beta\bar\beta$ deformation is completely trivial in the path integral, reducing the theory to a free-field system. This is not the case, however, because the change of variables does not trivialize the global data of the path integral, in particular the boundary condition \eqref{BTZ-quotient}.

Guided by this intuition, we define the deformed string theory in eq.~\eqref{eq:deformed-action-section4} starting from the auxiliary string theory of Section~\ref{sec:auxiliary-duality} and applying the change of variables \eqref{path-integral-change-variables}. A careful implementation of this field redefinition leads precisely to a worldsheet $\mathcal N=4$ algebra, which in turn defines the physical operators of the deformed theory. We will elaborate on this construction in the next section and study its consequences in the remainder of the paper. Finally, we note that, for $\mu\neq0$, the fields $\gamma_\mu$
and $\bar\gamma_\mu$ are no longer holomorphic and anti-holomorphic, respectively. This fact will play an important role below.

\smallskip

As mentioned in the Introduction, the single-trace $\TT$-deformation duality \eqref{ads3/cft2-deformed} is expected to hold only at $k=1$. Nevertheless, the change of variables in \eqref{path-integral-change-variables} is closely related to structures that have appeared in the literature on deformed AdS$_3$ strings at $k>1$. In the coset descriptions of \cite{Giveon:2017nie, Giveon:2017myj, Asrat:2017tzd}, analogous combinations arise as gauge-invariant variables. Similarly, in the TsT descriptions of \cite{Alday:2005ww, Araujo:2018rho, Apolo:2019zai, Cui:2023jrb}, the field redefinition \eqref{path-integral-change-variables} has a close counterpart. Our use of this change of variables should therefore be viewed as a $k=1$ implementation of a structure suggested by these earlier constructions, rather than as a direct import of the $k>1$ duality.

\subsection[\texorpdfstring{The $\mathcal N=2$ and $\mathcal N=4$ algebras on the worldsheet}{The N=2 and N=4 algebras on the worldsheet}]{The $\boldsymbol{\mathcal N=2}$ and $\boldsymbol{\mathcal N=4}$ algebras on the worldsheet}\label{N=2-and-N=4-deformed}

The analysis of the previous section suggests that the $\mu^2 \beta \bar \beta$ deformation of the auxiliary string theory discussed in Section~\ref{sec:auxiliary-duality} can be formulated in terms of the change of variables \eqref{path-integral-change-variables}. Motivated by this intuition, we consider the following $\mathcal N=2$ algebra
\begin{subequations} \label{eq:n=2-lambda-after}
	\begin{align}
        T &= T_{\mathfrak{psu}} + T_{\rho} + T_{\sigma} + T_{\mathbb T^4} +T_{\text{aux}}\ , \\
		G^+ &= e^{-\rho} Q_\mu + e^{i\sigma} T - \partial[e^{i\sigma} (2J-i \partial\sigma)] +G^+_{\mathbb T^4}+\pi \partial X^+_\mu + \gamma^{\text{(g)}} b^{\text{(g)}} \label{G+-after-lambda}\ , \\
		G^- &= e^{-i\sigma} \ , \\ 
		J &= \tfrac{1}{2}\partial(\rho+i\sigma)+J_{\mathbb T^4}+J_{\text{aux}}=\tfrac{1}{2}\partial(\rho+i\sigma+iH+u+i\chi)\ ,     \label{eq:n=2-lambda-after-J}     
	\end{align}
\end{subequations}
where we defined
\begin{equation} \label{eq:q-betabetabar}
    Q_\mu = p_1 p_2 \partial \gamma_\mu \ . 
\end{equation}
One may verify that the fields in \eqref{eq:n=2-lambda-after} are obtained by implementing the substitution
\begin{equation}
    \gamma \mapsto \gamma_\mu \ , \qquad \bar \gamma \mapsto \bar \gamma_\mu \ , \qquad X^\pm \mapsto X_\mu^\pm \ ,  
\end{equation}
see eq.~\eqref{path-integral-change-variables}, at the level of the $\mathcal N=2$ algebra \eqref{eq:n=2-lambda0-after}. More precisely, the stress-tensor, the supercurrent $G^-$ and the $\mathcal R$-charge $J$ remain invariant, whereas the operator $Q$ and the term $\pi \partial X^+$ entering the definition of $G^+$, see eqs.~\eqref{eq:q-free} and \eqref{G+-after-lambda0}, are mapped to \eqref{eq:q-betabetabar} and $\pi \partial X^+_\mu$, respectively. One may further check that the generators \eqref{eq:n=2-lambda-after} satisfy the OPEs of a topologically twisted $\mathcal{N}=2$ algebra with $\mathtt c=6$.

The right-moving $\mathcal{N}=2$ algebra is defined analogously: $\bar Q = \bar p_1 \bar p_2 \bar \partial \bar \gamma$ is mapped to $\bar Q_\mu =  \bar p_1 \bar p_2 \bar \partial \bar \gamma_\mu$, while $\bar \pi \bar \partial X^-$ is mapped to $\bar \pi \bar\partial X^-_\mu$, see eq.~\eqref{path-integral-change-variables}. The left- and right-moving $\mathcal N=2$ algebras (anti-)commute with one another.

The BRST charge is the generator of the BRST symmetry of the gauge-fixed string action \eqref{Bob-Sav-path-integral} and is given by the zero mode of the supercurrent $G^+$. This suggests that the BRST structure associated with the deformed action
\eqref{Bob-Sav-path-integral} can be obtained by studying how the BRST current is modified as a function of the deformation parameter $\mu$. Although we have not derived the expression \eqref{G+-after-lambda} directly from a first-principles analysis of the gauge-fixed action, the free field rewriting \eqref{eq:sbeta-free}, together with the change of variables
\eqref{path-integral-change-variables}, strongly suggests that \eqref{G+-after-lambda} is the appropriate BRST current of the deformed theory. We should note, however, that the BRST charge is defined only up to total derivatives, so the expression
\eqref{G+-after-lambda} may differ from the current obtained directly from the gauge-fixed action by such terms.

\paragraph{The $\boldsymbol{\mathcal{N}=4}$ algebra.} Since the hybrid physical state conditions are naturally formulated in terms of a topologically twisted $\mathcal N=4$ algebra, not just its $\mathcal N=2$ subalgebra, we introduce the $\mathfrak{su}(2)_1$ $\mathcal{R}$-symmetry currents 
\begin{align}
    J^{++} &= e^{\rho+i\sigma+iH+u+i\chi} \ , &  J^{--} &= -e^{-(\rho+i\sigma+iH+u+i\chi)} \ .
\end{align}
and extend the $\mathcal N=2$ algebra \eqref{eq:n=2-lambda-after} to a topologically twisted $\mathcal{N}=4$ algebra by adjoining the generators
\begin{subequations} \label{eq:n=4-lambda}
	\begin{align}
        \widetilde G^+ &= e^{\rho + i H+u+i\chi}\ , \label{eq:lambda-gtildep}\\
        \widetilde G^- & = -e^{-2 \rho - i \sigma - iH-u-i\chi}Q_\mu + Te^{- \rho - i H-u-i\chi} + \widetilde G^-_{\mathbb T^4} e^{- \rho - i \sigma-u-i\chi}  \nonumber \\ 
& \qquad \ \ + \partial X^+_\mu e^{-\rho-i\sigma-iH-u}  + b^{\text{(g)}} e^{-\rho-i\sigma-iH-2u-i\chi-iv}  \nonumber \\ 
& \qquad \ \ - [\partial(i \sigma) \partial(\rho + i H+u+i\chi) + \partial^2(\rho + i H+u+i\chi)]e^{- \rho - i H-u-i\chi} \ . 
	\end{align}
\end{subequations}
The right-moving sector is treated analogously.

We propose the $\mathcal N=4$ algebra discussed above as the central algebraic structure defining physical operators in the $\beta \bar \beta$ deformation of the auxiliary string theory introduced in Section~\ref{sec:auxiliary-duality}. Since the tensionless string is embedded into the auxiliary string theory, see Section~\ref{sec:embedding}, this proposal also furnishes a principle for defining physical vertex operators for the string theory \eqref{ads3/cft2-deformed}. While we did not provide a first-principles derivation of the proposed $\mathcal N=4$ algebra, and our proposal is instead motivated by the intuition developed in Section~\ref{sec:path-integral-N=2-N=4} and by the analogy between the transformation \eqref{path-integral-change-variables} and similar structures appearing in \cite{Alday:2005ww,Araujo:2018rho,Apolo:2019zai,Cui:2023jrb}, we will provide several checks in the following sections, where we discuss DDF operators, the symmetries of the space-time theory and correlation functions, and compare them with expectations from the $\TT$ literature.

\subsection{Physical vertex operators} 
\label{sec:vertex}

By a natural analogy with the physical state conditions \eqref{eq:physical-state-conditions} and \eqref{eq:auxiliary-physical-state-conditions}, we define the physical states by the conditions
\begin{equation} \label{eq:lambda-physical-state-conditions}
    G^+_0 \phi = \widetilde G^+_0 \phi = (J_0-\tfrac{1}{2})\phi = T_0 \phi = 0 \ , \qquad \phi \sim \phi + G^+_0 \widetilde G^+_0 \psi \ , 
\end{equation}
where the $\mathcal N=4$ generators are given by eqs.~\eqref{eq:n=2-lambda-after} and \eqref{eq:n=4-lambda}. 

Since we will be primarily interested in the subsector of physical states connected to the tensionless string, and which reduce to the corresponding subsector of the auxiliary string theory as $\mu \rightarrow 0$, we further restrict to physical states satisfying\footnote{These conditions essentially reproduce the twisted boundary conditions in eqs.~(4.3) and (4.4) of \cite{Cui:2023jrb}. Moreover, when the constraint \eqref{W0-condition} is satisfied, our $\gamma_\mu$ and $\overline{\gamma}_\mu$ can effectively be identified with $\hat{\gamma}$ and $\hat{\overline{\gamma}}$ in \cite{Cui:2023jrb}, see in particular their eq.~(4.5).}
\begin{equation} \label{W0-condition}
    (\partial \mathcal{X}^+_\mu)_0\, \phi = 0 \ , \qquad (\bar \partial \mathcal{X}^-_\mu)_0 \, \phi = 0 \ . 
\end{equation}
Here 
\begin{equation} \label{eq:winding-w-ddf}
    (\partial \mathcal{X}^+_\mu)_0 = \oint \text dz \, \partial X^+_\mu \ , \qquad (\bar \partial \mathcal{X}^-_\mu)_0 = \oint \text d \bar z \, \bar \partial X^-_\mu \ ,
\end{equation}
are DDF operators that naturally generalize those of the auxiliary string theory, see eqs.~\eqref{eq:ddf-aux1-X} and \eqref{eq:null-gauging-ST} for $n=0$. As we will discuss in the following section, the worldsheet $\mu^2 \beta \bar \beta$ deformation breaks the conformal symmetry of the space-time theory, and therefore one does not expect the DDF operators \eqref{eq:ddf-aux1-X} with $n \neq 0$ to admit analogous extensions once $\mu \neq 0$.\footnote{We note that the vertex operator \eqref{eq:deformed-ground-lambda} we are going to construct below also satisfies ${\mathcal N_0\mathcal V_\mu = (e^{- \mathcal Y})_{\frac{3}{2}}\mathcal V_\mu = \Pi_{\frac12}\mathcal V_\mu=0}$, and therefore appears to decouple the remaining auxiliary fields on the boundary as well.}

We would like to understand how the physical fields of the auxiliary string theory are deformed once $\mu \neq 0$, and what cohomology is defined by the conditions \eqref{eq:lambda-physical-state-conditions}.\footnote{As discussed, in this paper we restrict ourselves to states that, in the limit $\mu \rightarrow  0$, reduce to auxiliary string states which can be mapped to tensionless string states.} While this is, in general, a difficult problem, in the present paper we restrict attention to identifying the vertex operator dual to the space-time vacuum and the $\mathcal R$-symmetry generators in the single-trace deformed theory. We therefore seek a vertex operator that obeys eqs.~\eqref{eq:lambda-physical-state-conditions}, satisfies eq.~\eqref{W0-condition}, and reduces to the corresponding tensionless fields as $\mu \rightarrow 0$.

\paragraph{Momentum basis.} As briefly discussed in the Introduction, $\TT$-deformed CFTs are expected to exhibit non-local features. For this reason, it is often more natural to express vertex operators in the \emph{momentum basis} rather than in the $x$-basis \eqref{eq:x-basis-definition}. Given a field $V(x,\bar x,z,\bar z)$ in the $x$-basis, its momentum space
representation is defined by\footnote{We work in Euclidean signature, so $\bar p$ is the complex conjugate of $p$. Our convention for the integral measure is $\text{d}^2 x = \text{d} x \, \text{d} \bar x$.}
\begin{equation} \label{eq:fourier-def}
    V(p,\bar p,z,\bar z) = \int_{\mathbb{C}} \text d^2 x \, e^{ip x - i \bar p \bar x} \, V(x,\bar x,z,\bar z) \ .
\end{equation}
We will focus on the deformation of the space-time ground state, which lies in the $w=1$ sector and corresponds on the worldsheet to the vertex operator
\begin{equation} \label{eq:lambda=0-ground-state}
    \mathcal{V}_{\mu=0}(p,\bar p,z,\bar z)=\Omega^0_{\mu=0}(p,\bar p,z,\bar z) \ ,
\end{equation}
where for later convenience we have introduced the notation
\begin{equation} \label{eq:omega-n-def-lambda}
    \Omega^m_{\mu=0}(p,\bar p,z,\bar z) = \left(\Omega^m_{L}e^{u+i\chi}\right)(z) \left(\Omega^m_{R} e^{\bar u+i\bar \chi}\right)(\bar z) e^{ip\gamma-i\bar p \bar \gamma} \ ,
\end{equation}
see eq.~\eqref{eq:omega-n-L-R}. These expressions are obtained by applying the Fourier transform \eqref{eq:fourier-def} to the twisted sector ground states \eqref{twisted-ground-states-aux} and \eqref{eq:w-odd-tensionless}.

\paragraph{Deformed vertex operator.} Given the definition of $X^\pm_\mu$ in eq.~\eqref{path-integral-change-variables}, and in order to satisfy eq.~\eqref{W0-condition}, it is natural to replace the contribution $e^{ip \gamma-i\bar p \bar \gamma}$ entering \eqref{eq:omega-n-def-lambda} by
\begin{equation} \label{eq:momentum-change}
    \exp \left( ip \gamma-i\bar p \bar \gamma \right) \ \mapsto \, \exp \left(ip\gamma_\mu - i\bar p \bar \gamma_\mu \right) \ ,
\end{equation}
and to consider the ansatz, 
\begin{equation} \label{eq:deformed-ground-lambda}
    \mathcal{V}_{\mu}(p,\bar p,z,\bar z) = \Psi^L \Psi^R e^{ip\gamma_\mu - i\bar p \bar \gamma_\mu} \ ,
\end{equation}
where $\Psi^L$ and $\Psi^R$ denote, respectively, purely left- and right-moving vertex operators. Notice that the replacement \eqref{eq:momentum-change} induces a shift in the mass-shell condition,
\begin{equation}\label{eq:delta-change}
    \Delta_0(m) = 0 \quad \mapsto \quad \Delta(m) = \Delta_0(m) + \mu^2 p \bar p = 0 \ . 
\end{equation}
As discussed in more detail in the Introduction, it was shown in \cite{Apolo:2019zai} that this shift in the worldsheet mass-shell condition leads to a space-time two-point function that matches expectations from the $\TT$ literature. Although the details of our construction differ, since we work in the tensionless regime, we take this as encouraging evidence that we are on the right track. Indeed, the shift \eqref{eq:delta-change} will also play a role in our analysis. 

Our strategy is as follows: we begin by writing 
\begin{equation} \label{PsiL-PsiR}
    \Psi^L = \Omega^m_{L}e^{u+i\chi}+\cdots \ , \qquad     \Psi^R = \Omega_{R}^m e^{\bar u+i\bar \chi} +\cdots \ ,
\end{equation}
where $\Omega_{L}^m$ and $\Omega_{R}^m$ have been introduced in eq.~\eqref{eq:omega-n-L-R}. In eq.~\eqref{PsiL-PsiR}, the dots denote additional contributions required to make \eqref{eq:deformed-ground-lambda} physical. Note in particular the factor $(\partial \gamma)^{-m}$ in eq.~\eqref{eq:omega-n-L-R}, which mimics the structure of the tensionless string theory ground states \eqref{eq:w-odd-tensionless} and \eqref{eq:w-even-tensionless} and introduces a dependence on the as yet undetermined parameter $m$. This parameter will indeed be fixed a posteriori by the mass-shell condition \eqref{eq:delta-change} to be 
\begin{equation} \label{eq:nlambda-lambda}
    m = m_0 + \mu^2 p \bar p \ , 
\end{equation}
where $m_0$ is the undeformed space-time weight, in a way completely analogous to the treatment of twisted sector ground states in the AdS$_3$ literature \cite{Eberhardt:2018ouy,Dei:2020zui}.

Completing the ansatz \eqref{PsiL-PsiR} and determining the additional terms required for the vertex operator \eqref{eq:deformed-ground-lambda} to satisfy the physical state conditions is rather involved, and we relegate the detailed derivation to Appendix~\ref{app:vertex}. Here we present only the final result:\footnote{At this stage, the normalization of the vertex operator does not appear to be fixed. In particular, multiplying it by a smooth function of $\mu^2 p\bar p$ that reduces to one at $\mu=0$ would still give a physical operator. In the $x$-basis, this corresponds to the freedom to add to any physical operator $V(x,\bar x)$ its descendants $(\mu^2\partial_x\partial_{\bar x})^n V(x,\bar x)$, with $n\in\mathbb N$. One can check that the vertex operator \eqref{eq:deformed-ground-lambda} cannot be written as a sum of such contributions, making it a particularly natural choice of normalization. In any case, this freedom does not affect the leading logarithmic contribution in the $x$-basis.}
\begin{subequations} \label{eq:psil-psir}
\begin{multline}
        \Psi^L = e^{if_1-if_2} (\partial \gamma)^{-m} e^{2\rho+i\sigma+iH} e^{u+i\chi} \\
        + \mu\,  m \, e^{if_1-if_2} (\partial \gamma)^{-m-1} e^{2\rho+i\sigma+iH} e^{2u+2i\chi} \partial(e^{iv}) c^{\text{(g)}} \\
        -i \mu^2 \bar{p} (\partial \gamma)^{-m} e^{\rho+i\sigma+iH} e^{2u+i\chi} \partial(e^{iv}) c^{\text{(g)}}\ ,
\end{multline}
and similarly for $\Psi_R$
\begin{multline}
        \Psi^R = e^{i\bar f_1-i\bar f_2} (\bar\partial \bar\gamma)^{- m} e^{2\bar \rho+i\bar \sigma+i\bar H} e^{\bar u+i\bar \chi} \\
        - \mu \, m \, e^{i \bar f_1-i\bar f_2} (\bar\partial \bar\gamma)^{-m-1} e^{2\bar\rho+i\bar\sigma+i\bar H} e^{2\bar u+2i\bar \chi} \bar \partial(e^{i\bar v}) \bar{c}^{\text{(g)}} \\
        +i \mu^2 p \, (\bar \partial \bar \gamma)^{-m} e^{\bar\rho+i\bar \sigma+i\bar H} e^{2\bar u+i\bar \chi} \bar \partial(e^{i\bar v}) \bar{c}^{\text{(g)}}  \ . 
\end{multline}
\end{subequations}
In Appendix~\ref{app:vertex}, together with finding the physical fields corresponding to deformed $\mathcal R$-symmetry generators, we show that the vertex operator \eqref{eq:deformed-ground-lambda}, with $\Psi^L$ and $\Psi^R$ given in eqs.~\eqref{eq:psil-psir}, satisfies the conditions \eqref{eq:lambda-physical-state-conditions} and \eqref{W0-condition} and is therefore physical. In particular, the condition $J_0=\tfrac{1}{2}$ is satisfied as a combination of contributions from the tensionless fields and the auxiliary fields. Moreover, the derivations carried out in \cite{Giveon:2017myj,Apolo:2019zai} suggest that the space-time energy is deformed according to the $\TT$ square root formula, see \eqref{eq:nlambda-lambda}. It is therefore natural to identify \eqref{eq:deformed-ground-lambda} as the worldsheet vertex operator dual to the ground state of the deformed space-time theory.

\subsection{DDF operators and space-time global symmetries} 
\label{sec:global}

In a direct analogy with Sections~\ref{sec:tensionless-physical} and \ref{sec:physical-states}, DDF operators in the deformed string theory are defined by the conditions 
\begin{equation}
    [G^+_0,\mathcal{D}] = [\widetilde G^+_0,\mathcal{D}] = [J_0,\mathcal{D}] = [T_0,\mathcal{D}] = 0 \ , 
\end{equation}
where the $\mathcal N=4$ generators are now given by eqs.~\eqref{eq:n=2-lambda-after} and \eqref{eq:n=4-lambda}. In particular, by this definition one expects to capture all the space-time symmetry generators. 

While we do not exclude the possibility that the infinite dimensional symmetry algebra discussed in \cite{Georgescu:2022iyx,Chakraborty:2023wel,Georgescu:2025jlx} admits a realization in our string theory, in this paper we restrict ourselves to a discussion of the DDF operators associated with the global symmetries of the space-time theory. 

\paragraph{Global symmetries of the space-time theory.} From the $\TT$ literature, one expects this deformation to preserve the maximal symmetry group compatible with the breaking of conformal invariance \cite{Baggio:2018rpv, Chang:2018dge, Jiang:2019hux, Chang:2019kiu, Dei:2024sct}. It is therefore natural to expect that conformal symmetry is broken down to space-time translations, while the maximal amount of supersymmetry is preserved. Accordingly, one is led to conjecture that the symmetry algebra of the single-trace $\TT$ deformation of $\mathcal N=(4,4)$ symmetric orbifolds is the two-dimensional $\mathcal N=(4,4)$ Euclidean supersymmetry algebra.\footnote{We are working with Euclidean signature that can be obtained from the complexified Lie algebra imposing the reality conditions \eqref{eq:euclidean-reality-condition}.}

As we review in Appendix~\ref{app:super} in more detail, the $2$d $\mathcal N=(4,4)$ supersymmetry algebra is formed out of the supercurrents $\{\mathcal{Q}^{\alpha\beta},\overline{\mathcal{Q}}^{\alpha\beta}\}$ with $\alpha,\beta\in\{+,-\}$, and the bosonic generators $\{\mathcal{L},\mathcal{P},\overline{\mathcal{P}},\mathcal{K}^a,\overline{\mathcal{K}}^a\}$. $\mathcal{L}$ is the generator of Euclidean rotation $\text{SO}(2)$ while $\mathcal{P}$ and $\overline{\mathcal{P}}$ are the left- and right-moving translation generators. $\mathcal{K}^a$ and $\overline{\mathcal{K}}^a$ form $\mathfrak{su}(2) \times \overline{\mathfrak{su}(2)}$.\footnote{Together with two outer $\mathfrak{su}(2)_{\text{outer}}\times \overline{\mathfrak{su}(2)}_{\text{outer}}$ they form the $\mathcal{R}$-symmetry group $\text{SO}(4)\times \overline{\text{SO}(4)}$.}

Focusing on the left-moving fields, we can identify the corresponding DDF operators
\begin{subequations}\label{eq:symmetry-generators}
\begin{equation}
    \mathcal P = \oint \text dz \, \beta \ ,
\end{equation}
\begin{equation}
    \mathcal K^+ = \oint \text dz \, p_2 \theta^1 \ , \qquad \mathcal K^3 = \oint \text dz \Bigl[ -\tfrac{1}{2}(p_1\theta^1)+\tfrac{1}{2}(p_2\theta^2) \Bigr] \ , \qquad \mathcal K^- = \oint \text dz \, p_1 \theta^2 \ ,
\end{equation}
\begin{equation}
    \mathcal Q^{++} = \oint \text dz \, p_2 \ , \qquad \mathcal{Q}^{-+} = \oint \text dz \, p_1 \ ,
\end{equation}
\begin{equation}
    \mathcal Q^{+-} = \oint \text dz \, \Bigl[ \beta \theta^1 + p_2 \, e^{-\rho-i\sigma} - p_2 \, e^{-\rho} (G^-_{\mathbb{T}^4}+G^-_{\text{aux}}) \Bigr] \ ,
\end{equation}
\begin{equation}
    \mathcal Q^{--} = \oint \text dz \, \Bigl[ -\beta \theta^2 + p_1 \, e^{-\rho-i\sigma} - p_1 \, e^{-\rho} (G^-_{\mathbb{T}^4}+G^-_{\text{aux}}) \Bigr]\ .
\end{equation}
\end{subequations}
The barred generators are obtained analogously. Let us mention that the complicated expressions for $\mathcal{Q}^{+-}$ and $\mathcal{Q}^{--}$ is a well-known consequence of the fact that only this combination is a DDF operator in the BVW string \cite{Berkovits:1999im}.\footnote{Notice that $G^-_{\text{aux}}$ and $\overline{G}^-_{\text{aux}}$ are invariant under the change of variables \eqref{path-integral-change-variables}.} We have not succeeded in identifying a DDF operator corresponding to the $\text{SO}(2)$ rotation $\mathcal{L}$, partly because it is neither purely left- or right-moving. This reflects the standard fact that space-time rotation generators need not be represented by purely holomorphic worldsheet currents.\footnote{Similarly, we have not found expressions for the DDF operators of the outer $\mathfrak{su}(2)_{\text{outer}}$. As far as we are aware, the associated DDF operators have also not been found in the tensionless string: while the outer $\mathfrak{su}(2)_{\text{outer}}\times \overline{\mathfrak{su}(2)}_{\text{outer}}$ is a symmetry of the symmetric orbifold, its generators are neither holomorphic nor anti-holomorphic, see e.g.\ Section~5 in \cite{David:2002wn}.} The DDF operators in \eqref{eq:symmetry-generators} satisfy
\begin{subequations}\label{eq:zero-mode-left-global}
\begin{equation}
    [\mathcal K^3, \mathcal K^{\pm}] = \pm \mathcal K^{\pm} \ , \qquad [\mathcal K^+, \mathcal K^-] = 2 \mathcal K^3 \ ,
\end{equation}
\begin{equation}
    [\mathcal K^3, \mathcal Q^{\alpha\beta}] = \frac{\alpha}{2} \mathcal Q^{\alpha\beta} \ , \qquad [\mathcal K^\pm, \mathcal Q^{\mp\beta}] = \mathcal Q^{\pm\beta} \ , \qquad \alpha , \beta \in \{+,- \} \ , 
\end{equation}
\begin{equation}
    \{ \mathcal Q^{\alpha\beta}, \mathcal Q^{\mu\nu} \} = -\epsilon^{\alpha\mu} \epsilon^{\beta\nu} \mathcal P \ . 
\end{equation}
\end{subequations}
Comparing with the (anti-)commutation relations listed in Appendix~\ref{app:super}, this indeed provides evidence that the symmetry algebra of the deformed space-time theory is the two-dimensional $\mathcal N=(4,4)$ Euclidean supersymmetry algebra.

\section{\texorpdfstring{$\boldsymbol{T\overline T}$ correlation functions}{T Tbar correlation functions}}
\label{sec:TTbar-correlators}

Before turning into a detailed computation of correlation functions of worldsheet vertex operators in Section~\ref{sec:corr}, we adopt in this section a complementary boundary perspective and analyze correlators in $\TT$-deformed two-dimensional CFTs, with the aim of elucidating their general structures. In the Introduction, we listed various publications on this topic.

In \cite{Cardy:2019qao}, Cardy derived a partial differential equation in momentum space, which we will refer to as Cardy's Callan-Symanzik equation, that $\TT$-deformed correlators are expected to satisfy. One possible approach to the study of correlation functions in $\TT$ deformed two-dimensional CFTs is therefore to analyze the solutions of this differential equation. 

Among the results of \cite{Cui:2023jrb}, the authors observed that shifting the conformal weights of two-dimensional CFT correlation functions by $\mu^2 p \bar p$, precisely the shift that already appeared in the previous section, see eq.~\eqref{eq:nlambda-lambda}, yields a solution to Cardy's Callan-Symanzik equation for two-point functions.\footnote{This observation was further supported by an analysis of the worldsheet mass-shell condition, as well as by a perturbative study of the correlators, both in the bulk and in the space-time theory.} It is then natural to ask whether this structure persists for higher-point functions, and whether similar simple solutions to Cardy's Callan-Symanzik equation can be constructed in that case. This is not obvious, however, since unlike two-point functions, higher-point functions are not universal and depend on the specific CFT under consideration.

In this section, we begin by comparing the two-point function of \cite{Cui:2023jrb} against perturbative $\TT$ calculations. See \cite{Chen:2025jzb} for a related discussion. By emphasizing the similarity and interplay between momentum space Ward identities in two-dimensional CFTs \cite{Bzowski:2013sza} and Cardy's differential equation, we then show that a remarkably simple solution to the Callan-Symanzik equation can be constructed from an arbitrary correlator of an arbitrary two-dimensional CFT by shifting each conformal dimension by $\mu^2 p \bar p$. 

\subsection[\texorpdfstring{$T\overline T$-deformed two-point function in position and momentum space}{T Tbar-deformed correlators in position and momentum space}]{\texorpdfstring{$\boldsymbol{T\overline T}$-deformed two-point function in position and momentum space}{T Tbar-deformed correlators in position and momentum space}} \label{sec:TT-correlator-pos-and-mom}

We begin by specifying our notations. In a given two-dimensional CFT, we denote local quasi-primary operators of conformal dimension $(h_i^0,h_i^0)$ as\footnote{For simplicity we assume that the left and right conformal weights are equal.}
\begin{equation} \label{2d-CFT-x-operator}
    \mathcal{O}^0_{h_i^0}(x_i, \bar x_i) \ , 
\end{equation}
and the associated genus-zero $n$-point functions by 
\begin{equation} \label{2pt-undeformed-sec-5}
    \left\langle \prod_{i=1}^n \mathcal{O}^0_{h_i^0}(x_i, \bar x_i)\right\rangle_0 \ . 
\end{equation}
For later convenience, we also sometimes write
\begin{equation} \label{eq:energy-def-ttbar}
    E_i = 2h_i^0 \ .
\end{equation}
With a slight abuse of notation, we denote the momentum space analogue of \eqref{2pt-undeformed-sec-5} as 
\begin{equation}
     \left\langle \prod_{i=1}^n \mathcal{O}^0_{h_i^0}(p_i, \bar p_i)\right\rangle_0 \ . \label{2d-CFT-correlator-p}
\end{equation}
In the $\TT$-deformed theory, we assume that there exist renormalized momentum space operators 
\begin{equation}
    \mathcal{O}^\mu_{h_i^0}(p_i, \bar p_i)  \label{V-lambda}
\end{equation}
that reduce, upon taking the $\mu \rightarrow 0$ limit, to the undeformed two-dimensional CFT operators \eqref{2d-CFT-x-operator}, 
\begin{equation} \label{eq:limit-operator}
    \lim_{\mu \to 0}  \mathcal{O}^\mu_{h_i^0}(p_i, \bar p_i) =  \mathcal{O}^0_{h_i^0}(p_i, \bar p_i) \ .
\end{equation}
Since the $\TT$-deformed theory is no longer conformal, the deformed operators can no longer be assigned conformal dimensions in the usual sense. Nevertheless, it is sometimes convenient to retain the explicit dependence on $h_i^0$ in order to indicate the undeformed operators to which they reduce as $\mu \to 0$.

A precise definition of deformed operators in perturbation theory is scheme dependent. In particular, different renormalization prescriptions or normalization conventions may lead to different definitions of the operators $\mathcal O^\mu$, while still preserving the limit \eqref{eq:limit-operator}. Among these ambiguities though, at the $n$-th order in perturbation theory, the power $n$ logarithmic contribution is scheme independent \cite{Cardy:2019qao,Aharony:2023dod,He:2023kgq}. This implies that when comparing renormalized $\TT$ correlation functions between different renormalization schemes, the highest power of logarithmic divergences must agree, while the lowest terms depend on individual renormalization procedures.

Let us consider the two-point function of an operator $\mathcal{O}^\mu_{E/2}$ with undeformed energy $E=2h^0$. The perturbative leading logarithmic contributions to the deformed two-point function can be written as \cite{Cardy:2019qao,Hirano:2020nwq,Hirano:2024eab,Li:2026ecl}
\begin{equation} \label{eq:perturbative-formula-ttbar} 
    \langle \mathcal{O}^\mu_{E/2}(x) \mathcal{O}^\mu_{E/2}(0) \rangle_{\mu} \sim \sum_{n=0}^\infty \frac{\left(2\mu^2\right)^n}{n!} \left( \frac{\Gamma(E+n)}{\Gamma(E)} \right)^2 e^{n\partial_E} \partial_E^n \langle \mathcal{O}^0_{E/2}(x) \mathcal{O}^0_{E/2}(0) \rangle_{0} \ ,
\end{equation}
where
\begin{equation} \label{2ptf-x}
    \langle \mathcal{O}^0_{E/2}(x) \mathcal{O}^0_{E/2}(0) \rangle_{0} = \frac{1}{|x|^{2E}} \ ,
\end{equation}
and `$\sim$' denotes equality at the level of leading logarithmic contributions. At each order in perturbation theory, the renormalized two-point function is a polynomial in $\log|x|$ \cite{Cardy:2019qao,Chen:2025jzb}. As discussed above, only the highest power of the logarithm is scheme independent and eq.~\eqref{eq:perturbative-formula-ttbar} captures precisely these contributions. Let us also mention that the Fourier transform of eq.~\eqref{2ptf-x} is \cite{Bzowski:2013sza, Aharony:2023dod}
\begin{equation}
    \mathcal F[\langle \mathcal{O}^0_{E/2}(x) \mathcal{O}^0_{E/2}(0) \rangle_{0}] = U(E) (p \bar p)^{E-1} \ , 
\end{equation}
where
\begin{equation} \label{U(s)}
    U(s) = \frac{\pi \, 2^{1-2s} \, \Gamma(1-s)}{\Gamma(s)} \ . 
\end{equation}
Now we would like to show that, at the level of leading logarithmic contributions, the Fourier transform of eq.~\eqref{eq:perturbative-formula-ttbar} can be formally written as 
\begin{equation} \label{eq:cui-sec-5}
   \mathcal F[\langle \mathcal{O}^\mu_{E/2}(x) \mathcal{O}^\mu_{E/2}(0) \rangle_{\mu}] \equiv \langle \mathcal{O}^\mu_{E/2}(p) \mathcal{O}^\mu_{E/2}(-p) \rangle_{\mu} = U(E+2\mu^2 p \bar p) \, (p\bar p)^{E+2\mu^2 p \bar p-1} \ .
\end{equation}
Comparing with eqs.~\eqref{generic-two-pointf}-\eqref{two-point-Cui}, this will show that the momentum space two-point function proposed in \cite{Cui:2023jrb} agrees to all orders with the perturbative calculations of \cite{Cardy:2019qao, Hirano:2020nwq, Hirano:2024eab, Li:2026ecl}. Since the Fourier transform is involutive, it is sufficient to show that the Fourier transform of eq.~\eqref{eq:cui-sec-5} reproduces the leading logarithmic terms captured by eq.~\eqref{eq:perturbative-formula-ttbar}. 

To this end, notice that, since \eqref{U(s)} generically defines a locally analytic function in $E$, one may formally write
\begin{equation}
    U(E+2\mu^2 p \bar p) \, (p\bar p)^{E+2\mu^2 p \bar p-1} = e^{2 \mu^2 p\bar p \partial_E} \big[ U(E) (p\bar p)^{E-1} \big] \ . 
\end{equation}
The Fourier transform of eq.~\eqref{eq:cui-sec-5} therefore takes the form
\begin{equation} \label{eq:inverse-fourier-ttbar}
\begin{split}
        \mathcal{F}^{-1}[\langle \mathcal{O}^\mu_{E/2}(p) \mathcal{O}^\mu_{E/2}(-p) \rangle_{\mu}] &= e^{2\mu^2 \partial_x \partial_{\bar x} \partial_{E}} \left( \frac{1}{|x|^{2E}} \right) \\
        &= \sum_{n=0}^\infty \frac{(2\mu^2)^n}{n!} \partial_{E}^n (\partial_x \partial_{\bar x})^n  \left( \frac{1}{|x|^{2E}} \right) \ ,
\end{split}
\end{equation}
where in the second equality we used the fact that the derivatives commute. Using the identity,
\begin{equation}
\begin{split}
    (\partial_x \partial_{\bar x})^n  \left( \frac{1}{|x|^{2E}} \right) &= \left[E \times \cdots \times (E+n-1)\right]^2 \frac{1}{|x|^{2(E+n)}} \\ &= \left( \frac{\Gamma(E+n)}{\Gamma(E)} \right)^2 e^{n\partial_E} \left( \frac{1}{|x|^{2E}} \right) \ ,
\end{split}
\end{equation}
and retaining only leading logarithmic contributions, we obtain
\begin{equation}
    \mathcal{F}^{-1}[\langle \mathcal{O}^\mu_{E/2}(p) \mathcal{O}^\mu_{E/2}(-p) \rangle_{\mu}] \sim \sum_{n=0}^\infty \frac{(2\mu^2)^n}{n!} \left( \frac{\Gamma(E+n)}{\Gamma(E)} \right)^2 e^{n\partial_E} \partial_E^n \left( \frac{1}{|x|^{2E}} \right) \ , 
\end{equation}
which indeed agrees with eq.~\eqref{eq:perturbative-formula-ttbar}.

We emphasize that, in eq.~\eqref{eq:inverse-fourier-ttbar}, the operator $\partial_E^n$ also acts on the ratio $\Gamma(E+n)/\Gamma(E)$. However, the leading logarithmic contribution proportional to $\log^n|x|$ arises solely from the action of $\partial_E^n$ on $|x|^{-2E}$. Interestingly, the expression \eqref{eq:cui-sec-5} can be formally obtained from the perturbative expansion \eqref{eq:perturbative-formula-ttbar} by moving the ratio of $\Gamma$ functions to the right of the derivatives $\partial_E^n$.

\subsection{A simple solution to Cardy's equation}
\label{sec:KZ-solution}

Cardy showed that renormalized $n$-point correlation functions of $\TT$-deformed $2$d CFTs satisfy the Callan-Symanzik equation \cite{Cardy:2019qao}
\begin{equation}
   \left( \Lambda \partial_\Lambda + 2 \mu^2 \sum_{i = 1}^n p_i \bar p_i \right) \left\langle \prod_{i=1}^n \mathcal{O}^\mu_{h_i^0}(p_i, \bar p_i)\right\rangle_\mu =0 \ , 
   \label{KZ-compact}
\end{equation}
where $\Lambda$ denotes the renormalization scale and $\mu^2$ is the dimensionless $\TT$ deformation parameter. Note that the deformed correlation function depends on $\mu$ through the deformed operators \eqref{V-lambda} and through the deformed action, which we indicate by the additional subscript $\mu$. In terms of the path integral representation, this dependence can be made explicit as 
\begin{equation}
    \left\langle \prod_{i=1}^n \mathcal{O}^\mu_{h_i^0}(p_i, \bar p_i)\right\rangle_\mu = \int \mathcal DX e^{-S_\mu} \prod_{i=1}^n \mathcal{O}^\mu_{h_i^0}(p_i, \bar p_i) \ , \qquad \partial_\mu S_\mu = \int \text{d}^2 x \det(T_{ab}^\mu) \ ,  
\end{equation}
where $X$ collectively denotes the various fields entering the deformed theory. 

Assuming for simplicity that $h_i^0 = \bar h_i^0$, the $\Lambda$ dependence of the deformed correlation function can then be estimated by dimensional analysis, leading to\footnote{The correlator considered here is the full correlator including the momentum conservation delta function, cf.~\eqref{eq:double-bracket-notation}.}
\begin{equation}
    \left( \Lambda \partial_\Lambda + \sum_{i=1}^{n} p_i \frac{d}{dp_i} + \sum_{i=1}^{n} \bar p_i \frac{d}{d \bar p_i} - 2\mu^2 \partial_{\mu^2} - 2\sum_{i=1}^n(h_i^0 -1) \right) \left\langle \prod_{i=1}^n \mathcal{O}^\mu_{h_i^0}(p_i, \bar p_i)\right\rangle_\mu  = 0 \ . 
\end{equation}
Cardy's Callan-Symanzik equation \eqref{KZ-compact} can thus be rewritten in the form \cite{Cui:2023jrb, Hirano:2020nwq}
\begin{equation}
    \left( \sum_{i=1}^{n} p_i \frac{d}{dp_i} + \sum_{i=1}^{n} \bar p_i \frac{d}{d \bar p_i} - 2 \mu^2 \partial_{\mu^2} - 2 \sum_{i=1}^n(h_i -1) \right) \left\langle \prod_{i=1}^n \mathcal{O}^\mu_{h_i^0}(p_i, \bar p_i)\right\rangle_\mu  = 0 \ ,   
    \label{KZ}
\end{equation}
where we introduced the notation 
\begin{equation}
    h_i = h_i^0 + \mu^2 p_i \bar p_i \label{hp} \ . 
\end{equation}

\paragraph{A simple solution to Cardy's Callan-Symanzik equation.}
We now show that the function
\begin{equation}
    \left\langle \prod_{i=1}^n \mathcal{O}^0_{h_i}(p_i, \bar p_i)\right\rangle_0 \label{KZ-simple-solution}
\end{equation}
solves Cardy's equation \eqref{KZ}. Notice that the function \eqref{KZ-simple-solution} is simply the undeformed two-dimensional CFT correlator \eqref{2d-CFT-correlator-p}, with the original conformal weights $h_i^0$ replaced by the shifted $h_i$ defined in eq.~\eqref{hp}.\footnote{In particular, we do not assume that fields with weights $h_i$ exist in the original CFT. We are instead only assuming that \eqref{2pt-undeformed-sec-5} can be defined as a differentiable function of $h_i^0$ and satisfies global Ward identities.}

To follow the derivation below, it is important to keep in mind that the momentum dependence of \eqref{KZ-simple-solution} may either be explicit or arise implicitly through the dependence of $h_i$ on the momenta, see eq.~\eqref{hp}. Accordingly, total derivatives with respect to the momenta can be decomposed into partial derivatives as
\begin{equation}
 \begin{aligned}
     p_i \frac{d}{dp_i} &=  p_i \frac{\partial}{\partial p_i} + p_i \frac{\partial h_i}{\partial p_i} \frac{\partial}{\partial h_i}  = p_i \frac{\partial}{\partial p_i} + \mu^2 p_i \bar p_i \frac{\partial}{\partial h_i} \ , \label{pdp}\\
     \bar p_i \frac{d}{d \bar p_i} &=  \bar p_i \frac{\partial}{\partial \bar p_i} + \bar p_i \frac{\partial h_i}{\partial \bar p_i} \frac{\partial}{\partial h_i}  = \bar p_i \frac{\partial}{\partial \bar p_i} + \mu^2 p_i \bar p_i \frac{\partial}{\partial h_i} \ . 
 \end{aligned}
 \end{equation}
Since the dependence on $\mu$ enters only through $h_i$, one similarly obtains 
\begin{equation}
    - 2 \mu^2 \partial_{\mu^2} = -2 \mu^2 \sum_{i=1}^n \frac{\partial h_i}{\partial \mu^2} \frac{\partial}{\partial h_i} = - 2 \mu^2 \sum_{i=1}^n p_i \bar p_i \frac{\partial}{\partial h_i} \ . \label{lambda-d-lambda}
\end{equation}
Combining eqs.~\eqref{pdp} and \eqref{lambda-d-lambda}, one finds that \eqref{KZ-simple-solution} satisfies Cardy's differential equation \eqref{KZ} provided that
\begin{equation} \label{Ward-id}
    \left( \sum_{i=1}^{n} p_i \frac{\partial}{\partial p_i} + \sum_{i=1}^{n} \bar p_i \frac{\partial}{\partial \bar p_i} - 2 \sum_{i=1}^n(h_i -1) \right) \left\langle \prod_{i=1}^n \mathcal{O}^0_{h_i}(p_i, \bar p_i)\right\rangle_0  = 0 \ . 
\end{equation}
Eq.~\eqref{Ward-id} is satisfied for arbitrary values of the $h_i$'s, as it is the $\mathcal{L}_0$ momentum space Ward identity of a two-dimensional CFT \cite{Bzowski:2013sza}. This completes our derivation that \eqref{KZ-simple-solution} solves Cardy's differential equation.

\paragraph{General structure of $\boldsymbol{T \overline T}$-deformed correlators.} The result above is suggestive because, on general grounds, one does not expect operators in a $\TT$-deformed CFT to be obtained simply by shifting the conformal dimensions of quasi-primary operators of the undeformed theory. Rather, even if the operator $\mathcal O^0_{h_i^0}$ can be analytically continued in $h_i^0$ and one can formally provide a definition for $\mathcal O^0_{h_i}$, one expects that 
\begin{equation} \label{op-inequality}
    \mathcal{O}^\mu_{h_i^0}(p_i, \bar p_i) \neq \mathcal{O}^0_{h_i}(p_i, \bar p_i) \ . 
\end{equation}
Nevertheless, correlation functions constructed from the operators appearing on both sides of \eqref{op-inequality} satisfy Cardy's differential equation: correlators of operators on the left-hand-side do so by the results of \cite{Cardy:2019qao}, while correlators of operators on the right-hand-side satisfy it by the derivation presented above. This naturally raises the possibility that, despite the operator inequality \eqref{op-inequality}, the corresponding correlation functions may nevertheless coincide,
\begin{equation} \label{simple-correlator}
    \left\langle \prod_{i=1}^n \mathcal{O}^\mu_{h_i^0}(p_i, \bar p_i)\right\rangle_\mu \stackrel{?}{=} \left\langle \prod_{i=1}^n \mathcal{O}^0_{h_i}(p_i, \bar p_i) \right\rangle_0 \ . 
\end{equation}
The question mark in eq.~\eqref{simple-correlator} emphasizes that this is only a conjectural possibility, which we wish to explore without prejudice. Indeed, strictly speaking, the equality \eqref{simple-correlator} is far from being implied by the above  derivation. Cardy's equation is a partial differential equation and therefore admits multiple solutions, so the two sides of \eqref{simple-correlator} may differ even though they satisfy the same equation.

With this motivation in mind, in the next section we compute the two-point function of the worldsheet vertex operators \eqref{eq:deformed-ground-lambda}, which are dual to the $\TT$-deformed ground state. We will find that, although the vertex operator \eqref{eq:deformed-ground-lambda} is not obtained by the naive replacement of \eqref{eq:nlambda-lambda} into the undeformed vertex operators \eqref{eq:omega-n-def-lambda}, the relation~\eqref{simple-correlator} nevertheless holds for two-point functions when the undeformed boundary CFT is taken to be the auxiliary CFT of Section \ref{sec:auxiliary-space-time-CFT}. 

\section{String correlation functions}
\label{sec:corr}

This section is devoted to a worldsheet computation of the two-point function of the deformed space-time ground state identified in Section~\ref{sec:vertex}. We begin in Section~\ref{sec:correlators-def-and-checks} by discussing the definition of tree-level correlation functions in the auxiliary string theory introduced in Section \ref{sec:auxiliary-string}, as well as in the deformed string theory of Section~\ref{sec:deforming}. We then turn, in Section \ref{eq:npt-twisted-ground-states}, to a warm-up computation of correlators of twisted sector ground states in the auxiliary string theory, deriving in particular some of the results anticipated in Section~\ref{sec:physical-states}. Finally, in Section~\ref{sec:two-pt-deformed}, we arrive at one of the results of this paper: the worldsheet computation of the two-point function of the $\TT$-deformed space-time ground state.

\subsection{Definition and checks of correlation functions}
\label{sec:correlators-def-and-checks}

In Section~\ref{sec:tensionless-physical}, we discussed the definition of tree-level correlation functions in the tensionless string. The string theories considered in Sections~\ref{sec:auxiliary-string} and \ref{sec:deforming} contain additional worldsheet fields carrying non-vanishing background charges, see the discussion below \eqref{eq:bosonizations-added-ghosts}. It is therefore necessary to ensure that correlation functions can still be defined consistently in the precesence of these auxiliary fields. Since the prescriptions for the $n$-point functions in the two theories are almost identical, we discuss them together.

\smallskip 

Consider $n$ physical vertex operators $V_j$, with $j = 1, \dots, n$, satisfying either eqs.~\eqref{eq:auxiliary-physical-state-conditions} or \eqref{eq:lambda-physical-state-conditions}, depending on whether we are considering the auxiliary string theory of Section~\ref{sec:auxiliary-string} or the deformed string theory of Section~\ref{sec:deforming}. The associated tree-level $n$-pt function is defined as\footnote{Similar to \eqref{eq:corr}, there is a sum over $M$ and $M^\prime$ with the corresponding chemical potentials. In the correlators considered here, however, the sum receives contributions from only a single value of $M$ and a single value of $M^\prime$.}
\begin{align} \label{eq:corr-lambda}
    \int \mathrm{d}\Omega \,
    \langle
    \prod_{\ell,\ell^{\prime}=1}^{M,M^{\prime}} D\bar D(\zeta_\ell) Z\overline Z(\nu_{\ell^{\prime}}) \,
    \widetilde{V}_1(z_1,\bar z_1)\,
    V_2(z_2,\bar z_2)\,
    V_3(z_3,\bar z_3)
    \prod_{j=4}^n
    G^-_{-1}\bar{G}^-_{-1}V_j(z_j,\bar z_j)
    \rangle \ , 
\end{align}
where we have introduced the shorthand notation
\begin{equation}
    \mathrm{d}\Omega \equiv \mathrm{d}^{2M^{\prime}}\nu \ \mathrm{d}^{2M}\zeta_\ell  \   \prod_{j=4}^n \mathrm{d}^2 z_j \ , 
\end{equation}
and for brevity we have suppressed the dependence of the measure on $n,M$ and $M^{\prime}$. 

Let us explain various ingredients entering eq.~\eqref{eq:corr-lambda}.  For any $j\in\{1,\cdots,n\}$, the operator $\widetilde{V}_j$ is defined, exactly as in \eqref{eq:vtilde}, by solving 
\begin{equation} \label{eq:vtilde-lambda}
    V_j = \widetilde{G}^+_0 \overline{\widetilde{G}^+_0} \widetilde{V}_j \ , 
\end{equation}
where left- and right-moving $\mathcal N=4$ generators are now given by eqs.~\eqref{eq:lambda0-gtildep} or \eqref{eq:n=4-lambda}. A convenient solution of eq.~\eqref{eq:vtilde-lambda} is
\begin{equation} \label{eq:vtilde-explicit-lambda}
    \widetilde{V}_1 = \xi_0 \bar\xi_0 V_1 
\end{equation}
with
\begin{equation} \label{eq:xi-lambda}
    \xi = e^{-\rho-iH-u-i\chi} \ , \qquad \bar \xi = e^{-\bar \rho-i\bar H-\bar u-i\bar \chi} \ , 
\end{equation}
which satisfies
\begin{equation}
    \{\widetilde{G}^+_n,\xi_m \} = \delta_{n+m,0} \ , 
\end{equation}
and a similar equation in the right-moving sector. 

The operators $D$ and $Z$ are screening operators needed to saturate background charges of tensionless and auxiliary fields, respectively. The operator $D$ is defined as in \eqref{eq:def-d}, while $Z$ is given by 
\begin{equation} \label{eq:z-def}
    Z=c^{\text{(g)}} \, \partial \gamma^{\text{(g)}} \delta(\gamma^{\text{(g)}}) \ .
\end{equation}
The number $M$ of insertions of $D$ is determined by eq.~\eqref{eq:N-covering-cond}, while the number $M'$ of insertions of $Z$ depends on the specific vertex operators under consideration and is determined by background charge conservation of auxiliary fields. 
The operators $D$ and $Z$, together with their right-moving fields, are exactly marginal. To see this, it is sufficient to notice that 
\begin{equation}
    \mathcal{D} = D \overline{D} \, e^{i\sigma+i\bar\sigma}  \qquad \text{and} \qquad \mathcal{Z}=Z\overline{Z} \, e^{i\sigma+i\bar \sigma} 
\end{equation}
are unintegrated physical operators. The associated integrated physical operators 
\begin{equation}
    \int \text d^2 z \,  G^-_{-1} \overline{G}^-_{-1} \mathcal D \ , \qquad \int \text d^2 z \,  G^-_{-1} \overline{G}^-_{-1} \mathcal Z
\end{equation}
are therefore exactly marginal by the argument of \cite{Berkovits:1994vy}. Finally, the picture raising operator is defined by 
\begin{equation} \label{eq:picture raising-def-6}
    P_+ V_j = G^+_0 \xi_0 V_j \ , 
\end{equation}
where $\xi$ is defined in eq.~\eqref{eq:xi-lambda}, while $G^+_0$ is given by either eq.~\eqref{G+-after-lambda0} or \eqref{G+-after-lambda}, depending on the string theory under consideration. As in the tensionless string, the sum of the picture numbers of the physical vertex operators is assumed to satisfy eq.~\eqref{picture-sum}. We explain in Appendix~\ref{app:corr} that the correlation function in \eqref{eq:corr-lambda} is well-defined, namely that exact fields decouple and the definition is independent of the distribution of picture numbers.

Similar to the tensionless string, see \cite{Berkovits:1994vy,Maldacena:2001km,Erbin:2019uiz,Gaberdiel:2021njm} and the discussion around eq.~\eqref{eq:two-point-free}, the two-point function must be defined separately. We define it as 
\begin{equation} \label{eq:two-point-lambda-0}
    \int \text d^{2M}\zeta_\ell \, \text{d}^{2M^{\prime}} \nu_{\ell^{\prime}} \, \big< \prod_{\ell,\ell^{\prime}=1}^{M,M^{\prime}} D\overline D(\zeta_\ell) Z\overline Z(\nu_{\ell^{\prime}}) [(e^{i\sigma+i\bar \sigma})_0 \widetilde{V}_1](z_1,\bar z_1) \, V_2(z_2,\bar z_2) \big> \ .
\end{equation}
This is the prescription that we will use to compute two-point functions of twisted sector ground states in the next section, and of the $\TT$-deformed space-time ground state \eqref{eq:deformed-ground-lambda} in Section~\ref{sec:two-pt-deformed}.

\subsection{Correlation functions of the auxiliary string theory} 
\label{eq:npt-twisted-ground-states}

As a warm-up exercise, we compute the $n$-point function of the fields given in eq.~\eqref{eq:degeneracy-twisted-ground-states} in the auxiliary string theory of Section~\ref{sec:auxiliary-duality}. To keep the notation compact, let us introduce the shorthand notation
\begin{equation} \label{eq:l-pt-function-notation}
\begin{aligned}
    \mathcal{V}^{q,w}(x,z) & = \left( e^{f(q)(\mathcal Y + i \mathcal K)+f(q)(\overline{\mathcal Y} + i \overline{\mathcal K})}\right)_0 \ket{0}_w \\
    &=  \Phi_w(x,z) \overline{\Phi}_w(\bar x,\bar z) e^{(q+1)(u+i\chi)} e^{(q+1)(\bar u+i\bar \chi)} \ ,
\end{aligned}
\end{equation}
where, as discussed below \eqref{eq:degeneracy-twisted-ground-states}, $f(q)$ is an integer-valued function of $q$. We wish to compute the correlator 
\begin{equation} \label{eq:npt-vwn}
     \int \text d \Omega \,  \big< \prod_{\ell,\ell^{\prime}=1}^{M,M^{\prime}} D\bar D(\zeta_\ell) Z\overline Z(\nu_{\ell^{\prime}}) \widetilde{\mathcal V}^{q_1,w_1} \, \mathcal V^{q_2,w_2} \, P_+ \overline{P}_+ \mathcal V^{q_3,w_3} \prod_{j=4}^n G^-_{-1} \bar{G}^-_{-1} P_+ \overline{P}_+ \mathcal V^{q_j,w_j} \big> \ , 
\end{equation}
for an arbitrary collection of $q_j\in\mathbb{Z}_{\geq -1}$ and positive odd $w_j$'s.\footnote{The calculation in the case where some of the $w_j$'s are even proceeds similarly, and again agrees with the space-time result. See \cite{Dei:2023ivl} for a discussion of this point. } 

\paragraph{Picture raised vertex operators.} As discussed in Appendix~\ref{app:corr}, the correlation function is independent of how the picture is distributed among the vertex operators. For this reason, we choose $\mathcal V^{q_1,w_1}$ and $\mathcal V^{q_2,w_2}$ in picture $P=-2$, so that the associated $\Phi_{w_1}$ and $\Phi_{w_2}$ are given by eq.~\eqref{eq:w-odd-tensionless}, while for the remaining vertex operators we apply the picture raising operator \eqref{eq:picture raising-def-6}. More explicitly, suppressing the right-moving dependence and using the definition \eqref{eq:xi-lambda}, one finds that $\widetilde{\mathcal V}^{q,w}$ for $w$ odd is given by
\begin{equation} \label{eq:w-odd-tilde-lambda}
\begin{aligned} 
    \widetilde{\mathcal V}^{q,w} &= \xi_0 \Phi_w e^{(q+1)(u+i\chi)} = (e^{-\rho - iH})_0 \Phi_w e^{q(u+i\chi)} \\
    &= \exp{\left[\tfrac{w+1}{2}(if_1-if_2)\right]} \Bigl(\frac{\partial^w \gamma}{w!}\Bigr)^{-m_w} \delta_w(\gamma-x) \, e^{\rho+i\sigma} e^{q(u+i\chi)} \ . 
\end{aligned}
\end{equation}
The picture-raised vertex operators depend on the value of $q$. Suppressing again the right-moving dependence, $P_+ \mathcal{V}^{q,w}$ for odd $w$ and $q\geq 0$ reads, up to irrelevant numerical factors,
\begin{equation} \label{eq:w-odd-picture-raised}
\begin{aligned}
    P_+ \mathcal{V}^{q,w} &= P_+ \Phi_w e^{(q+1)(u+i\chi)} \\
    &= \exp{\left[\tfrac{w-1}{2}(if_1-if_2)\right]} \Bigl(\frac{\partial^w \gamma}{w!}\Bigr)^{-m_w+1} \delta_w(\gamma-x) \, e^{i\sigma} e^{q(u+i\chi)} \ ,
\end{aligned}
\end{equation}
with $m_w$ given by eq.~\eqref{eq:mw}. The case $q=-1$ is slightly more involved, as the picture raising operator \eqref{eq:picture raising-def-6} acts in a more complicated fashion on $e^{-u-i\chi}$. In this case,
\begin{align} \label{eq:w-odd-picture-raised-q=-1}
    P_+ \mathcal{V}^{-1,w} & = \exp{\left[\tfrac{w-1}{2}(if_1-if_2)\right]} \left(\frac{\partial^w \gamma}{w!}\right)^{-m_w+1} \delta_w(\gamma(z)-x) \, e^{i\sigma} e^{-(u+i\chi)} \nonumber \\ 
    & \quad - \exp{\left[\tfrac{w+1}{2}(if_1-if_2)\right]} \left(\frac{\partial^w \gamma}{w!}\right)^{-m_w} \delta_w(\gamma(z)-x) \, e^{\rho+i\sigma} e^{-u} \partial X^+ \\ 
    & \quad +\exp{\left[\tfrac{w+1}{2}(if_1-if_2)\right]} \left(\frac{\partial^w \gamma}{w!}\right)^{-m_w} \delta_w(\gamma(z)-x) \, e^{\rho+i\sigma} e^{-2u-iv}b^{\text{(g)}} e^{-i\chi} \ ,\nonumber
\end{align}
where we used the bosonization formulae~\eqref{eq:bosonizations-added-ghosts}.

\paragraph{Two-point function.} Let us begin by computing the two-point function 
\begin{align} \label{Vq1w1-Vq2w2}
   \big< \mathcal{V}^{q_1,w_1} \mathcal{V}^{q_2,w_2} \big>= \int \text d^{2M} \zeta_\ell \, \text{d}^{2M^{\prime}} \nu \, \big< \prod_{\ell,\ell^{\prime}=1}^{M,M^{\prime}} D\bar D(\zeta_\ell) Z\overline Z(\nu_{\ell^{\prime}}) [(e^{i\sigma+i\bar \sigma})_0\widetilde{\mathcal V}^{q_1,w_1}] \, \mathcal V^{q_2,w_2} \big> \ .
\end{align}
The background charges of each tensionless field, listed in Table~\ref{table:tensionless-background-charges} of Appendix~\ref{app:details-tensionless}, are saturated provided that $M$ is chosen as in eq.~\eqref{eq:N-covering-cond}. Moreover, the background charge of $c^{\text{(g)}}$ forces $M^{\prime}=1$. It is then straightforward to see that the two-point function \eqref{Vq1w1-Vq2w2} factorizes into a product of a tensionless string correlator and a correlator involving only the auxiliary fields. One finds
\begin{equation} \label{eq:two-point-factorization-aux}
    \langle \mathcal{V}^{q_1,w_1} \mathcal{V}^{q_2,w_2} \rangle = \langle \mathcal{V}^{q_1,w_1} \mathcal{V}^{q_2,w_2} \rangle_{\text{tens}} \langle \mathcal{V}^{q_1,w_1} \mathcal{V}^{q_2,w_2} \rangle_{\text{aux}} \ ,
\end{equation}
where suppressing the right-moving dependence,
\begin{align}
    \big< \mathcal{V}^{q_1,w_1} \mathcal{V}^{q_2,w_2} \big>_{\text{tens}} & = \int \text{d}^{2M}\zeta_\ell \, \Bigl< \prod_{\ell=1}^M D(\zeta_\ell) \, (e^{i\sigma})_0[(e^{-\rho-iH})_0 \Phi_{w_1}] \Phi_{w_2} \Bigr> \ , \\ 
   \big< \mathcal{V}^{q_1,w_1} \mathcal{V}^{q_2,w_2} \big>_{\text{aux}} & = \int \text{d}^2 \nu \, \big< Z(\nu) \, e^{q_1(u+i\chi)} \, e^{(q_2+1)(u+i\chi)} \big> \ .
\end{align}
The tensionless contribution is precisely the two-point function of twisted sector ground states and it vanishes unless
\begin{equation}
    w_1=w_2 \ .
\end{equation}
The fields $e^{\alpha(u+i\chi)}$ have vanishing worldsheet conformal weight for any $\alpha \in \mathbb{Z}$. Moreover, as explained in detail in Appendix~\ref{app:corr}, $Z$ has a trivial OPE with $(u+i\chi)$. Therefore, the auxiliary correlator carries no dependence on $\{x_j,z_j\}$ and reduces to a non-vanishing constant, provided that the background charges of $u$ and $i \chi$ are saturated. This occurs when
\begin{equation}
    q_1+q_2 = 0 \ . 
\end{equation}
If one considers the two-point function of a field with itself, so that $q_1 = q_2$, this implies that only for $q_1=q_2=0$ one obtains a non-vanishing correlator, and that in this case the result is constant. This proves the statement made below eq.~\eqref{twisted-ground-states-aux}.

\paragraph{$\boldsymbol n$-point correlation functions.} More generally, let us consider $n$-point functions of the vertex operators \eqref{eq:l-pt-function-notation}. The background charge of $c^{\text{(g)}}$ (together with $\rho$) once more enforces $M^{\prime}=1$. Despite the seemingly complicated form of some of the expressions involved, the correlation function can be computed relatively straightforwardly. Since $\widetilde{\mathcal V}^{q_1,w_1}$ and $\mathcal V^{q_2,w_2}$ alone saturate the background charge of $\rho$, see Table~\ref{table:tensionless-background-charges} in Appendix~\ref{app:details-tensionless}, if $q=-1$ only the first term in eq.~\eqref{eq:w-odd-picture-raised-q=-1} contributes. Similar to the two-point function computation above, the $n$-pt function \eqref{eq:npt-vwn} factorizes as 
\begin{equation}
    \big< \prod_{j=1}^{n} \mathcal{V}^{q_j,w_j} \big>= \big< \prod_{j=1}^{n} \mathcal{V}^{q_j,w_j} \big>_{\text{tens}} \big< \prod_{j=1}^{n} \mathcal{V}^{q_j,w_j} \big>_{\text{aux}} \ .
\end{equation}
The tensionless correlator is exactly the $n$-point function of twisted ground states, while the auxiliary field contribution is given by  
\begin{equation} \label{n-point-auxiliary-warmup}
    \big< \prod_{j=1}^{n} \mathcal{V}^{q_j,w_j} \big>_{\text{aux}} = \int \text{d}^2 \nu \, \big< Z(\nu) \, e^{q_1(u+i\chi)} \, e^{(q_2+1)(u+i\chi)} \, \prod_{j=3}^n e^{q_j(u+i\chi)} \big> \ . 
\end{equation}
Saturation of the background charges of $u$ and $i\chi$ implies that the correlator \eqref{n-point-auxiliary-warmup} vanishes unless 
\begin{equation}
    \sum_{j=1}^{n} q_j = 0 \ ,
\end{equation}
and in this case, it does not depend on the insertion points $\{z_j\}$ on the worldsheet, or $\{x_j\}$ in space-time. We therefore conclude that when 
\begin{equation}
    \sum_{j=1}^{n} f(q_j) = 0 \ , 
\end{equation}
as the vertex operators $e^{f(q)(\mathcal Y + i \mathcal K)+f(q)(\overline{\mathcal Y} + i \overline{\mathcal K})}$ have trivial OPEs among each other, the $n$-point function of the vertex operators in the first line of eq.~\eqref{eq:l-pt-function-notation} reduces to the $n$-point function of twisted ground states. This concludes our derivation of eq.~\eqref{eq:q-n-conditions}, which as discussed there, suggests the identification $f(q)=q$.

\subsection[{\texorpdfstring{Two-point function of $T \overline T$-deformed fields}{Two-point function of T Tbar-deformed fields}}]{\texorpdfstring{Two-point function of $\boldsymbol{T \overline T}$-deformed fields}{Two-point function of T Tbar-deformed fields}}
\label{sec:two-pt-deformed}

We now have all the ingredients needed to compute the two-point function of deformed physical fields. The calculation for the deformed space-time $\mathcal{R}$-symmetry generators given in Appendix~\ref{app:vertex} is almost identical to the two-point function of the deformed ground state given in \eqref{eq:deformed-ground-lambda}, and for this reason we focus on the latter. According to the prescription \eqref{eq:two-point-lambda-0}, the two-point function reads\footnote{Saturation of the background charge of $c^{\text{(g)}}$ again selects $M^{\prime}=1$, while eq.~\eqref{eq:N-covering-cond} implies $M=1$.}
\begin{equation} \label{eq:two-point-lambda}
    \left< \mathcal{V}_{\mu} \mathcal{V}_{\mu} \right> = \int \text d^2 \zeta \, \text{d}^{2} \nu \, \big< D\overline D(\zeta) Z\overline Z(\nu) [(e^{i\sigma+i\bar \sigma})_0 \widetilde{\mathcal{V}}_{\mu}](p_1,\bar p_1,z_1,\bar z_1) \, \mathcal{V}_{\mu}(p_2,\bar p_2,z_2,\bar z_2) \big> \ .
\end{equation}
Here, using eqs.~\eqref{eq:deformed-ground-lambda}, \eqref{eq:vtilde-explicit-lambda} and \eqref{eq:xi-lambda}, we have 
\begin{subequations} \label{Vmu-tildeVmu}
\begin{align}
    \widetilde{\mathcal{V}}_{\mu}(z_1, \bar z_1) &= (e^{-\rho-iH-u-i\chi})_0 (e^{-\bar \rho-i\bar H-\bar u-i\bar \chi})_0 \Psi^L \Psi^R e^{ip\gamma_\mu - i\bar p \bar \gamma_\mu}(z_1, \bar z_1) \ , \\
    \mathcal{V}_{\mu}(z_2, \bar z_2) &= \Psi^L \Psi^R e^{ip\gamma_\mu - i\bar p \bar \gamma_\mu}(z_2, \bar z_2) \ , 
\end{align}
\end{subequations}
where $\Psi^L$ and $\Psi^R$ are given in eq.~\eqref{eq:psil-psir}. 

The computation proceeds along lines similar to those discussed in the previous section. Using eqs.~\eqref{eq:psil-psir}, each of the vertex operators \eqref{Vmu-tildeVmu} can be written as a sum of nine different terms, giving rise to $9^2$ contributions to the two-point function. Writing all the terms explicitly would not be very illuminating. Instead, by imposing background charge conservation for all the fields entering the correlator, we will argue that only one of these $81$ terms contributes. In fact, saturation of the background charge of $\rho$ and $u$ implies that, for both $\widetilde{\mathcal{V}}_{\mu}(z_1, \bar z_1)$ and $\mathcal{V}_{\mu}(z_2, \bar z_2)$ only the first term in each of the expressions in \eqref{eq:psil-psir} contributes. This is a special feature of the two-point function. In higher-point functions, the background-charge constraints are different and we expect that they do not in general decouple the additional contributions in \eqref{eq:psil-psir}. Introducing the notation
\begin{equation}
    \omega^{\mu}(p,\bar p,z,\bar z) = \left(\Omega^m_{L} e^{u+i\chi} \right)(z) \left(\Omega^m_{R} e^{\bar u+i\bar \chi} \right)(\bar z) \, e^{ip\gamma_\mu-i\bar p \bar \gamma_\mu} \ ,
\end{equation}
where $\Omega^m_{L}$, $\Omega^m_{R}$ and $m$ are defined in eqs.~\eqref{eq:omega-n-L-R} and \eqref{eq:nlambda-lambda} respectively, the correlator \eqref{eq:two-point-lambda} reduces to 
\begin{equation} \label{eq:two-point-lambda-reduced}
   \left< \mathcal{V}_{\mu} \mathcal{V}_{\mu} \right> =  \int \text d^2\zeta d^2 \nu \, \big< D\overline{D}(\zeta) Z\overline{Z}(\nu) [(e^{i\sigma+i\bar \sigma})_0 \widetilde{\omega}^{\mu}_1](p_1,\bar p_1, z_1,\bar z_1) \, \omega^{\mu}_2(p_2,\bar p_2,z_2,\bar z_2) \big> \ . 
\end{equation}
Similar to the computation of the previous section, the correlation function \eqref{eq:two-point-lambda-reduced} factorizes into the product of a tensionless string correlator and a free-field correlator involving only auxiliary fields: 
\begin{equation} \label{eq:two-point-factorization-aux-lambda}
    \llangle \mathcal{V}_{\mu} \mathcal{V}_{\mu} \rrangle = \llangle \mathcal{V}_{\mu} \mathcal{V}_{\mu} \rrangle_{\text{tens}} \llangle \mathcal{V}_{\mu} \mathcal{V}_{\mu} \rrangle_{\text{aux}} \ , 
\end{equation}
where, 
\begin{align} 
\left< \mathcal{V}_{\mu} \mathcal{V}_{\mu} \right>_{\text{aux}} &= \int \text{d}^2 \nu \, \Bigl< Z(\nu) [e^{-i\mu p_1 X^- -i\mu \bar p_1 X^+}](z_1,\bar z_1) [e^{u+i\chi+\bar u+i\bar \chi} e^{-i\mu p_2 X^- -i\mu \bar p_2 X^+}](z_2,\bar z_2) \Bigr> \ ,  
\end{align}
and,  omitting right-moving fields,
\begin{align}
\left< \mathcal{V}_{\mu} \mathcal{V}_{\mu} \right>_{\text{tens}} & = \int \text{d}^{2}\zeta \, \Bigl< D(\zeta) \, [(e^{i\sigma})_0 (e^{-\rho-iH})_0 \, \Omega^m_L e^{i p_1 \gamma}](z_1) [\Omega^m_L e^{i p_2 \gamma}](z_2) \Bigr>  \label{eq:tensionless-contribution-lambda}\ . 
\end{align}
In eq.~\eqref{eq:two-point-factorization-aux-lambda}, we denoted with double brackets correlation functions stripped off of the canonical momentum conservation delta function. For instance,\footnote{Strictly speaking, because of the zero modes of the fields, the full correlator in \eqref{eq:two-point-lambda-reduced} does not fully factorize into tensionless and auxiliary CFT correlators. However, once the zero mode integration is removed via \eqref{eq:double-bracket-notation}, eq.~\eqref{eq:two-point-factorization-aux-lambda} holds.} 
\begin{equation} \label{eq:double-bracket-notation}
    \langle \mathcal{V}_{\mu}(p_1,\bar p_1) \mathcal{V}_{\mu}(p_2,\bar p_2) \rangle = \llangle \mathcal{V}_{\mu}(p_1,\bar p_1) \mathcal{V}_{\mu}(p_2,\bar p_2) \rrangle \, \delta^{(2)}(p_1+p_2) \ ,
\end{equation}
The auxiliary field contribution reads
\begin{equation} \label{eq:two-point-lambda-v-v-aux}
    \left< \mathcal{V}_{\mu} \mathcal{V}_{\mu} \right>_{\text{aux}} = C^\prime |z_1-z_2|^{2\mu^2 p_1 \bar p_2 + 2\mu^2 \bar p_1 p_2} \delta^{(2)}(p_1+p_2) = C^\prime |z_1-z_2|^{-4\mu^2 p_1 \bar p_1} \delta^{(2)}(p_1+p_2) \ ,
\end{equation}
where $C^\prime$ is an overall $\mu$-independent normalization constant.
In deriving the first equality in \eqref{eq:two-point-lambda-v-v-aux}, we used the fact that the OPE of $Z$ with $e^{u+i\chi}$ is regular, see Appendix~\ref{app:corr}, while in deriving the second equality we imposed the delta function momentum conservation.

In order to calculate the tensionless correlator in \eqref{eq:tensionless-contribution-lambda}, we first pass to the $x$-basis \eqref{eq:x-basis-definition} using the definition \eqref{eq:fourier-def}. The corresponding two-point function in the $x$-basis is given in
eq.~\eqref{eq:free-two-point-xbasis}. In order to write down the final answer in terms of momenta, we then Fourier transform \eqref{eq:free-two-point-xbasis},\footnote{The integral in the LHS is divergent when $m>0$ and has poles at $2m\in \mathbb{Z}_{\geq 1}$. We are writing the physical answer through a renormalization discussed e.g.~in Appendix~C of \cite{Aharony:2023dod}.}
\begin{equation}
    \int_{\mathbb{C}} \text d^2x \, \frac{e^{i p x - i \bar p \bar x}}{|x|^{4m}} = \frac{\pi \, 2^{1-4m} \, \Gamma(1-2m)}{\Gamma(2m)}  \, (p\bar p)^{2m - 1} \,.
\end{equation}
The tensionless contribution is then simply given as
\begin{equation} \label{eq:two-point-lambda-v-v-tensionless}
    \left< \mathcal{V}_{\mu} \mathcal{V}_{\mu} \right>_{\text{tens}} = C \frac{2^{-4m} \, \Gamma(1-2m)}{\Gamma(2m)}  \, (p_1\bar p_1)^{2m - 1} |z_1-z_2|^{-4\Delta_0} \delta^{(2)}(p_1+p_2) \ ,
\end{equation}
where
\begin{equation} \label{eq:m-def-lambda}
    m - m_0 = - \Delta_0 = \mu^2 p_1 \bar p_1 \ , 
\end{equation}
see eqs.~\eqref{eq:delta-change} and \eqref{eq:nlambda-lambda}, while $C$ is an overall $\mu$-independent normalization constant. Assembling eqs.~\eqref{eq:two-point-lambda-v-v-aux} and \eqref{eq:two-point-lambda-v-v-tensionless} and reabsorbing normalization factors, we finally obtain
\begin{equation} \label{eq:two-point-function}
    \left< \mathcal{V}_{\mu} \mathcal{V}_{\mu} \right> = C \frac{2^{-4m} \, \Gamma(1-2m)}{\Gamma(2m)}  \, (p_1\bar p_1)^{2m - 1} \delta^{(2)}(p_1+p_2) \ , \qquad m = m_0 +\mu^2 p \bar p \ ,
\end{equation}
where $m_0 = 0$ for the deformed ground state and $m_0=1$ for the deformed space-time $\mathcal R$-symmetry operators. 

Notice that eq.~\eqref{eq:two-point-function} has precisely the form \eqref{generic-two-pointf}--\eqref{two-point-Cui}. Thus, our computation reproduces the proposal of \cite{Cui:2023jrb} for the two-point function of $\TT$-deformed correlators.  

Let us also consider the deformed ground state with $m_0=0$. Using
\begin{equation}
    \frac{1}{2 \pi} \lim_{x\to 0^+} \frac{\Gamma(1-x)}{\Gamma(x)} (p \bar p)^{x-1} = \delta^{(2)}(p) \ ,
\end{equation}
one recovers the undeformed correlator in the limit $\mu\to 0$. This agrees with eq.~\eqref{eq:free-two-point-xbasis} for $m=0$, as expected.

\vspace{3cm}

\acknowledgments We thank
Jan De Boer,
Shai Chester, 
Lorenz Eberhardt, 
Silvia Georgescu,
Monica Guica, 
Bob Knighton,
Emil Martinec, 
Savdeep Sethi
and 
Arkady Tseytlin
for useful discussions and feedback on a draft version of this manuscript. AD in particular thanks Emil Martinec for many conversations and for collaboration on related projects. The work of AD is supported by the DOE grant DE-SC0009924. The work of KN is supported by the UK Engineering and Physical Sciences grant EP/Z000106/1. KN also acknowledges support from the Swiss National Science Foundation (SNSF) and the NCCR SwissMAP, where part of this work was carried out. During proofreading, we found the feedback provided by Refine \cite{Refine} and Get Physics Done \cite{GetPhysicsDone} useful.

\newpage

\appendix

\section{Details on the tensionless duality}
\label{app:details-tensionless}

In this appendix, we review several aspects of the tensionless duality.

\subsection{The boundary CFT} \label{app:boundary-cft}

The boundary CFT is the symmetric orbifold of $\mathbb{T}^4$. Here we focus on fixing our conventions for the fields of the \textit{seed} theory $\mathbb{T}^4$. This is built out of four free compact bosons,
\begin{equation} \label{fourfreebos}
    \partial \mathcal X^i(z) \partial \bar{\mathcal X}^j(w) \sim \frac{\delta^{ij}}{(z-w)^2} \ , \qquad i, j \in \{ 1,2\} \ , 
\end{equation}
and four free fermions,
\begin{equation} \label{fourfreefer}
    \Lambda^{\alpha, j}(z) \Lambda^{\beta, \ell}(w) \sim \frac{\epsilon^{\alpha \beta} \epsilon^{\ell j}}{z-w} \ , \qquad i, j \in \{ 1,2\} \ , \quad \alpha, \beta \in \{ +,-\} \ , 
\end{equation}
where our conventions for the Levi-Civita tensors are $\epsilon^{+-} = \epsilon^{12} = 1$ and similar equations can be written in the anti-holomorphic sector. The spectrum of the symmetric orbifold of $\mathbb{T}^4$ is then generated by the action of fractional modes of the free fields introduced above on $w$-twisted ground states. 

These fields form a small $\mathcal{N}=4$ superconformal algebra with $\mathtt c=6$. Below, we first spell out our conventions for the $\mathcal N=2$ and $\mathcal N=4$ superconformal algebras and then express their generators in terms of the $\mathbb T^4$ fields.

\paragraph{$\boldsymbol{\mathcal N=2}$ and $\boldsymbol{\mathcal N=4}$ superconformal algebras.} In our conventions, an untwisted $\mathcal N=2$ superconformal algebra of central charge $\mathtt c$ is defined by the OPEs
\begingroup
\allowdisplaybreaks
\begin{subequations} \label{N=2-conventions-untwisted}
\begin{align}
    T(z) \, T(w) &\sim  \frac{\mathtt c}{2(z-w)^4}+\frac{2 T(w)}{(z-w)^2} + \frac{\partial T(w)}{(z-w)} \ , \label{n=2-TT-OPE-untwisted} \\
    T(z) \, G^{\pm}(w) &\sim \frac{3G^{\pm}(w)}{2(z-w)^2} + \frac{\partial G^{\pm}(w)}{(z-w)} \ , \\
    T(z) \, J(w) &\sim \frac{J(w)}{(z-w)^2} + \frac{\partial J(w)}{(z-w)} \ , \\
    J(z) \, J(w) &\sim \frac{\mathtt c/12}{(z-w)^2} \ , \label{eq:n=2-jj-ope-untwisted} \\
    J(z) \, G^\pm(w)  &\sim \pm \frac{1}{2} \frac{G^\pm(w)}{(z-w)} \ , \\
    G^+(z) \, G^-(w) &\sim \frac{\mathtt c/3}{(z-w)^3} + \frac{2 J(w)}{(z-w)^2} + \frac{T(w)+\partial J(w)}{(z-w)} \ ,  \label{eq:g+g--conventions-untwisted}
\end{align}
\end{subequations}
\endgroup
where $\sim$ denotes equality up to regular terms. A small $\mathcal N=4$ untwisted superconformal algebra, in addition to $T$, $G^\pm$, $J$, contains the fields $\widetilde G^\pm$, $J^{\pm \pm}$. The remaining non-trivial OPEs are given by
\begingroup
\allowdisplaybreaks
\begin{subequations}
\label{N=4-conventions-untwisted}
\begin{align}
    T(z) \, \widetilde G^{\pm}(w) &\sim \frac{3\widetilde G^{\pm}(w)}{2(z-w)^2} + \frac{\partial \widetilde G^{\pm}(w)}{(z-w)} \ , \\
    T(z) \, J^{\pm\pm}(w) &\sim \frac{J^{\pm\pm}(w)}{(z-w)^2} + \frac{\partial J^{\pm\pm}(w)}{(z-w)} \ , \\
    J(z) \, J^{\pm \pm}(w)  &\sim \pm \frac{J^{\pm \pm}(w)}{(z-w)} \ , \\
    J(z) \, \widetilde G^\pm(w)  &\sim \pm \frac{1}{2} \frac{\widetilde G^\pm(w)}{(z-w)} \ , \\
    J^{++}(z) \, J^{--}(w) &\sim \frac{\mathtt c/6}{(z-w)^2} + \frac{2 J(w)}{z-w} \ , \\
    J^{\pm \pm}(z) \, G^{\mp}(w) &\sim \pm \frac{\widetilde G^{\pm}(w)}{z-w} \ , \\
    J^{\pm \pm}(z) \, \widetilde G^{\mp}(w) &\sim \mp \frac{G^{\pm}(w)}{z-w} \ , \\
    \widetilde G^+(z) \, \widetilde G^-(w) &\sim \frac{\mathtt c/3}{(z-w)^3} + \frac{2 J(w)}{(z-w)^2} + \frac{T(w)+\partial J(w)}{(z-w)} \ , \\
    G^{\pm}(z) \widetilde G^{\pm}(w) &\sim \mp \frac{2 J^{\pm\pm}(w)}{(z-w)^2} \mp  \frac{\partial J^{\pm\pm}(w)}{(z-w)} \ . 
\end{align}
\end{subequations}
\endgroup

\paragraph{$\boldsymbol{\mathcal N=2}$ and $\boldsymbol{\mathcal N=4}$ superconformal algebras of $\mathbb{T}^4$.} The untwisted $\mathcal{N}=2$ superconformal generators of $\mathbb{T}^4$ can be written in terms of the free fields \eqref{fourfreebos} and \eqref{fourfreefer} as
\begingroup
\allowdisplaybreaks
\begin{subequations}
\label{N=2-T4-untwisted}
\begin{align}
    \mathcal{T}_{\mathbb T^4} &= \partial \mathcal X^j \partial \bar{\mathcal X}^j +\frac{1}{2} \epsilon^{\alpha\beta} \epsilon^{kl} \Lambda^{\alpha,k} \partial \Lambda^{\beta,l}  \ , \\
    \mathcal{G}^+_{\mathbb T^4} &=  \partial \bar{\mathcal X}^j \Lambda^{+,j} \ , \\
    \mathcal{G}^-_{\mathbb T^4} &= - \epsilon^{jk}  \partial \mathcal{X}^j \Lambda^{-,k} \ ,  \\
    \mathcal{J}_{\mathbb T^4} &= - \tfrac{\epsilon^{ij}}{2} \Lambda^{+,i} \Lambda^{-,j} \ .  
\end{align}
\end{subequations}
\endgroup
The $\mathcal N=2$ superconformal algebra \eqref{N=2-T4-untwisted} can be extended to a small untwisted $\mathcal N=4$ superconformal algebra with $\mathtt c=6$ where the additional fields are given by
\begingroup
\allowdisplaybreaks
\begin{subequations}
\label{N=4-T4-untwisted}
\begin{align}
    \mathcal J^{++}_{\mathbb T^4} &= \Lambda^{+,1} \Lambda^{+,2} \ , \\
    \mathcal J^{--}_{\mathbb T^4} &= \Lambda^{-,2} \Lambda^{-,1} \ , \\
    \widetilde{\mathcal G}^+_{\mathbb T^4} &= -\epsilon^{jk} \partial \mathcal X^j \Lambda^{+,k} \ , \\
    \widetilde{\mathcal G}^-_{\mathbb T^4} &= - \partial \bar{\mathcal X}^j \Lambda^{-,j} \ . 
\end{align}
\end{subequations}
\endgroup

\subsection[\texorpdfstring{The $\text{AdS}_3 \times \text{S}^3 \times \mathbb T^4$ tensionless string}{The AdS3 x S3 x T4 tensionless string}]{The $\boldsymbol{\text{AdS}_3 \times \text{S}^3 \times \mathbb T^4}$ tensionless string}

Let us also spell out our conventions for the worldsheet fields entering the tensionless string in \eqref{F} and review various details entering its definition.

\paragraph{The $\boldsymbol{\mathfrak{psu}(1,1|2)_1}$ algebra and its free field realization.} We adopt the conventions of \cite{Dei:2023ivl}, in which the $\mathfrak{psu}(1,1|2)_1$ OPEs read 
\begin{subequations}
\label{psu-comm-rel}
\begingroup
\allowdisplaybreaks
\begin{align}
    J^3(z) J^3(w) &= -\frac{1}{2 (z-w)^2} \ , \\
    J^3(z) J^\pm(w) &= \pm \frac{J^\pm(w)}{z-w} \ , \\
    J^+(z) J^-(w) &= \frac{1}{(z-w)^2} - \frac{2 J^3(w)}{z-w} \ , \\
    K^3(z) K^3(w) &= \frac{1}{2 (z-w)^2}  \ , \\
    K^3(z) K^\pm(w) &= \pm \frac{K^\pm(w)}{z-w}  \ , \\
    K^+(z) K^-(w) &= \frac{1}{(z-w)^2} + \frac{2 K^3(w)}{z-w} \ , \\
    J^a(z) S^{\alpha \beta  \gamma}(w) &= \frac{c_a (\sigma^a)^\alpha{}_\mu S^{\mu \beta \gamma}(w)}{2(z-w)} \ , \\
    K^a(z) S^{\alpha \beta  \gamma}(w) &= \frac{(\sigma^a)^\beta{}_\nu S^{\alpha \nu \gamma}(w)}{2(z-w)} \ , \\
    S^{\alpha \beta \gamma}(z) S^{\mu \nu \rho}(w) &=- \frac{ \epsilon^{\alpha \mu} \epsilon^{\beta \nu} \epsilon^{\gamma \rho}}{(z-w)^2} + \frac{\epsilon^{\beta \nu} \epsilon^{\gamma \rho} c_a (\sigma_a)^{\alpha \mu}J^a(w) - \epsilon^{\alpha \mu} \epsilon^{\gamma \rho} (\sigma_a)^{\beta \nu} K^a(w)}{z-w} \ . 
\end{align}
\endgroup
\end{subequations}
Greek indices denote spinor indices taking values in $\{ +, - \}$, while $a \in \{ +, -, 3 \}$. The Levi-Civita tensor reads $\epsilon^{+-} = - \epsilon^{-+} = 1$, the constant $c_a$ is defined by
\begin{equation}
    c_- = -1 \ , \qquad c_3 = 1 \ , \qquad c_+ = 1 \ , 
\end{equation}
and Pauli matrices are given by 
\begin{subequations}
\label{pauli}
\begin{align}
(\sigma_-)^{\alpha \beta} &= \bigg( \begin{matrix} 1 & 0 \\ 0 & 0 \ \end{matrix} \bigg) \ , &
(\sigma_3)^{\alpha \beta} &= \biggl( \begin{matrix} 0 & 1 \\ 1 & 0 \\ \end{matrix} \bigg) \ , &
(\sigma_+)^{\alpha \beta} &= \bigg( \begin{matrix} 0 & 0 \\ 0 & -1 \\ \end{matrix} \bigg) \ , \\
(\sigma^-)^\alpha{}_\beta &= \bigg( \begin{matrix} 0 & 0 \\ 2 & 0 \ \end{matrix} \bigg) \ , & 
(\sigma^3)^\alpha{}_\beta &= \bigg( \begin{matrix} -1 & 0 \\ 0 & 1 \ \end{matrix} \bigg) \ , &
(\sigma^+)^\alpha{}_\beta &= \bigg( \begin{matrix} 0 & 2 \\ 0 & 0 \ \end{matrix} \bigg) \ , & \\
(\sigma^-)_{\alpha \beta} &= \bigg( \begin{matrix} 2 & 0 \\ 0 & 0 \ \end{matrix} \bigg) \ , & 
(\sigma^3)_{\alpha \beta} &= \bigg( \begin{matrix} 0 & 1 \\ 1 & 0 \ \end{matrix} \bigg) \ , &
(\sigma^+)_{\alpha \beta} &= \bigg( \begin{matrix} 0 & 0 \\ 0 & -2 \ \end{matrix} \bigg) \ , & \\
(\sigma_-)^\alpha{}_\beta &= \bigg( \begin{matrix} 0 & 1 \\ 0 & 0 \ \end{matrix} \bigg) \ , & 
(\sigma_3)^\alpha{}_\beta &= \bigg( \begin{matrix} -1 & 0 \\ 0 & 1 \ \end{matrix} \bigg) \ , &
(\sigma_+)^\alpha{}_\beta &= \bigg( \begin{matrix} 0 & 0 \\ 1 & 0 \ \end{matrix} \bigg) \ , & \\
(\sigma_-)_{\alpha \beta} &= \bigg( \begin{matrix} 0 & 0 \\ 0 & -1 \ \end{matrix} \bigg) \ , & 
(\sigma_3)_{\alpha \beta} &= \bigg( \begin{matrix} 0 & 1 \\ 1 & 0 \ \end{matrix} \bigg) \ , &
(\sigma_+)_{\alpha \beta} &= \bigg( \begin{matrix} 1 & 0 \\ 0 & 0 \ \end{matrix} \bigg) \ .   & 
\end{align}
\end{subequations}
The free fields \eqref{betagamma} and \eqref{ptheta} generate the $\mathfrak{psu}(1,1|2)_1$ affine superalgebra \cite{Beem:2023dub, Dei:2023ivl} through
\begin{subequations}
\label{psu-free-field}
    \begin{align}
        J^+ &= \beta \ , &  K^+ &= p_2 \theta^1 \ , \\
        J^3 & = \gamma \beta + \tfrac12 p_a \theta^a \ , & K^3 &= - \tfrac12 p_1 \theta^1 + \tfrac12 p_2 \theta^2 \ , \\
        J^- & = (\beta \gamma) \gamma + (p_a \theta^a) \gamma \ , & K^- &= p_1 \theta^2 \ , 
    \end{align}
\ \vspace{-25pt}
    \begin{align}
        S^{+++} &=  p_2 \ , & S^{-++} &= - \gamma  p_2 \ , & S^{+--} &= - \beta \theta^2 \ , & S^{-+-} &=  -(\beta \gamma +  p_a \theta^a) \theta^1  \ , \\ 
        S^{+-+} &= p_1 \ , & S^{--+} &= - \gamma p_1 \ , & S^{++-} &= \beta \theta^1 \ , & S^{---} &= (\beta \gamma +  p_a \theta^a) \theta^2 \ .        
    \end{align}
\end{subequations}

\paragraph{Topologically twisted algebras.} In the main text, at various places we make use of topologically twisted algebras. A topologically twisted $\mathcal{N}=4$ algebra is obtained from an $\mathcal{N}=4$ superconformal algebra, see Appendix~\ref{app:boundary-cft}, via topologically twisting the stress-tensor as $T\mapsto T+ \partial J$.\footnote{As well-known, more generally one can topologically twist as $T\mapsto T-\partial J$. This relative choice of sign is important between the left- and right-moving fields. For the Type IIB string theory that we are considering, both sectors have the same twist and we have chosen the plus sign for convenience.} In our conventions, an $\mathcal N=2$ topologically twisted algebra of central charge $\mathtt c$ is defined by the OPEs
\begingroup
\allowdisplaybreaks
\begin{subequations} \label{N=2-conventions}
\begin{align}
    T(z) \, T(w) &\sim  \frac{2 T(w)}{(z-w)^2} + \frac{\partial T(w)}{(z-w)} \ , \label{n=2-TT-OPE} \\
    T(z) \, G^+(w) &\sim \frac{G^+(w)}{(z-w)^2} + \frac{\partial G^+(w)}{(z-w)} \ , \\
    T(z) \, G^-(w) &\sim \frac{2 G^-(w)}{(z-w)^2} + \frac{\partial G^-(w)}{(z-w)} \ , \\
    T(z) \, J(w) &\sim -\frac{\mathtt c/6}{(z-w)^3} + \frac{J(w)}{(z-w)^2} + \frac{\partial J(w)}{(z-w)} \ , \\
    J(z) \, J(w) &\sim \frac{\mathtt c/12}{(z-w)^2} \ , \label{eq:n=2-jj-ope} \\
    J(z) \, G^\pm(w)  &\sim \pm \frac{1}{2} \frac{G^\pm(w)}{(z-w)} \ , \\
    G^+(z) \, G^-(w) &\sim \frac{\mathtt c/3}{(z-w)^3} + \frac{2 J(w)}{(z-w)^2} + \frac{T(w)}{(z-w)} \ .  \label{eq:g+g--conventions}
\end{align}
\end{subequations}
\endgroup
Topologically twisted (small) $\mathcal N=4$ algebras contain, in addition to $T$, $G^\pm$, $J$, the generators $\widetilde G^\pm$, $J^{\pm \pm}$, and the OPEs \eqref{N=2-conventions} are supplemented by  
\begingroup
\allowdisplaybreaks
\begin{subequations}
\label{N=4-conventions}
\begin{align}
    T(z) \, \widetilde G^+(w) &\sim \frac{\widetilde G^+(w)}{(z-w)^2} + \frac{\partial \widetilde G^+(w)}{(z-w)} \ , \\
    T(z) \, \widetilde G^-(w) &\sim \frac{2 \widetilde G^-(w)}{(z-w)^2} + \frac{\partial \widetilde G^-(w)}{(z-w)} \ , \\
    T(z) \, J^{++}(w) &\sim  \frac{\partial J^{++}(w)}{(z-w)} \ ,  \\
    T(z) \, J^{--}(w) &\sim \frac{2J^{--}(w)}{(z-w)^2} + \frac{\partial J^{--}(w)}{(z-w)} \ , \\
    J(z) \, J^{\pm \pm}(w)  &\sim \pm \frac{J^{\pm \pm}(w)}{(z-w)} \ , \\
    J(z) \, \widetilde G^\pm(w)  &\sim \pm \frac{1}{2} \frac{\widetilde G^\pm(w)}{(z-w)} \ , \\
    J^{++}(z) \, J^{--}(w) &\sim \frac{\mathtt c/6}{(z-w)^2} + \frac{2 J(w)}{z-w} \ , \\
    J^{\pm \pm}(z) \, G^{\mp}(w) &\sim \pm \frac{\widetilde G^{\pm}(w)}{z-w} \ , \label{OPE-J-G-generating}\\
    J^{\pm \pm}(z) \, \widetilde G^{\mp}(w) &\sim \mp \frac{G^{\pm}(w)}{z-w} \ , \\
    \widetilde G^+(z) \, \widetilde G^-(w) &\sim \frac{\mathtt c/3}{(z-w)^3} + \frac{2 J(w)}{(z-w)^2} + \frac{T(w)}{(z-w)} \ , \\
    G^{\pm}(z) \widetilde G^{\pm}(w) &\sim \mp \frac{2 J^{\pm\pm}(w)}{(z-w)^2} \mp  \frac{\partial J^{\pm\pm}(w)}{(z-w)} \ . 
\end{align}
\end{subequations}
\endgroup

\paragraph{Topologically twisted algebra on $\boldsymbol{\mathbb T^4}$.} The topologically twisted algebra on the worldsheet $\mathbb T^4$ is built out of four free bosons 
\begin{equation}
    \partial X^i(z) \partial \bar{X}^j(w) \sim \frac{\delta^{ij}}{(z-w)^2} \ , \qquad i, j \in \{ 1,2\} \ , 
\end{equation}
and four topologically twisted fermions 
\begin{equation}
    \Psi^{\alpha, j}(z) \Psi^{\beta, \ell}(w) \sim \frac{\epsilon^{\alpha \beta} \epsilon^{\ell j}}{z-w} \ , \qquad i, j \in \{ 1,2\} \ , \quad \alpha, \beta \in \{ +,-\} \ , 
\end{equation}
where $\epsilon^{+-} = \epsilon^{12} = 1$. It is often useful to bosonize the twisted fermions as
\begin{align}
    \Psi^{+,1} &= e^{i H^1} \ , & \Psi^{+,2} &= e^{i H^2} \ , &\Psi^{-,1} &= e^{-i H^2} \ , & \Psi^{-,2} &= - e^{-i H^1} \,,
    \label{T4-fermions}
\end{align}
where the free bosons $H^1$ and $H^2$ obey eq.~\eqref{Hij-OPE}. 

The four bosons and twisted fermions form an $\mathcal N=2$ topologically twisted algebra of central charge $\mathtt c = 6$, with generators
\begingroup
\allowdisplaybreaks
\begin{subequations}
\label{N=2-T4}
\begin{align}
    T_{\mathbb T^4} &= \partial X^j \partial \bar{X}^j - \epsilon^{ij} \Psi^{-,i} \partial \Psi^{+,j}  \ , \\
    G^+_{\mathbb T^4} &=  \partial \bar{X}^j \Psi^{+,j} \ , \\
    G^-_{\mathbb T^4} &= - \epsilon^{jk}  \partial X^j \Psi^{-,k} \ ,  \\
    J_{\mathbb T^4} &= - \tfrac{\epsilon^{ij}}{2} \Psi^{+,i} \Psi^{-,j} = \tfrac{i}{2} \partial H \ .  
\end{align}
\end{subequations}
\endgroup
The $\mathcal N=2$ algebra \eqref{N=2-T4} is extended to an $\mathcal N=4$ algebra with the same central charge by the additional generators
\begingroup
\allowdisplaybreaks
\begin{subequations}
\label{N=4-T4}
\begin{align}
    J^{++}_{\mathbb T^4} &= \Psi^{+,1} \Psi^{+,2} = e^{iH} \ , \\
    J^{--}_{\mathbb T^4} &= \Psi^{-,2} \Psi^{-,1} =  e^{-iH} \ , \\
    \widetilde G^+_{\mathbb T^4} &= -\epsilon^{jk} \partial X^j \Psi^{+,k} \ , \\
    \widetilde G^-_{\mathbb T^4} &= - \partial \bar{X}^j \Psi^{-,j} \ . 
\end{align}
\end{subequations}
\endgroup

\paragraph{Background charge conventions. } In our conventions, a chiral boson $X$, obeying the OPE
\begin{equation}
    X(z) X(w) \sim - \epsilon \ln(z-w) 
\end{equation}
with $\epsilon=\pm 1$, has a stress-tensor
\begin{equation}\label{eq:background-charge-definition}
    T = -\frac{\epsilon}{2} (\partial X)^2 + \frac{\Lambda}{2} \partial^2 X \ ,
\end{equation}
and we refer to $\Lambda$ as the background charge of $X$. For example, both the background charges of $\rho$ and $i\sigma$ in \eqref{eq:n=2-free-before} are $\Lambda_\rho=\Lambda_{i\sigma}=3$. In Table~\ref{table:tensionless-background-charges} we list the free bosons appearing in the worldsheet CFT of the tensionless string, together with their associated background charges.

\begin{table}[htbp]
\centering
\begin{tabular}{|c|c|c|c|c|c|c|}
\hline
Field             & $\rho$ & $i\sigma$ & $iH^1$ & $iH^2$ & $if_1$ & $if_2$ \\ \hline
Background charge & $3$    & $3$       & $1$    & $1$    & $1$    & $-1$   \\ \hline
\end{tabular}
\caption{Background charges of fields in tensionless string theory.}
\label{table:tensionless-background-charges}
\end{table}

\paragraph{$\boldsymbol{\mathcal{N}=4}$ fields.} The fields of the tensionless string that we have discussed so far form an $\mathcal{N}=2$ topologically twisted algebra with $\mathtt c=6$, see eqs.~\eqref{eq:n=2-free-before} \cite{Berkovits:1999im}. For completeness, let us write down the additional fields that extend it to a small $\mathcal{N}=4$ algebra. There are two additional $\mathfrak{su}(2)_1$ fields,
\begin{equation}
    J^{\pm\pm}=e^{\pm(\rho+i\sigma+iH)} \ ,
\end{equation}
that using \eqref{N=4-conventions}, generate two additional supercurrents
\begin{subequations} \label{eq:before-additioan-n=4}
    \begin{equation}
        \widetilde{G}^+ = e^{\rho+iH}+e^{\rho+i\sigma}\widetilde{G}^+_{\mathbb{T}^4} \ ,
    \end{equation}
    \begin{equation}
        \widetilde{G}^- = e^{-2\rho-i\sigma-iH} Q - e^{-\rho-iH} \left( T_{\mathfrak{psu}} - \frac{1}{2}[\partial(\rho+i\sigma)]^2+\frac{1}{2}\partial^2(\rho+i\sigma) \right) + e^{-\rho-i\sigma}\widetilde{G}^-_C \ .
    \end{equation}
\end{subequations}

\section{Conventions for the auxiliary boundary CFT} 
\label{app:dual}

In this appendix, we collect our conventions for the auxiliary boundary CFT \eqref{eq:symm-dual-C0}. As in Appendix~\ref{app:boundary-cft}, we work at the level of the seed theory. The conventions for the $\mathbb T^4$ fields and their $\mathcal N=4$ generators are summarized there. The fields defining the $\mathcal{A}_0$ sector are introduced in Section~\ref{sec:auxiliary-space-time-CFT}. The seed theory $\mathbb{T}^4 \times \mathcal{A}_0$ forms an $\mathcal{N}=2$ algebra with $\mathtt c=6$ with the following generators
\begingroup
\allowdisplaybreaks
\begin{subequations}
\label{N=2-ST-untwisted}
\begin{align}
    \mathcal{T} &= \partial \mathcal X^j \partial \bar{\mathcal X}^j +\frac{1}{2} \epsilon^{\alpha\beta} \epsilon^{kl} \Lambda^{\alpha,k} \partial \Lambda^{\beta,l}  - \mathcal{B} \partial \mathcal{C} - \frac{1}{2} \Sigma \partial \Gamma + \frac{1}{2}\Gamma\partial \Sigma \\ 
    & \quad -\partial \mathcal{X}^+ \partial \mathcal{X}^- -\frac{1}{2}\Upsilon \partial \Pi -\frac{1}{2} \Pi \partial \Upsilon  \ , \nonumber \\
    \mathcal{G}^+ &=  \partial \bar{\mathcal X}^j \Lambda^{+,j} + \Gamma \mathcal{B} + \partial \mathcal{X}^+ \Pi \ , \\
    \mathcal{G}^- &= - \epsilon^{jk}  \partial \mathcal{X}^j \Lambda^{-,k} - \Sigma \partial \mathcal{C} - \partial \mathcal{X}^- \Upsilon\ ,  \\
    \mathcal{J} &= - \tfrac{\epsilon^{ij}}{2} \Lambda^{+,i} \Lambda^{-,j} - \tfrac12 \Sigma\Gamma -\tfrac12 \Upsilon \Pi \,.
\end{align}
\end{subequations}
\endgroup
The generators are simply the sum of the $\mathcal{N}=2$ generators of $\mathbb{T}^4$ given in eqs.~\eqref{N=2-T4-untwisted} and the $\mathcal{N}=2$ generators of $\mathcal{A}_0$ given in eqs.~\eqref{auxiliary-n=2-space-time}. See eqs.~\eqref{N=2-conventions-untwisted} for our conventions on the OPEs of an $\mathcal{N}=2$ superconformal algebra.

Together with the additional generators
\begingroup
\allowdisplaybreaks
\begin{subequations}
\label{N=4-ST-untwisted}
\begin{align}
    \mathcal J^{++} &= \Lambda^{+,1} \Lambda^{+,2} e^{\mathcal Y+i\mathcal K} \ , \\
    \mathcal J^{--} &= \Lambda^{-,2} \Lambda^{-,1} e^{-i\mathcal K-\mathcal Y} \ , \\
    \widetilde{\mathcal G}^+ &= -\epsilon^{jk} \partial \mathcal X^j \Lambda^{+,k} e^{\mathcal Y+i\mathcal K} + \Lambda^{+,1}\Lambda^{+,2} e^{2\mathcal{Y}+i\mathcal{K}} \partial \Xi \partial \mathcal{C} - \Lambda^{+,1}\Lambda^{+,2} e^{\mathcal{Y}} \partial \mathcal{X}^- \ , \\
    \widetilde{ \mathcal G}^- &= - \partial \bar{\mathcal X}^j \Lambda^{-,j} e^{-i\mathcal{K}-\mathcal{Y}} - \Lambda^{-,2} \Lambda^{-,1} e^{-i\mathcal K-2\mathcal Y} \mathcal{N} \mathcal{B} + \Lambda^{-,2} \Lambda^{-,1} e^{-\mathcal Y} \partial \mathcal{X}^+ \,,
\end{align}
\end{subequations}
\endgroup
the fields \eqref{N=2-ST-untwisted} form an $\mathcal{N}=4$ superconformal algebra with $\mathtt c=6$. In eq.~\eqref{N=4-ST-untwisted}, we used the bosonization formulae given in eq.~\eqref{bosonized-auxiliary-ghosts}. Our conventions for the OPEs of the $\mathcal N=4$ superconformal algebra are summarized in eqs.~\eqref{N=4-conventions-untwisted}.

\section{Similarity transformation} \label{app:similarity-transformation}

In this appendix, we spell out our conventions for the $\mathcal N=2$ generators of the hybrid string, both before and after the similarity transformations. The discussion is slightly more general than the $\mathbb T^4$ case considered above. Namely, the formulae we review below remain valid if the $\mathbb T^4$ generators $F_{\mathbb T^4}$ are replaced by any set of generators $F_C \in \{ T_{C}, G^{+}_{C}, G^{-}_{C}, J_{C} \}$ closing into a topologically twisted $\mathcal N=2$ algebra with $\mathtt c=6$, provided that this sector (anti-)commutes with the $\mathfrak{psu}(1,1|2)_1$ fields, as well as with $\rho$ and $\sigma$.

For this reason, we consider the generators, 
\begin{subequations} \label{eq:n=2-free-before-appendix}
	\begin{align}
        T &= T_{\mathfrak{psu}}-\frac{1}{2} \left( (\partial \rho)^2 + (\partial \sigma)^2 \right) + \frac{3}{2} \partial^2 (\rho + i \sigma) + T_{C}\ , \label{eq:appendix-stress-tensor-n=2-similarity}\\
		G^{+} &=e^{-\rho} Q + e^{i \sigma} \left[ T_{\mathfrak{psu}}-\frac{1}{2} (\partial \rho+i \partial \sigma)^2 +\frac{1}{2} (\partial^2 \rho+i \partial^2 \sigma) \right] + G^{+}_{C} \ , \label{Gp-appendix}\\
		G^{-} &=e^{-i \sigma}+G^{-}_{C} \ , \\ 
		J &= \tfrac{1}{2}\partial(\rho+i\sigma) + J_{C} \,,
	\end{align}
\end{subequations}
where $Q$ is defined in eq.~\eqref{eq:q-free}. These are simply the
generators in eq.~\eqref{eq:n=2-free-before}, with the replacement
\begin{equation} \label{F-to-F}
    F_{\mathbb T^4} \to F_C \, , 
\end{equation}
highlighting that the discussion below applies more generally to any topologically twisted $\mathcal N=2$ sector with $\mathtt c=6$.\footnote{As mentioned in Section~\ref{sec:tensionless}, the normal ordering we use here is the conformal normal ordering, see \cite{Polchinski:2001tt, Gaberdiel:2022als}, but the same computation can be carried out in terms of the radial normal ordering. See footnote~\ref{footnote-normal}.} We refer to the generators \eqref{eq:n=2-free-before-appendix} as the $\mathcal{N}=2$ generators \textit{before} the similarity transformation.  

We now consider the operator
\begin{equation}
\label{R-C3}
    R = \oint \text d z \, e^{i\sigma} G^-_C \ ,
\end{equation}
and compute the generators after the similarity transformation by
applying
\begin{equation}
    F \longmapsto e^{R} F e^{-R}
\end{equation}
to each $F\in\{T,G^+,G^-,J\}$. For completeness, we discuss the four
transformations in turn.

\paragraph{$\boldsymbol T$.} The operator $R$ is the zero mode of the primary field $e^{i\sigma} G^-_C$ of weight $1$. Therefore, it is straightforward to conclude that the stress-tensor is not changed, i.e.\
\begin{equation}
    e^{R} T e^{-R} = T \ .
\end{equation}

\paragraph{$\boldsymbol G^{\boldsymbol +}$.} The integrand $e^{i\sigma} G^-_C$ in eq.~\eqref{R-C3} has non-singular OPEs with $e^{-\rho} Q$ and $e^{i\sigma} T_{\mathfrak{psu}}$. Nonetheless the other terms in \eqref{Gp-appendix} contribute and we find 
\begin{align}
	& \oint \text d z \, e^{i\sigma(z)} G^-_C(z) e^{i \sigma(w)} \left[-\frac{1}{2} (\partial \rho+i\partial \sigma)^2 +\frac{1}{2} (\partial^2 \rho+i\partial^2 \sigma) \right]=+e^{2i\sigma} G^-_C \ , \\[10pt]
	& \oint \text d z \, e^{i\sigma(z)} G^-_C(z) G^+_C(w)=\oint \text dz \, e^{i\sigma(z)} \left[ \frac{2}{(z-w)^3} - \frac{2 J_C(z)}{(z-w)^2} + \frac{T_C(w)}{z-w} \right] \ , \\[7pt]
    & \hspace{115pt} = \partial^2(e^{i\sigma})-2\partial(e^{i\sigma}J_C)+e^{i\sigma} T_C \ .
\end{align}
where we have used eq.~\eqref{eq:g+g--conventions} with $\mathtt c=6$ and swapped $G^+_C$ and $G^-_C$. Still the last term $e^{i\sigma}T_C$ has a singular OPE with $R$ (but note that the term $e^{i\sigma} J_C$ is regular with $R$) which gives
\begin{equation}
	\frac{1}{2!} \oint \text d z \, e^{i\sigma(z)} G^-_C(z) e^{i\sigma(w)} T_C(w)= - e^{2i\sigma} G^-_C \ ,
\end{equation}
where the extra minus sign comes from exchanging $e^{i\sigma}$ with $G^-_C$. In the end, we get
\begin{align}
	e^R G^{+} e^{-R} =e^{-\rho} Q + e^{i \sigma} &\left[ T_{\mathfrak{psu}}-\frac{1}{2} (\partial \rho+i\partial \sigma)^2 +\frac{1}{2} (\partial^2 \rho+i\partial^2 \sigma) + T_C \right] \\&+ G^{+}_C -2\partial(e^{i\sigma} J_C)+\partial^2(e^{i\sigma}) \ .\nonumber
\end{align}
Simplifying terms multiplying $e^{i\sigma}$ we get
\begin{equation}
	e^{i\sigma} \left[T_{\mathfrak{psu}}-\frac{1}{2} \left( (\partial \rho)^2 + (\partial \sigma)^2 \right) + \frac{3}{2} \partial^2 (\rho + i \sigma) + T_C\right] \ ,
\end{equation}
which is
\begin{equation}
	e^{i\sigma} T \ ,
\end{equation}
see eq.~\eqref{eq:appendix-stress-tensor-n=2-similarity}. Altogether, we have
\begin{equation}
	e^R G^+ e^{-R} = e^{-\rho} Q + e^{i \sigma} T + G^{+}_C - \partial(e^{i\sigma} [2J-\partial(i\sigma)]) \ .
\end{equation}

\paragraph{$\boldsymbol{G}^{\boldsymbol{-}}$.} It is easy to verify that
\begin{equation}
	\oint \text d z \, e^{i\sigma(z)} G^-_C(z) e^{-i\sigma(w)}=-G^-_C \ ,
\end{equation}
and so
\begin{equation}
e^R G^- e^{-R} = e^{-i\sigma} \ .
\end{equation}

\paragraph{$\boldsymbol J$.} One can see that
\begin{equation}
	J(z) e^{i\sigma(w)} G^-_C(w) \sim 0 \ ,
\end{equation}
and hence $J$ does not get modified.

\paragraph{Final result.} Assembling all the contributions, we find that the generators \eqref{eq:n=2-free-before-appendix} are mapped to 
\begin{subequations} \label{eq:n=2-free-after-similarity-appendix}
	\begin{align}
        T &= T_{\beta \gamma} + T_{p \theta} + T_{\rho} + T_{\sigma} + T_{C} \ , \\
		G^+ &= e^{-\rho} Q + e^{i\sigma} T - \partial[e^{i\sigma} (2J-i \partial\sigma)] +G^+_{C} \ , \\
		G^- &= e^{-i\sigma} \ , \\ 
		J &= \tfrac{1}{2}\partial(\rho+i\sigma)+J_{C} \ ,          
	\end{align}
\end{subequations}
where for simplicity we adopted the same notation for generators before and after the similarity transformation. We refer to the generators \eqref{eq:n=2-free-after-similarity-appendix} as the $\mathcal{N}=2$ generators \textit{after} the similarity transformation.

\section{Deformed vertex operators}
\label{app:vertex}

In this appendix, we study two examples of non-trivial physical vertex operators in the auxiliary  string theory of Section~\ref{sec:deforming}. Recall that the physical state conditions are given in \eqref{eq:lambda-physical-state-conditions}, with the $\mathcal N=4$ generators defined in eq.~\eqref{eq:n=2-lambda-after}. The two examples we consider are the deformations of the space-time ground state vertex operator and of the $\mathcal R$-symmetry generators. Since the calculations are similar in the two cases, we discuss them in parallel. As explained in Section~\ref{sec:vertex}, we consider the ansatz 
\begin{equation} \label{eq:vertex-appendix-final}
    \mathcal{V}_{\mu} = \Psi^L \Psi^R e^{ip\gamma_\mu- i \bar{p} \bar \gamma_\mu} \ ,
\end{equation}
where $\gamma_\mu$ and $\bar \gamma_\mu$ are defined in eq.~\eqref{path-integral-change-variables}, and determine $\Psi^L$ and $\Psi^R$ such that $\mathcal V_\mu$ is physical. Let us introduce the notation
\begin{equation}
    \Psi^L = \Psi^L_0 + \Psi^L_1 + \Psi^L_2 \ , \qquad 
    \Phi^L_j := \Psi^L_j e^{ip\gamma_\mu- i \bar{p} \bar \gamma_\mu} \ , \quad j = 0, 1, 2 \ .  
\end{equation}
We will see that with the expressions given below for $\Psi^L_0$, $\Psi^L_1$ and $\Psi^L_2$ (and their anti-holomorphic analogues) the vertex operator \eqref{eq:vertex-appendix-final} is physical. We begin with
\begin{equation} \label{eq:psil0}
    \Psi^L_0 = \mathcal{E} \, e^{if_1-if_2} (\partial \gamma)^{-m} e^{2\rho+i\sigma+iH} e^{u+i\chi} \ ,
\end{equation}
where $\mathcal{E}\in\{1,K^3,K^+,K^-\}$. This is the undeformed tensionless vertex operator, with the value of $m$ fixed by the deformed mass-shell condition to be
\begin{equation} \label{m-m0}
    m = m_0 + \mu^2 p \bar p \ ,
\end{equation}
where $m_0$ is the undeformed space-time weight.

We observe that $G^+_0$, as defined in \eqref{eq:n=2-lambda-after}, does not annihilate $\Phi_0^L$. Carefully taking into account cocycle factors and the minus signs arising from fermionic exchanges we find   
\begin{align}
    G^+_0 \Phi_0^L &= (e^{-\rho} Q_\mu)_0 \Phi^L_0  +  (e^{i\chi} \partial X^+_\mu)_0 \Phi^L_0 \\
    & = i \mu^2 \bar{p} \, \mathcal{E} \, e^{ip\gamma_\mu- i \bar{p} \bar \gamma_\mu} (\partial \gamma)^{-m} e^{\rho+i\sigma+iH} e^{u+i\chi}  \nonumber \\
    & \quad + \mu \, m \, \mathcal{E} \, e^{if_1-if_2} e^{ip\gamma_\mu- i \bar{p} \bar \gamma_\mu} (\partial \gamma)^{-m-1} e^{2\rho+i\sigma+iH} e^{u+2i\chi} \ . \label{G^+_0-on-Phi^L_0}
\end{align}
In order to compensate the non-vanishing action of $G^+_0$ on $\Phi_0^L$, we consider\footnote{Although, for convenience, we express the vertex operators in terms of bosonized fields, we have checked that each of them can be rewritten without relying on bosonization. In particular, $e^{2u} \partial(e^{iv}) = \beta^{\text{(g)}}_0 \delta(\gamma^{\text{(g)}})$ and $e^{2i\chi} = (\partial \pi) \pi$. This shows, in particular, that $\Psi^L_j$ lies in the small Hilbert space of the $\beta\gamma$ system $(\beta^{\text{(g)}},\gamma^{\text{(g)}})$.} 
\begin{equation}
    \Psi^L_1 = x_1 \, \mathcal{E} \, e^{if_1-if_2} (\partial \gamma)^{-m-1} e^{2\rho+i\sigma+iH} e^{2u+2i\chi} \partial(e^{iv}) c^{\text{(g)}} \ ,
\end{equation}
where $x_1$ is a yet-to-be determined complex coefficient. Let us compute the action of $G^+_0$ on $\Phi^L_1$. Noticing that 
\begin{equation}
    (e^{i\chi} \partial X^+_\mu)_0 \Phi^L_1 = 0 \ , 
\end{equation}
we find 
\begin{align}
    G^+_0 \Phi_1^L &= (e^{-\rho} Q_\mu)_0 \Phi^L_1 +  (\gamma^{\text{(g)}} b^{\text{(g)}})_0 \Phi^L_1 \\
    & = i\mu^2 \bar{p} \, x_1 \, \mathcal{E} \, e^{ip\gamma_\mu- i \bar{p} \bar \gamma_\mu} (\partial \gamma)^{-m-1} e^{\rho+i\sigma+iH} e^{2u+2i\chi} \partial(e^{iv}) c^{\text{(g)}} \nonumber \\
    & \quad - x_1 \, \mathcal{E} \, e^{if_1-if_2} e^{ip\gamma_\mu- i \bar{p} \bar \gamma_\mu} (\partial \gamma)^{-m-1} e^{2\rho+i\sigma+iH} e^{u+2i\chi} \ . \label{G^+_0-on-Phi^L_1}
\end{align}
We observe that for 
\begin{equation} \label{x1}
    x_1 = \mu \, m
\end{equation}
the second term in eq.~\eqref{G^+_0-on-Phi^L_0} cancels against the second term in eq.~\eqref{G^+_0-on-Phi^L_1}, 
\begin{equation}
    (e^{i\chi} \partial X^+_\mu)_0 \Phi^L_0 + (\gamma^{\text{(g)}} b^{\text{(g)}})_0 \Phi^L_1 = 0 \ . 
\end{equation}
To sum up, fixing $x_1$ as in eq.~\eqref{x1}, we are left with the non-vanishing terms 
\begin{align}
    G^+_0 (\Phi^L_0 + \Phi^L_1) &= (e^{-\rho} Q_\mu)_0 \Phi^L_0 + (e^{-\rho} Q_\mu)_0 \Phi^L_1 \\
    & = i \mu^2 \bar{p} \, \mathcal{E} \, e^{ip\gamma_\mu- i \bar{p} \bar \gamma_\mu} (\partial \gamma)^{-m} e^{\rho+i\sigma+iH} e^{u+i\chi}  \nonumber \\
    & \quad + i\mu^3 m \, \bar{p} \, \mathcal{E} \, e^{ip\gamma_\mu- i \bar{p} \bar \gamma_\mu} (\partial \gamma)^{-m-1} e^{\rho+i\sigma+iH} e^{2u+2i\chi} \partial(e^{iv}) c^{\text{(g)}} \ .  \label{G+0-0-and-1}
\end{align}
In order to cancel these contributions, we furthermore introduce 
\begin{equation}
    \Psi_2^L = x_2 \, \mathcal{E} \, (\partial \gamma)^{-m} e^{\rho+i\sigma+iH} e^{2u+i\chi} \partial(e^{iv}) c^{\text{(g)}} \ ,
\end{equation}
where $x_2$ is an additional complex parameter which we are going to fix below. It is straightforward to see that
\begin{equation}
    (e^{-\rho} Q_\mu)_0 \Phi^L_2 = 0 \ ,
\end{equation}
and hence we obtain 
\begin{align}
    G^+_0 \Phi^L_2 &= (e^{i\chi} \partial X^+_\mu)_0 \Phi^L_2 + (\gamma^{\text{(g)}} b^{\text{(g)}})_0 \Phi^L_2 \\
    &= x_2 \, \mu \, m \, \mathcal{E} \, e^{ip\gamma_\mu- i \bar{p} \bar \gamma_\mu} (\partial \gamma)^{-m-1} e^{\rho+i\sigma+iH} e^{2u+2i\chi} \partial(e^{iv}) c^{\text{(g)}} \nonumber \\
    & \quad + x_2 \, \mathcal{E} \, e^{ip\gamma_\mu- i \bar{p} \bar \gamma_\mu} (\partial \gamma)^{-m} e^{\rho+i\sigma+iH} e^{u+i\chi} \ .  \label{G+0-on-PhiL2}
\end{align}
We notice that the choice  
\begin{equation}
    x_2 =- i\mu^2 \bar{p} \ ,
\end{equation}
simultaneously makes the second term in the right-hand-side of eq.~\eqref{G+0-0-and-1} cancel against the first term in the right-hand-side of eq.~\eqref{G+0-on-PhiL2}, 
\begin{equation}
    (e^{-\rho} Q_\mu)_0 \Phi^L_1 + (e^{i\chi} \partial X^+_\mu)_0 \Phi^L_2 = 0 \ , 
\end{equation}
and the first term in the right-hand-side of eq.~\eqref{G+0-0-and-1} cancel against the second term in the right-hand-side of eq.~\eqref{G+0-on-PhiL2}, 
\begin{equation}
    (e^{-\rho} Q_\mu)_0 \Phi^L_0 + (\gamma^{\text{(g)}}b^{\text{(g)}})_0 \Phi^L_2 = 0 \ .
\end{equation}
To summarize, for
\begin{equation}
    \begin{aligned}
        \Psi^L & = \Psi^L_0 + \Psi^L_1 + \Psi^L_2 \ ,   \\
        \Psi^L_0 &= \mathcal{E} \, e^{if_1-if_2} (\partial \gamma)^{-m} e^{2\rho+i\sigma+iH} e^{u+i\chi} \ , \\
        \Psi^L_1 &= \mu\, m \, \mathcal{E} \, e^{if_1-if_2} (\partial \gamma)^{-m-1} e^{2\rho+i\sigma+iH} e^{2u+2i\chi} \partial(e^{iv}) c^{\text{(g)}} \ , \\
        \Psi^L_2 &= -i \mu^2 \bar{p} \, \mathcal{E} \,(\partial \gamma)^{-m} e^{\rho+i\sigma+iH} e^{2u+i\chi} \partial(e^{iv}) c^{\text{(g)}}\ ,
    \end{aligned}
\end{equation}
and 
\begin{equation}
    \begin{aligned}
        \Psi^R &= \Psi^R_0 + \Psi^R_1 + \Psi^R_2 \ , \\
        \Psi^R_0 &= \overline{\mathcal{E}} \, e^{i\bar f_1-i\bar f_2} (\bar\partial \bar\gamma)^{-m} e^{2\bar \rho+i\bar \sigma+i\bar H} e^{\bar u+i\bar \chi} \ , \\
        \Psi^R_1 & = - \mu \, m \, \overline{\mathcal{E}} \, e^{i \bar f_1-i\bar f_2} (\bar\partial \bar\gamma)^{-m-1} e^{2\bar\rho+i\bar\sigma+i\bar H} e^{2\bar u+2i\bar \chi} \bar \partial(e^{i\bar v}) \bar{c}^{\text{(g)}}  \ , \\
        \Psi^R_2 &= i \mu^2 p \, \overline{\mathcal{E}} \, (\bar \partial \bar \gamma)^{-m} e^{\bar\rho+i\bar \sigma+i\bar H} e^{2\bar u+i\bar \chi} \bar \partial(e^{i\bar v}) \bar{c}^{\text{(g)}} \ , 
    \end{aligned}
\end{equation}
we have shown that the vertex operator \eqref{eq:vertex-appendix-final} is  annihilated by $G^+_0$.

Notice also that \eqref{eq:vertex-appendix-final} satisfies the conditions \eqref{W0-condition}. This follows from the fact that the OPE of $\beta$ with $(\partial \gamma)$ (and similarly for the right-moving fields) has no single pole. Moreover, the dependence on $\gamma$ and $\bar\gamma$ also enters through the deformed combinations $\gamma_\mu$ and $\bar\gamma_\mu$, which are chosen so as to be neutral under the corresponding zero mode constraints.

Moreover, one can verify that \eqref{eq:vertex-appendix-final} has $J_0=\frac{1}{2}$ eigenvalue and satisfies $T_0=0$. The first condition is relatively non-trivial: the total $J_0$ eigenvalue receives contributions from different fields in the various terms entering \eqref{eq:vertex-appendix-final}. The second condition amounts to the mass-shell condition and fixes $m$ as in eq.~\eqref{m-m0}.  

Therefore, all the physical state conditions in \eqref{eq:physical-state-conditions} are satisfied, and we conclude that \eqref{eq:vertex-appendix-final} is physical.

\section{\texorpdfstring{Two-dimensional $\boldsymbol{\mathcal{N}=(4,4)}$ supersymmetry algebra}{Two-dimensional N=(4,4) supersymmetry algebra}}
\label{app:super}

In this appendix, we briefly review the $2$d $\mathcal{N}=(4,4)$ supersymmetry algebra \cite{Ivanov:1994er,Hull:2008de} in both Euclidean and Lorentzian signature. As discussed in \cite{Hull:2008de}, these two signatures are both obtained by considering the complexification of the super Lie algebra and imposing an appropriate reality condition. For this reason, we first discuss the complexified super Lie algebra and later write the reality conditions that must be imposed. We will refer to these algebras as the $2$d $\mathcal{N}=(4,4)$ Euclidean or Poincar\'{e} supersymmetry algebras.

Let us begin by listing the generators. Following Section~2 of both \cite{Ivanov:1994er} and \cite{Hull:2008de}, this super Lie algebra contains the odd generators $\{\mathcal{Q}^{\alpha\beta},\overline{\mathcal{Q}}^{\alpha\beta}\}$ with $\alpha,\beta\in\{+,-\}$, and the bosonic generators $\{\mathcal{L},\mathcal{P},\overline{\mathcal{P}},\mathcal{K}^a,\overline{\mathcal{K}}^a\}$. All  generators except $\mathcal{L}$ have left- and right-moving counterparts, which we denote by placing a bar over the corresponding left-moving generator. Depending on the space-time signature, $\mathcal{L}$ generates either the Euclidean rotation group $\text{SO}(2)$ or the Lorentz group $\text{SO}(1,1)$. The generators $\mathcal{K}^a$ and $\overline{\mathcal{K}}^a$ form two commuting $\mathfrak{su}(2)$ and $\overline{\mathfrak{su}(2)}$ respectively,
\begin{subequations}
    \begin{equation}
        [\mathcal{K}^+,\mathcal{K}^-]=2\mathcal{K}^3 \ , \qquad [\mathcal{K}^3,\mathcal{K}^\pm]= \pm \mathcal{K}^\pm \ ,
    \end{equation}
    \begin{equation}
        [\overline{\mathcal{K}}^+,\overline{\mathcal{K}}^-]=2\overline{\mathcal{K}}^3 \ , \qquad [\overline{\mathcal{K}}^3,\overline{\mathcal{K}}^\pm]=\pm \overline{\mathcal{K}}^\pm \ ,
    \end{equation}
\end{subequations}
where we only list non-zero (anti-)commutation relations here and throughout the rest of this appendix. The supercurrents $\mathcal{Q}^{\alpha\beta}$ and $\overline{\mathcal{Q}}^{\alpha\beta}$ form a doublet with respect to these $\mathfrak{su}(2)$ and $\overline{\mathfrak{su}(2)}$, specified by the index $\alpha$, i.e.\
\begin{subequations}
    \begin{equation}
        [\mathcal{K}^\pm,\mathcal{Q}^{\mp\beta}]=\mathcal{Q}^{\pm\beta} \ , \qquad [\mathcal{K}^3,\mathcal{Q}^{\alpha\beta}]=\frac{\alpha}{2}\mathcal{Q}^{\alpha\beta} \ ,
    \end{equation}
    \begin{equation}
        [\overline{\mathcal{K}}^\pm,\overline{\mathcal{Q}}^{\mp\beta}]=\overline{\mathcal{Q}}^{\pm\beta} \ , \qquad [\overline{\mathcal{K}}^3,\overline{\mathcal{Q}}^{\alpha\beta}]=\frac{\alpha}{2}\overline{\mathcal{Q}}^{\alpha\beta} \ .
    \end{equation}   
\end{subequations}
The other index $\beta$ denotes a doublet representation with respect to two outer $\mathfrak{su}(2)_{\text{outer}}$ and $\overline{\mathfrak{su}(2)}_{\text{outer}}$, respectively. The algebra $\mathfrak{su}(2) \times \mathfrak{su}(2)_{\text{outer}}$ forms $\mathfrak{so}(4)$ rotating the left-moving odd generators of $\mathcal{N}=(4,4)$ into each other. The same holds for the right-moving generators, so that altogether we obtain the $\mathcal{R}$-symmetry $\mathfrak{so}(4)\times \overline{\mathfrak{so}(4)}$.

The generators $\mathcal{P}$ and $\overline{\mathcal{P}}$ are the left- and right-moving translations, which satisfy
\begin{equation}
    \{\mathcal{Q}^{\alpha\beta},\mathcal{Q}^{\nu\eta}\}= - \epsilon^{\alpha\nu} \epsilon^{\beta\eta} \mathcal{P}\ , \qquad \{\overline{\mathcal{Q}}^{\alpha\beta},\overline{\mathcal{Q}}^{\nu\eta}\}= - \epsilon^{\alpha\nu} \epsilon^{\beta\eta} \overline{\mathcal{P}} \ .
\end{equation}
The generator $\mathcal{L}$ is inherently neither left-moving nor right-moving. The supercurrents transform under $\mathcal{L}$ as
\begin{equation}
    [\mathcal{L},\mathcal{Q}^{\alpha\beta}] = \frac{1}{2}\mathcal{Q}^{\alpha\beta} \ , \qquad [\mathcal{L},\overline{\mathcal{Q}}^{\alpha\beta}] = - \frac{1}{2}\overline{\mathcal{Q}}^{\alpha\beta} \ ,
\end{equation}
and
\begin{equation}
    [\mathcal{L},\mathcal{P}] = \mathcal{P}\ , \quad [\mathcal{L},\overline{\mathcal{P}}]=-\overline{\mathcal{P}} \ .
\end{equation}
The $2$d $\mathcal{N}=(4,4)$ Euclidean supersymmetry algebra is obtained by imposing\footnote{We are using the conventions that $[A,B]^\dagger = - [A^\dagger,B^\dagger]$ and $\{A,B\}^\dagger=\{A^\dagger,B^\dagger\}$.}
\begin{equation} \label{eq:euclidean-reality-condition}
    \mathcal{L}^{\dagger} = \mathcal{L} \ , \qquad (\mathcal{Q}^{\alpha\beta})^{\dagger} = \overline{\mathcal{Q}}^{\alpha\beta} \ , \qquad (\mathcal{K}^a)^{\dagger} = -\overline{\mathcal{K}}^a \ , \qquad \mathcal{P}^{\dagger} = \overline{\mathcal{P}} \ ,
\end{equation}
where $\dagger$ denotes conjugation. On the other hand, the two-dimensional $\mathcal{N}=(4,4)$ Poincar\'{e} supersymmetry algebra is obtained by setting
\begin{subequations}
\begin{equation}
    \mathcal{L}^{\dagger} = - \mathcal{L} \ ,
\end{equation}
\begin{equation}
    (\mathcal{Q}^{\alpha\beta})^{\dagger} = \mathcal{Q}^{\alpha\beta} \ , \qquad (\mathcal{K}^a)^{\dagger} = -\mathcal{K}^a \ , \qquad \mathcal{P}^{\dagger} = \mathcal{P}\ ,
\end{equation}
\end{subequations}
and similarly for the barred fields.

\section{Consistency of correlation functions} 
\label{app:corr}

In this appendix, following \cite{Berkovits:1994vy}, we review the definition of correlators in Berkovits-Vafa topological string theory and show that the correlation functions of the string theory constructed in Section~\ref{sec:deforming} are well-defined. We separate the discussion into tree-level and higher genus correlation functions.

\subsection{Tree-level correlators}
\label{app:tree-level}

We consider the correlation function \eqref{eq:corr-lambda} and show that BRST exact terms decouple and that the result is independent of the distribution of picture number. The discussion for \eqref{eq:corr} is almost identical.

\paragraph{Decoupling of BRST exact states.} We begin with the decoupling of BRST-exact states. Consider
\begin{equation} \label{eq:tree-level-brst-exactness-arg}
    V_1 = G^+_0 \widetilde{G}^+_0 \Lambda \ , \qquad \widetilde{V}_1 = G^+_0\Lambda \ ,
\end{equation}
where the second relation is a solution of eq.~\eqref{eq:vtilde}. We will show that the corresponding correlation function vanishes.

The first ingredient is that $G^+_0$, with $G^+$ given in eq.~\eqref{G+-after-lambda}, commutes with the zero mode of the screening operator \eqref{eq:z-def},
\begin{equation} \label{eq:gp-z-check}
    [G^+_0,Z_0] = 0 \ . 
\end{equation}
To verify this explicitly, we bosonize the fields as in eq.~\eqref{eq:bosonizations-added-ghosts}, together with
\begin{equation} \label{eq:bos-pi}
    c^{\text{(g)}} = e^{i\varphi} \ ,
\end{equation}
where $i\varphi$ is a boson with background charge $1$, see eq.~\eqref{eq:background-charge-definition}. The commutator \eqref{eq:gp-z-check} follows term by term. The $e^{-\rho}Q_\mu$ term has a regular OPE with $Z$. The zero mode of $e^{i\sigma}T$ gives a total derivative, which vanishes after integration. It remains only to check the term $\gamma^{\text{(g)}}b^{\text{(g)}}$. Using the bosonizations above, one finds
\begin{equation} \label{eq:ope-check-corr}
    \gamma^{(g)} b^{\text{(g)}} = e^{-iv-u-i\varphi} \ , \qquad c^{\text{(g)}} \partial \gamma^{(g)} \delta(\gamma^{(g)}) = e^{i\varphi-iv} \ . 
\end{equation}
These two operators have a regular OPE, and therefore the $\gamma^{\text{(g)}}b^{\text{(g)}}$ term also commutes with $Z_0$. This proves eq.~\eqref{eq:gp-z-check}. Using 
\begin{equation}
    c^{\text{(g)}} \, \partial \gamma^{(g)} \delta(\gamma^{(g)}) = e^{i\varphi-iv} \ ,
\end{equation}
one similarly verifies that
\begin{equation}
    [\widetilde{G}^+_0 , Z_0 ] = 0 \ ,
\end{equation}
where $\widetilde G^+$ is given in eq.~\eqref{eq:lambda-gtildep}. In this case the commutator vanishes because the relevant bosonized expressions involve mutually independent fields and therefore have regular OPEs.

Similarly, one can check that
\begin{equation}
    [G^+_0,D_0]=[\widetilde{G}^+_0,D_0] = 0 \ ,
\end{equation}
see eq.~\eqref{eq:def-d}. After inserting $\widetilde V_1$ as given in eq.~\eqref{eq:tree-level-brst-exactness-arg} into eq.~\eqref{eq:corr-lambda}, the action of $G^+_0$ can be represented by a contour integral around $\Lambda$. The contour may then be deformed away from $\Lambda$ and moved around the remaining insertions. Since $G^+_0$ annihilates all fields appearing in eq.~\eqref{eq:corr-lambda}, including the screening operators, the correlation function vanishes.

Now let us assume that
\begin{equation}
    V_j = G^+_0 \widetilde{G}^+_0 \Lambda \ ,
\end{equation}
for some $j\geq 2$, and show that the corresponding correlation function vanishes. Deforming the contour for $\widetilde{G}^+_0$ around
the other insertions annihilates all fields except $\widetilde V_1$, which is mapped to
\begin{equation} \label{eq:vtilde1-transform}
    \widetilde{V}_1 \mapsto V_1 \ ,
\end{equation}
see \eqref{eq:vtilde}. One can then deform the remaining $G^+_0$ contour away from $V_j$; it annihilates all remaining insertions, and the correlator therefore vanishes. Thus BRST-exact fields decouple.

\paragraph{Picture number.} Before we continue, let us be more accurate on how the picture number is defined. For a vertex operator $V$, the picture number is defined as the eigenvalue under
\begin{equation} \label{eq:picture-accurate}
    P^\prime = (\partial \rho)_0 - \frac12 (p_a \theta^a)_0 \ , 
\end{equation}
and similarly for $\overline P^\prime$. The reason for this particular choice is that only with the inclusion of both terms, $G^+$ has picture number zero. The tree-level correlation function is non-zero provided that
\begin{equation}
    \sum_{j} P^\prime_j = \sum_{j} \bar P^\prime_j = -3 \ .
\end{equation}
As the background charges of $if_1$ and $(-if_2)$ are both $1$, see eq.~\eqref{eq:pa-thetaa-bosonization} and Table~\ref{table:tensionless-background-charges}, this translates into the following condition
\begin{equation}
    \sum_{j} P_j = \sum_{j} \bar P_j = -4 \ ,
\end{equation}
where $P_j$ are defined as the eigenvalue of $(\partial \rho)_0$ \cite{Gaberdiel:2021njm}.

\paragraph{Independence of the distribution of picture changing.} It is also straightforward to see that the correlator is independent of how picture number is distributed among the vertex operators. To show this consider any
$j\neq 1$ and write
\begin{equation}
    V_j = \widetilde{G}^+_0 \widetilde{V}_j \ ,
\end{equation}
see eq.~\eqref{eq:vtilde}. Such a solution always exists because the cohomology of $\widetilde{G}^+_0$ is trivial. Deforming the
$\widetilde{G}^+_0$ contour around the remaining insertions annihilates all fields except $\widetilde V_1$, which is mapped to $V_1$ as in eq.~\eqref{eq:vtilde1-transform}. Thus the tilded operator can be moved from one vertex operator to another without changing the correlator.

We now show that the same is true for the picture raising operator. Suppose that the correlator contains $P_+ V_k$, with $P_+$ defined in
eq.~\eqref{eq:picture raising-def}, and that we want to move $P_+$ from $V_k$ to another vertex operator $V_j$, with $j\neq k$. By the argument
above, we may first place the tilde on $V_j$. The definition of $P_+$ contains a $G^+_0$ contour. Deforming this contour away from $\widetilde V_k$ and around the other insertions, it annihilates all fields except $\widetilde V_j$. The latter is then mapped to $P_+ V_j$, see eqs.~\eqref{eq:vtilde} and \eqref{eq:picture raising-def}. Repeating this argument shows that the correlator is independent of the distribution of picture number among the vertex operators, provided the total picture condition \eqref{picture-sum} is satisfied.

\subsection{Higher genus correlators}

We denote the worldsheet genus by $g$. In the formulae below, we do not write the screening operators $D\overline D$ explicitly. For $g=1$, we
define the correlation functions as
\begin{subequations} \label{eq:app-corr-higher-genus}
\begin{equation} \label{eq:n=2-amplitudes-torus}
    \left< \int_{\mathcal{F}} \frac{d^2\tau}{\tau_2} \prod_{j=1}^{L} P_+(u_j) \left( \int J \right) \prod_{j=1}^n \int G^-_{-1} V_j \right> \ ,
\end{equation}
whereas for $g\geq 2$ we define
\begin{equation} \label{eq:n=2-amplitudes-higher-genus}
    \left< \prod_{j=1}^{3g-3} \int_{\mathcal M_g} G^-(\mu_j) 
    \prod_{j=1}^{2g-2+L} P_+(u_j) \left(\int \widetilde{G}^+ \right)^{g-1} \left( \int b^{\text{(g)}} \right)^{g-1} \left( \int J \right) \prod_{j=1}^n \int G^-_{-1} V_j \right> \ ,
\end{equation}
\end{subequations}
Let us clarify the ingredients entering these expressions. First, $\mathcal F$ denotes the fundamental domain of the torus. For
$g\geq 2$, the $3g-3$ insertions of $G^-$ are contracted with Beltrami differentials $\mu_j$ and integrated over the (even) moduli space $\mathcal M_g$ of genus-$g$ Riemann surfaces. Second, since we are considering a superstring theory, $2g-2$ picture-raising operators
arise from the integration over supermoduli space; see \cite{Witten:2012bh}.\footnote{We thank Bob Knighton for related
discussions.} In our conventions these are
\begin{equation} \label{eq:picture raising-def-p}
    P_+(u) = V(G^+_0 \xi_0 \ket{0},u) \ ,
\end{equation}
with
\begin{equation} \label{eq:xi-lambda-app}
    \xi = e^{-\rho-iH-u-i\chi} \ .
\end{equation}
Equivalently, the same number of insertions is required by the Berkovits-Vafa $\mathcal N=4$ topological-string prescription. We have
rewritten this prescription in terms of $\mathcal N=2$ fields and picture-raising operators; see also the last equation on page~3 of \cite{Berkovits:1993xq}. 

The $g-1$ insertions of $\widetilde G^+$ are required by the background-charge conservation associated with $\widetilde G^+$.
Similarly, we insert $g-1$ zero modes of $b^{\text{(g)}}$ in order to balance the background charge of the auxiliary fields appearing in
eq.~\eqref{auxiliary-n=2}. Fourth, following \cite{Berkovits:1994vy}, the insertion of $\int J$ is required for a consistent definition of the correlators; in particular, it plays a role in the decoupling of BRST-exact states discussed below. Finally, we assume that the picture numbers of the vertex operators obey
\begin{equation} \label{eq:picture-sum-higher-genus}
    \sum_{j=1}^{n} P_j = -2 L \ .
\end{equation}
As mentioned above, we have not written explicitly the screening operators relevant for higher-genus correlators of the tensionless string.

\paragraph{Decoupling of BRST exact states.} For a fixed $j\in\{1,\cdots,n\}$, suppose that
\begin{equation} \label{eq:v-higher-genus-brst-def}
    V_j = G^+_0 \widetilde{G}^+_0 \phi_j \ .
\end{equation}
We first deform the $\widetilde G^+_0$ contour around the remaining insertions. The picture-raising operators are annihilated by both
$G^+_0$ and $\widetilde G^+_0$. Indeed, $P_+(u)$, as defined in eq.~\eqref{eq:picture raising-def-p}, is annihilated by $G^+_0$ because $(G^+_0)^2=0$. It is also annihilated by $\widetilde G^+_0$, since 
\begin{equation}
    \widetilde G^+_0 P_+(u) = V(\widetilde{G}^+_0 G^+_0 \xi_0 \ket{0},u) = -V(G^+_0 \widetilde{G}^+_0 \xi_0 \ket{0},u) = - V(G^+_0 \ket{0},u) = 0 \ ,
\end{equation}
where we used 
\begin{equation}
    \{\widetilde{G}^+_0,\xi_0\}=1 \ ,
\end{equation}
see eqs.~\eqref{eq:lambda-gtildep} and \eqref{eq:xi-lambda}, together with the fact that $G^+$ is a primary field of conformal weight one. It then follows that the only non-vanishing contribution from deforming the
$\widetilde G^+_0$ contour is
\begin{equation} \label{eq:higher-genus-app-change}
    \int J \mapsto \int \widetilde{G}^+ \ .
\end{equation}
We then deform the remaining $G^+_0$ contour in eq.~\eqref{eq:v-higher-genus-brst-def} and let it act on the insertions in eqs.~\eqref{eq:app-corr-higher-genus}, with the replacement \eqref{eq:higher-genus-app-change} understood. For the torus ($g=1$), the result vanishes, and hence the BRST exact state decouples. For $g\geq2$, the contour annihilates the additional insertions as well, since $G^+$ has a regular OPE with $b^{\text{(g)}}$. Thus BRST exact states decouple.\footnote{The contour also annihilates the insertions involving $G^-$ up to boundary terms, which vanish after integration; see \cite{Berkovits:1994vy}.}

\paragraph{Independence of the distribution of picture changing.}In order to see that the correlator is independent of how picture number is distributed among the vertex operators, it is enough to show that replacing a picture-changing insertion by its worldsheet derivative gives a vanishing contribution. This follows from the fact that the derivative of the picture-changing operator is BRST exact. Indeed,
\begin{equation} \label{eq:derivative-picture}
    \partial_{u} P_+(u) = V(L_{-1} G^+_0 \xi_0 \ket{0},u) = V(G^+_0 \xi_{-1} \ket{0},u) = V(G^+_0 \widetilde{G}^+_0 \xi_0 \xi_{-1} \ket{0},u) \ .
\end{equation}

\end{document}